\documentclass[lettersize,journal]{IEEEtran} %
\usepackage{amsmath,amsfonts,amssymb}
\usepackage{algorithm}
\usepackage{array}
\usepackage[caption=false,font=normalsize,labelfont=sf,textfont=sf]{subfig}
\usepackage{textcomp}
\usepackage{stfloats}
\usepackage{url}
\usepackage{verbatim}
\usepackage{graphicx}
\usepackage{cite}

\usepackage{pgfplots}
\usepgfplotslibrary{fillbetween}
\usepgfplotslibrary{patchplots}
\pgfplotsset{compat=1.16}

\usepackage{tikz}
\usetikzlibrary{patterns,calc,intersections}
\usetikzlibrary{arrows,shapes.misc,chains,scopes,shadows,arrows.meta}
\usetikzlibrary{positioning,decorations.pathreplacing,calc}
\usetikzlibrary{decorations.text,spy}
\usetikzlibrary{positioning,quotes,calc,fit,shapes,shadows,arrows}
\usepgflibrary{arrows}

\usepackage{pgfplotstable}
\usepackage{verbatim,enumitem}
\usepackage{tabularx}

\usepackage{trfsigns} %
\usepackage{nccmath}
\usepackage{nicefrac}
\usepackage{siunitx}
\usepackage{booktabs}
\usepackage{dsfont}
\usepackage{multirow}
\usepackage{pgfmath}
\usepackage{stmaryrd}
\usepackage{algpseudocode}
\usepackage{empheq}
\usepackage{eqparbox}
\usepackage{bbm}

\usepackage{hyperref}
\hypersetup{unicode=true,pdfborder = {2 0 0.1}}

\newcommand\IfTwoCol[1]{\if@twocolumn\IfTwoColAUX{#1}\fi}
\newcommand\IfTwoColAUX[1]{#1}

\newcommand\IfOneCol[1]{\if@twocolumn\else\IfOneColAUX{#1}\fi}
\newcommand\IfOneColAUX[1]{#1}
\DeclareMathOperator{\diag}{diag}

\DeclareMathOperator{\trace}{tr}

\newcommand{\bm}[1]{\ensuremath{\mathbf{#1}}}

\newcommand{\chig}[5]{f_{\chi,#3}\left(#1; \, \,#2, \,#4, \,#5\right)}

\DeclareMathOperator*{\rank}{rank}
\DeclareMathOperator*{\tr}{tr}

\newcommand*{\tran}{^{\mkern-1.5mu\mathsf{T}}}
\newcommand*{\herm}{^{\mkern-1.5mu\mathsf{H}}}

\DeclareMathOperator*{\argmin}{argmin}

\DeclarePairedDelimiter\idxset{\llbracket}{\rrbracket}
\DeclareMathOperator*{\sinc}{sinc}
\DeclareMathOperator*{\Proj}{\mathcal{P}}
\DeclareMathOperator*{\CProj}{\mathcal{CP}}

\DeclareMathOperator*{\rect}{rect}
\DeclareMathOperator*{\blkdiag}{blkdiag}
\DeclareMathOperator*{\var}{var}

\newcommand{\modone}[1]{\ensuremath{#1\,\mathrm{mod}\,1}}

\DeclareMathOperator{\E}{\mathbb{E}}
\newcommand{\dimC}[1]{\ensuremath{\in \mathbb{C}^{#1}}}
\newcommand{\dimR}[1]{\ensuremath{\in \mathbb{R}^{#1}}}

\newcommand{\mident}[1]{\ensuremath{\mathrm{\textbf{I}}_{#1}}}
\newcommand{\mnull}[1]{\ensuremath{\mathrm{\textbf{0}}_{#1}}}

\tikzstyle{input} =  [coordinate]
\tikzstyle{output} = [coordinate]
\pgfmathsetmacro{\blockwidth}{30}
\pgfmathsetmacro{\blockheight}{17}
\tikzstyle{comblock} = [thick,black,fill=white,draw=black, rectangle,minimum height=\blockheight pt,minimum width=\blockwidth pt]
\tikzstyle{comtriag} = [regular polygon, regular polygon sides=3,
              draw=black, black,  text width=0.90em,
              inner sep=0.9mm, outer sep=0mm,
              shape border rotate=-90]

\tikzstyle{sum} = [thick,draw, circle,inner sep=-1pt,outer sep=0pt,font=\normalsize]

\tikzset{block/.style={draw,very thick,text width=1.5cm,minimum height=3cm,align=center},
         line/.style={-latex}}
\tikzset{blockV/.style={draw,very thick,text width=2cm,minimum height=2cm, minimum width=4cm,align=center},
         line/.style={-latex}}
\tikzset{blockExt/.style={draw,very thick,minimum height=1cm, minimum width=1cm,align=center},
         line/.style={-latex}}

\algnewcommand\algorithmicforeach{\textbf{for each}}
\algdef{S}[FOR]{ForEach}[1]{\algorithmicforeach\ #1\ \algorithmicdo}

\newdimen{\algindent}
\algnewcommand\LeftComment[2]{%
\hspace{#1\algindent}$\triangleright$ \eqparbox{COMMENT}{#2} \hfill %
}
\algnewcommand\LeftCommentNoTriangle[2]{%
\hspace{#1\algindent} \eqparbox{COMMENT}{#2} \hfill %
}
\algnewcommand\LeftCommentNoIntent[1]{%
$\triangleright$ \eqparbox{COMMENT}{#1} \hfill %
}

\newcommand{\Ad}{\mathbf{A}_\mathrm{d}}
\newcommand{\Adbar}{\bar{\mathbf{A}}_\mathrm{d}}

\newcommand{\Ac}{\mathbf{A}_\mathrm{c}}

\DeclarePairedDelimiter{\paren}{\lparen}{\rparen}
\newcommand{\Nos}{N_\text{os}}
\newcommand{\B}{B}
\newcommand{\Bd}{B_\text{D}}
\newcommand{\Brx}{B_\text{rx}}
\newcommand{\Bprime}{B'}
\newcommand{\Bsam}{B_\text{sam}}
\newcommand{\atx}{\alpha_\text{tx}}
\newcommand{\grx}[1]{g_\text{rx}\paren{#1}}
\newcommand{\gtx}[1]{g_\text{tx}\paren{#1}}

\newcommand{\vX}{\mathbf{X}}
\newcommand{\vY}{\mathbf{Y}}

\newcommand{\mA}{\mathbf{A}}
\newcommand{\Prx}{P_\text{rx}}
\newcommand{\Ptx}{P_\text{tx}}

\newcommand{\sicindex}{t} 
\newcommand{\siclength}{N}
\newcommand{\sics}{\ell}

\newcommand{\Usym}{\mathbf{U}}
\newcommand{\Sint}{\mathbf{S}}
\newcommand{\usym}{{\mathbf{u}}}
\newcommand{\sint}{\mathbf{s}}

\newcommand{\vdIh}{{\hat{\mathbf{u}}}_{1}}
\newcommand{\vdIIh}{{\hat{\mathbf{u}}}_{2}}

\newcommand{\z}{\mathbf{w}}
\newcommand{\Z}{\mathbf{W}}
\newcommand{\zIh}{\hat{\mathbf{w}}_1}
\newcommand{\zIIh}{\hat{\mathbf{w}}_2}

\makeatletter
\DeclareRobustCommand{\rvdots}{%
  \vbox{
    \baselineskip4\p@\lineskiplimit\z@
    \kern-\p@
    \hbox{.}\hbox{.}\hbox{.}
  }}
\makeatother

\newcolumntype{M}{>{\hfil$\displaystyle}X<{$\hfil}} %
\newcolumntype{L}{>{\collectcell\AddLabel}r<{\endcollectcell}}

\newlength{\AlgoWidth}

\begin{document}
\usepgfplotslibrary{colorbrewer}
\colorlet{colA}{Set1-A} %
\colorlet{colB}{Set1-B} %
\colorlet{colC}{Set1-C} %
\colorlet{colD}{Set1-D} %
\colorlet{colE}{Set1-E} %
\colorlet{colF}{Set1-G} %
\colorlet{colG}{Set1-H} %
\definecolor{colJ}{rgb}{0.3, 0.3, 0.3} %

\pgfkeys{/pgf/number format/.cd,fixed,precision=1}

\definecolor{mycolor1}{HTML}{0072bd}%
\definecolor{mycolor6}{HTML}{FF0000}%
\definecolor{mycolor5}{HTML}{27ae60}%
\definecolor{mycolorrnn}{HTML}{F5B400}%
\definecolor{vampopt}{HTML}{9D44B5}%

\pgfmathsetmacro{\gridwidth}{1.85in}
\pgfmathsetmacro{\gridheight}{1.7in}
\pgfmathsetmacro{\lwidth}{0.45pt}

\IfOneCol{
\pgfmathsetmacro{\figwidth}{4.00in}
\pgfmathsetmacro{\figheight}{3in}
\pgfmathsetmacro{\OneColMul}{1.2}
\setlength{\AlgoWidth}{0.6\columnwidth} %
}

\IfTwoCol{
\pgfmathsetmacro{\figwidth}{3.00in}
\pgfmathsetmacro{\figheight}{2.3in}
\pgfmathsetmacro{\OneColMul}{1.0}
\setlength{\AlgoWidth}{\columnwidth} %
}

\tikzstyle{arr} = [-latex,thick]
\tikzstyle{doubarr} = [latex-latex,thick]
\tikzstyle{doubarrsmall} = [thin,{Stealth}-{Stealth}]

\pgfplotsset{
    STDPLOT/.style={
        xticklabel style={
          /pgf/number format/fixed,
        },
        yticklabel style={
          /pgf/number format/fixed,
        },
        scaled y ticks=false,
        width=\figwidth,
        height=\figheight,
        grid=both,
        xminorgrids,
        yminorgrids,
        ymin=0,
        minor tick num=1,
        enlarge x limits={0},
        enlarge y limits={0}, %
        legend cell align={left},
        title style={yshift=-1ex}, %
      },
  Rate_VS_PTX_NuN1_HALVE/.style={
    xlabel={$\Ptx/ (B \nu_{1}/2)$ [dB]},
    ylabel={Rate [bpcu]},
  },
  Rate_VS_PRX_NuN1_HALVE/.style={
    xlabel={$\Prx/ (B \nu_{1}/2)$ [dB]},
    ylabel={Rate [bpcu]},
  },
  Rate_VS_PTX_NuN2/.style={
    xlabel={$\Ptx/\nu_{N_2}$ [dB]},
    ylabel={Rate [bpcu]},
  },
  Rate_VS_PRX_NuN2/.style={
    xlabel={$\Prx/\nu_{N_2}$ [dB]},
    ylabel={Rate [bpcu]},
  },
   PAM/.style={
    mycolor1,
    line width=\lwidth,
   },
   ASK/.style={
    mycolor6,
    line width=\lwidth,
    },
    QAM/.style={
       mycolor5,
    line width=\lwidth,
    },
    VAMPOPT/.style={
       vampopt,
        line width=\lwidth,
        mark=star,
    },
    EXIT/.style={
       densely dashdotted,
       line width=0.6pt,
    },
    RNN/.style={
        mycolorrnn,
        mark=*,mark size={1.0pt},mark options={solid,fill=white},line width=0.5pt
    },
    GS/.style={
        mycolor5,
        mark=*,mark size={1.0pt},mark options={solid,fill=white},line width=0.5pt
    },
    FBA/.style={
        mark=none,mark size={1.1pt},line width=0.5pt, densely dashdotted
    },
    UB/.style={
       line width=\lwidth,
       mark=none,
       black,
       solid
    },
    table/OPTSNRreal/.style={x expr={\thisrow{SNR}+6}},
    abovecurve/.style n args = {4}{ %
    postaction={decorate,
      decoration={text effects along path,
        text={|\normalsize|#1}, %
        raise=#3,  %
        text align={left, left indent=#4}, %
        text along path,
        text color={#2},  %
        every character/.style={double=white,shape=rectangle, fill=blue!20, draw=blue!40}
      }
    },
  },
}

\tikzset{
    arr/.style={thick,{Stealth[length=1.2mm, width=1.2mm]}-{Stealth[length=1.2mm, width=1.2mm]}},
    arrmagn/.style={line width=0.4pt,{Stealth[length=1.2mm, width=1.2mm]}-{Stealth[length=1.2mm, width=1.2mm]}},
    arrR/.style={thick,-{Stealth[length=1.2mm, width=1.2mm]}},
    opacitylabel/.style={
       fill=white, fill opacity=0.8,text opacity=1, draw opacity=1, inner sep=0.5pt
    },
    shadowed/.style={
    preaction={transform canvas={shift={(-0.1pt,0.1pt)}},draw=white,very thick}},
    shadowedb/.style={
    preaction={transform canvas={shift={(0.1pt,-0.1pt)}},draw=white,very thick}}
    }

\tikzset{
  opacitylabel/.style={
    fill=white, fill opacity=0.7,text opacity=1, draw opacity=1,
    inner sep=0.5pt,
    ,
  },
  MEASURE/.style={
    font=\small,
    opacitylabel,
    xshift=0.15cm
  },
}

\newcommand{\MeasureXDistance}[4]{%

  \path[name path global=tline]
  (axis cs:\pgfkeysvalueof{/pgfplots/xmin}, #1) --
  (axis cs:\pgfkeysvalueof{/pgfplots/xmax}, #1);

  \path[
    name intersections={of=tline and #2, name=pA},
    name intersections={of=tline and #3, name=pB}
  ];

  \draw[doubarrsmall,]
  let
  \p1 = (pA-1),
  \p2 = (pB-1)
  in
  (pA-1) -- (pB-1)
  node[MEASURE,#4] {%
    \pgfplotsconvertunittocoordinate{x}{\x1}%
    \pgfplotscoordmath{x}{datascaletrafo inverse to fixed}{\pgfmathresult}%
    \edef\valueA{\pgfmathresult}%
    \pgfplotsconvertunittocoordinate{x}{\x2}%
    \pgfplotscoordmath{x}{datascaletrafo inverse to fixed}{\pgfmathresult}%
    \pgfmathparse{\pgfmathresult - \valueA}%
    \pgfmathprintnumber[fixed, precision=1]{\pgfmathresult}\,dB%
  };
}

\newcommand{\MeasureYDistance}[4]{%

  \path[name path global=tline]
  (axis cs:\pgfkeysvalueof{/pgfplots/xmin}, #1) --
  (axis cs:\pgfkeysvalueof{/pgfplots/xmax}, #1);

  \path[name intersections={of=tline and #2, name=p1}];

  \path let \p1 = (p1-1) in
  \pgfextra{
    \pgfplotsconvertunittocoordinate{x}{\x1}
    \pgfplotscoordmath{x}{datascaletrafo inverse to fixed}{\pgfmathresult}
    \xdef\SavedXCoord{\pgfmathresult} %
  };

  \path[name path global=temp_vline]
  (axis cs:\SavedXCoord, \pgfkeysvalueof{/pgfplots/ymin}) --
  (axis cs:\SavedXCoord, \pgfkeysvalueof{/pgfplots/ymax});

  \path[
    name intersections={of=temp_vline and #2, name=pA},
    name intersections={of=temp_vline and #3, name=pB}
  ];

  \draw[doubarrsmall,]
  let
  \p1 = (pA-1),
  \p2 = (pB-1)
  in
  (pA-1) -- (pB-1)
  node[midway, MEASURE, #4 ] {%
    \pgfplotsconvertunittocoordinate{y}{\y1}%
    \pgfplotscoordmath{y}{datascaletrafo inverse to fixed}{\pgfmathresult}%
    \edef\valueA{\pgfmathresult}%
    \pgfplotsconvertunittocoordinate{y}{\y2}%
    \pgfplotscoordmath{y}{datascaletrafo inverse to fixed}{\pgfmathresult}%
    \pgfmathparse{\valueA - \pgfmathresult}%
    \pgfmathprintnumber[fixed, precision=1]{\pgfmathresult}\,bpcu%
  };
}

\newcommand{\MeasureXDistancePrecise}[4]{%

  \path[name path global=tline]
  (axis cs:\pgfkeysvalueof{/pgfplots/xmin}, #1) --
  (axis cs:\pgfkeysvalueof{/pgfplots/xmax}, #1);

  \path[
    name intersections={of=tline and #2, name=pA},
    name intersections={of=tline and #3, name=pB}
  ];

  \draw[doubarrsmall,]
  let
  \p1 = (pA-1),
  \p2 = (pB-1)
  in
  (pA-1) -- (pB-1)
  node[MEASURE,#4] {%
    \pgfplotsconvertunittocoordinate{x}{\x1}%
    \pgfplotscoordmath{x}{datascaletrafo inverse to fixed}{\pgfmathresult}%
    \edef\valueA{\pgfmathresult}%
    \pgfplotsconvertunittocoordinate{x}{\x2}%
    \pgfplotscoordmath{x}{datascaletrafo inverse to fixed}{\pgfmathresult}%
    \pgfmathparse{\pgfmathresult - \valueA}%
    \pgfmathprintnumber[fixed, precision=2]{\pgfmathresult}\,dB%
  };
}

\newcommand{\MeasureYDistancePrecise}[4]{%

  \path[name path global=tline]
  (axis cs:\pgfkeysvalueof{/pgfplots/xmin}, #1) --
  (axis cs:\pgfkeysvalueof{/pgfplots/xmax}, #1);

  \path[name intersections={of=tline and #2, name=p1}];

  \path let \p1 = (p1-1) in
  \pgfextra{
    \pgfplotsconvertunittocoordinate{x}{\x1}
    \pgfplotscoordmath{x}{datascaletrafo inverse to fixed}{\pgfmathresult}
    \xdef\SavedXCoord{\pgfmathresult} %
  };

  \path[name path global=temp_vline]
  (axis cs:\SavedXCoord, \pgfkeysvalueof{/pgfplots/ymin}) --
  (axis cs:\SavedXCoord, \pgfkeysvalueof{/pgfplots/ymax});

  \path[
    name intersections={of=temp_vline and #2, name=pA},
    name intersections={of=temp_vline and #3, name=pB}
  ];

  \draw[doubarrsmall,]
  let
  \p1 = (pA-1),
  \p2 = (pB-1)
  in
  (pA-1) -- (pB-1)
  node[midway, MEASURE, #4 ] {%
    \pgfplotsconvertunittocoordinate{y}{\y1}%
    \pgfplotscoordmath{y}{datascaletrafo inverse to fixed}{\pgfmathresult}%
    \edef\valueA{\pgfmathresult}%
    \pgfplotsconvertunittocoordinate{y}{\y2}%
    \pgfplotscoordmath{y}{datascaletrafo inverse to fixed}{\pgfmathresult}%
    \pgfmathparse{\valueA - \pgfmathresult}%
    \pgfmathprintnumber[fixed, precision=2]{\pgfmathresult}\,bpcu%
  };
}
\title{Information Rates of Approximate Message Passing for Bandlimited Direct-Detection Channels}

\author{Daniel Plabst, Mohamed Akrout, Tobias Prinz, Amine Mezghani and Gerhard Kramer%
\thanks{Date of current version \today. This work was supported by the German Research Foundation (DFG) under project 509917421, and by the Discovery Grants Program and a Doctoral Research Scholarship of the Natural Sciences and Engineering Research Council of Canada (NSERC).
\emph{(Corresponding author: Daniel Plabst.)}}
\thanks{D.\ Plabst and G.\ Kramer are with the Institute for Communications Engineering, School of Computation, Information and Technology, Technical University of Munich (TUM), Germany (e-mail: daniel.plabst@tum.de; gerhard.kramer@tum.de).
T.\ Prinz was with the Institute for Communications Engineering at TUM; he is now with 59engineers GmbH, Munich, Germany.
M.\ Akrout was with the Price Faculty of Engineering, University of Manitoba, Winnipeg, Manitoba, Canada. He is now with the Min H.\ Kao Department of Electrical Engineering and Computer Science, Tickle College of Engineering, University of Tennessee, Knoxville, USA (e-mail: makrout@utk.edu). A.\ Mezghani is with the Department of Electrical and Computer Engineering, Price Faculty of Engineering, University of Manitoba, Winnipeg, Manitoba, Canada (e-mail: amine.mezghani@umanitoba.ca).}}

\maketitle

\begin{abstract}
The capacity of bandlimited direct-detection channels is challenging to compute or approach due to the receiver non-linearity. A generalized vector approximate message passing (GVAMP) detector is designed to achieve high rates at a reasonable level of complexity. The rates increase by using multi-level coding and successive interference cancellation. The methods are applied to fiber-optic channels with intersymbol interference caused by spectrally efficient pulse shapes, chromatic dispersion, and receiver sampling at twice the baud rate. Bipolar modulation operates within 0.26 bits per channel use (bpcu) of the real-alphabet coherent capacity for optically amplified links, reducing the best-known theoretical gap of 1 bpcu. Remarkably, bipolar modulation achieves 6 dB and 3 dB of power gain over unipolar modulation with and without optical amplification, respectively. Simulations with polar-coded modulation confirm the gains.
The GVAMP complexity, measured in multiplications per information bit (mpib), is proportional to the number of iterations and to the logarithm of the block length, and is substantially lower than that of other equalizers. For example, a system with 64-ary bipolar modulation and a root-raised cosine pulse with a 1\% roll-off factor was simulated over 4 km of optically amplified standard single-mode fiber in the C-band. The GVAMP receiver requires 93 mpib to achieve 5 bpcu at 300 gigabaud.

\end{abstract}


\IEEEpeerreviewmaketitle

\section{Introduction}
Intensity detection is used if coherent detection is expensive or infeasible. Examples include direct detection (DD) for fiber-optic channels \cite{AgrawalThirdEdFiberOptics}, and phase retrieval for crystallography \cite{sayre1952crystall}, astronomy \cite{fienup1982phase,fienup1987phase}, microscopy \cite{miao2008extending}, X-ray \cite{bunk2007diffractive} and optical imaging, and quantum problems \cite{corbett2006pauli}.

We focus on fiber-optic DD channels. Consider the model
\begin{equation}
    \mathbf{Y} = f(\mathbf{A}\mathbf{X} + \mathbf{N}_1) + \mathbf{N}_2
    \label{eq:phase-retrieval_general}
\end{equation}
where the vector $\mathbf{X} \in \mathbb{C}^n$ is the input of a linear channel modeled by the matrix $\mathbf{A} \in \mathbb{C}^{m \times n}$ and optical noise $\mathbf{N}_1 \in \mathbb{C}^{m}$. A DD receiver $f(\cdot)$ with a photodetector (PD) performs optical to electrical conversion modeled by entry-wise squaring. The vector $\mathbf{N}_2 \in \mathbb{R}^{m}$ represents electrical noise from the PD and a low-noise amplifier. One should sample faster than the baud rate, i.e., $m > n$, to obtain sufficient statistics because the DD doubles the bandwidth \cite{mecozzi2018information,secondini2020direct,tasbihi2020capacity,tasbihi2021direct,plabst2022achievable,prinz2023successive,plabst2024neural}. The goal is to approach the mutual information rate $I(\mathbf{X}; \mathbf{Y})/n$ for large $n$.

\subsection{Reconstruction Algorithms}

Reconstructing $\mathbf{X}$ is challenging due to the non-linearity,  band-limitations, and noise \cite{dong2023phasetutorial}. For example, suppose $\mathbf{X}$ is real-valued, and there is no noise. The paper \cite[Thm.~2.2]{balan2006phase} argues that $m \geq 2n-1$ suffices to reconstruct $\mathbf{X}$ up to a global sign ambiguity by choosing the entries of $\mathbf{A}$ as independent and identically distributed (iid) standard Gaussian random variables (RVs). This number is necessary in general. Similarly, $m \geq 4n - \mathcal{O}(n)$ suffices for complex-valued $\mathbf{X}$ \cite{heinosaari2013quantum}. 
The paper \cite{maillard2020phase} performs a replica-based analysis of ``full recovery'' thresholds, i.e., the smallest $m/n$ that achieve the minimum mean square error (MMSE) \cite[Sec.~4]{maillard2020phase}.
This threshold is $m/n = 1$ for noiseless phase retrieval of real Gaussian inputs with iid real Gaussian and uniformly-sampled column-orthogonal $\mathbf{A}$ \cite{candes2006near}. Similarly, the threshold is $m/n = 2$ for complex Gaussian inputs with iid circularly-symmetric complex Gaussian (CSCG) and uniformly-sampled column-unitary $\mathbf{A}$ \cite[Table~I]{maillard2020phase}.

The papers \cite{gerchberg1972practical,fienup1982phase} perform greedy phase retrieval by iterative projections under magnitude or support constraints \cite[Sec.~2]{waldspurger2015phase}. However, greedy approaches can stall in local minima. The paper \cite{netrapalli2013phase} uses a careful initialization to guarantee convergence and reduce complexity. 

Suppose the noise $\mathbf{N}_2$ is Gaussian and much stronger than $\mathbf{N}_1$. Gradient-based methods \cite{candes2015wirtinger,guizarsicairos2008phase} approximate the maximum likelihood metric $p(\mathbf{y}|\mathbf{x})$, where $\mathbf{Y}=\mathbf{y}$ is fixed. Equivalently, one may minimize $\lVert \mathbf{y} - |\mathbf{A} \mathbf{x}|^2\rVert^2$, called a squared loss. The optimization is generally non-convex, and gradient-based algorithms can get stuck in local minima. The paper \cite{davis2020nonsmooth} uses subgradients for a non-smooth formulation in the noiseless setting, and \cite{dong2023phasetutorial} lists acceleration strategies and emphasizes good initialization. Coordinate-descent algorithms reduce complexity and achieve exact reconstruction with high probability for sufficiently large sampling factors \cite{zeng2020coordinate}.

The papers \cite{candes2013phaselift,waldspurger2015phase} solve a relaxed convex problem in higher dimensions with off-the-shelf optimizers. However, the complexity is high. Convex relaxations can reduce complexity \cite{goldstein2018phasemax,bahmani2017phase}; we refer to \cite{Fannjiang2020numerics} for a list of algorithms. Careful initialization improves performance and reduces complexity; for example, the initial guess might be the principal eigenvector of a covariance matrix built from the intensities, a technique known as spectral initialization \cite{dong2023phasetutorial,mondelli2022gampspectral}.

\subsection{Message Passing on Graphs}
Message passing on graphs can approximate Bayesian estimation. The main ideas were developed to decode low-density parity-check codes using extrinsic messages \cite{Gallager60,Hagenauer-IT96}. The invention of turbo codes \cite{berrou1996near} led to new methods to approach the capacity of noisy channels with intersymbol interference \cite{Douillard:95:iterative}. Complexity is reduced by quantizing messages, even to a single bit per message \cite{Gallager60,Lechner-C12}.
Similarly, one may pass only second-order statistics \cite{Rusmevichientong-Roy-IT01,minka2001family,seeger2005expectation}.

Extrinsic message passing is typically performed between two or three modules that collect messages, process them, and pass new messages to their neighboring modules. A helpful tool to track the convergence of such algorithms is the extrinsic information transfer (EXIT) chart that has one EXIT function per module \cite{tenbrink01}; see also \cite[p.~45-47]{Gallager60} and \cite{tenbrink2001multiEXIT,tenBrink-C04,ashikhmin04}. We derive EXIT functions in Secs.~\ref{subsec:vEXIT-functions} and~\ref{subsec:EXIT-Charts} below.

\subsection{Expectation Propagation}
\label{sec:EP}
A more general approach to probabilistic message passing is called expectation propagation (EP) \cite{minka2001family,seeger2005expectation,bishop2006pattern}. Consider a random vector $\boldsymbol{\Theta}=(\Theta_1,\dots,\Theta_{N})$ with probability distribution (or density) function that factors as
\begin{align}
    p(\boldsymbol{\theta}) = \prod\nolimits_{f=1}^F t_f(\boldsymbol{\theta}_{\mathcal{S}_f})
\end{align}
where each $t_f$ is a (conditional) probability distribution (or density) that is a function of a selection $\boldsymbol{\theta}_{\mathcal{S}_f}=(\theta_k:k\in\mathcal{S}_f)$ of variables, and $\mathcal{S}_f\subseteq \{1,\dots,N\}$ has cardinality $|\mathcal{S}_f|$.
For instance, $t_1$ might be a prior on data $\mathbf{X}:=(\Theta_1,\dots,\Theta_n)$ where $n=|\mathcal S_1|$, $t_2$ an indicator function specifying that $\mathbf{Z}:=(\Theta_{n+1},\dots,\Theta_{n+m})$ is a function of $\mathbf{X}$, e.g., $\mathbf{Z}=\mathbf{A}\mathbf{X}$, and $t_3$ a Gaussian density $p_{\mathbf{Y}|\mathbf{Z}}\sim\mathcal{N}(\mathbf{z},\mathbf{\Sigma})$ with mean $\mathbf{z}$ and covariance matrix $\mathbf{\Sigma}$, and where $\mathbf{Y}$ has dimension $m$. The problem is to compute \emph{a posteriori} distributions on $\mathbf{X}$.

EP replaces each $t_f$ by a surrogate density or distribution, often a Gaussian density $\widetilde{t}_f=\mathcal{N}(\widetilde{\boldsymbol{\mu}},\mathbf{\widetilde{\Sigma}})$ subject to constraints. For example, a simple constraint is $\mathbf{\widetilde{\Sigma}}=\widetilde{\sigma}^2 \mident{}$ where $\mident{}$ is an identity matrix and $\widetilde{\sigma}^2$ is a shared variance. The number of parameters describing $\widetilde{t}_f$ is then the dimension of $\widetilde{\boldsymbol{\mu}}$ plus one, which is typically a significant compression of the true \emph{a posteriori} densities. EP iteratively updates the $\widetilde{t}_f$ via
\begin{align}
    \widetilde{t}_f = \frac{\argmin
    D\left( t_f \prod_{g\ne f} \widetilde{t}_g / c \, \big\| \, \mathcal{N}\big(\widetilde{\boldsymbol{\mu}},\mathbf{\widetilde{\Sigma}}\big) \right)}{\prod_{g\ne f} \widetilde{t}_g}
    \label{eq:EP-general}
\end{align}
where $D(p\|q)$ is the informational divergence of the densities $p$ and $q$, the constant $c$ normalizes to a density, and the minimization is over the specified class of Gaussians. The module that performs the update \eqref{eq:EP-general} is called a \emph{denoiser}.

\subsection{Approximate Message Passing}
Approximate message passing (AMP) was introduced for linear systems, i.e., $f(x) = x$ in \eqref{eq:phase-retrieval_general}, to solve high-dimensional estimation problems via message passing \cite{donoho2009message,donoho2010message,meng2015expderiv}. 
AMP performance can be predicted by a scalar recursion, called state evolution (SE), similar to EXIT charts. The prediction is accurate if the dimensions of $\mathbf{A}$ are large and its entries are sampled from an appropriate distribution.
A common choice is iid Gaussian RVs \cite{bayati2011dynamics}; we refer to such matrices as iid Gaussian. SE allows proving Bayesian optimality of AMP with iid Gaussian matrices \cite{reeves2019replica,barbier2020mutual}. However, AMP may perform poorly for structured matrices encountered in practice.

The paper \cite{ma2015turbo} proposes an improved AMP-based detector for partial discrete Fourier transform (DFT) matrices $\mathbf{A}$. The papers \cite{rangan2019VAMP,ma2017orthogonal} propose a vector-AMP (VAMP) algorithm that is replica Bayes optimal under random right-rotationally-invariant $\mathbf{A}$ and a unique SE fixed point. The replica prediction is accurate for certain random rotationally-invariant $\mathbf{A}$ \cite{barbier2018mutualbeyond,li2024random}.
Another generalization, called GAMP, was developed for generalized linear systems (GLS) \cite{rangan2011generalized} with non-linear measurements such as $f(\cdot) = |\cdot|^2$ in \eqref{eq:phase-retrieval_general}. GAMP is Bayes optimal for iid Gaussian $\mathbf{A}$ and a unique SE fixed point \cite{barbier2019optimal}. The paper \cite{schniter2016VAMPGLM} proposes a generalized VAMP (GVAMP) that is replica Bayes optimal \cite{pandit2020inference} under random bi-rotationally invariant $\mathbf{A}$ and a unique SE fixed point. Empirically, GAMP and GVAMP work well for other classes of random $\mathbf{A}$ \cite{schniter2012phase,rangan2019VAMP}; see \cite{liu2023achievable,chi2024gampgvamp} for a summary.

The paper \cite{schniter2012phase} performs GAMP
phase retrieval for sparse signals and Gaussian $\mathbf{N}_1$ via iid Gaussian $\mathbf{A}$ and randomly masked Fourier $\mathbf{A}$. SE extends to randomly sub-sampled Hadamard-Walsh $\mathbf{A}$ and Gaussian priors \cite{dudeja2022universality}. The paper \cite{zhu2019phase} studies GAMP for quantized phase retrieval with iid Gaussian $\mathbf{A}$.
GVAMP for amplitude measurements, i.e., $f(\cdot) = |\cdot|$, under Gaussian $\mathbf{N}_1$ and $\mathbf{N}_2$ was studied in \cite{wang2020decentralized}. The paper \cite{ma2021spectral} analyses spectral initializers for uniformly sampled column-unitary matrices $\mathbf{A}$, coded diffraction patterns, and where $\mathbf{A}$ is a partial discrete Fourier transform matrix. The paper \cite{maillard2020phase} considers GVAMP for noiseless phase retrieval with complex Gaussian priors, uniformly sampled column-unitary matrices, and randomly subsampled DFT matrices. 
GVAMP achieves the MMSE at $m/n \approx 2.27$, which is larger than the replica-analysis threshold $m/n = 2$; see \cite[Table~I, Fig.~2]{maillard2020phase}. Likewise, there is a gap for real Gaussian inputs in uniformly sampled column-orthogonal channels, where GVAMP achieves the MMSE at $m/n \approx 1.58$, while the replica-based threshold is $m/n = 1$. 

Real-valued image phase retrieval with shot-noise $\mathbf{N}_2$, coded diffraction patterns, and oversampled Fourier measurements has been assisted by neural networks \cite{metzler2018prdeep,shastri2023deep,shastri2025fastvampimage}. While \cite{metzler2018prdeep} struggles under oversampled Fourier $\mathbf{A}$, \cite{shastri2023deep,shastri2025fastvampimage} improve accuracy and computational cost by stochastic damping and refining calculations with neural networks.

\subsection{Fiber-Optic Communications with DD}
We study \eqref{eq:phase-retrieval_general} for optical fiber communication, which puts engineering constraints on $\mathbf{A}$ and $\mathbf{X}$. We aim to design low-cost transmitters and receivers that operate near capacity, i.e., the system achieves nearly error-free performance at data rates approaching the maximal mutual information rate.

Early DD systems had low baud rates and treated residual intersymbol interference (ISI) as noise, effectively making $\mA$ an $n \times n$ identity matrix. Modern systems use time-frequency resources more efficiently by signaling faster, which leads to ISI. The receiver must now consider the combined linear and non-linear effects of ISI and DD. For example, a typical receiver structure pairs a channel shortening filter \cite{rusek2012optimal} with a low-memory trellis decoder \cite{wettlin2020dsp}.

A more sophisticated method effectively separates the linear and non-linear effects by using minimum-phase signals and the Kramers-Kronig (KK) relations for detection \cite{mecozzi2016kramers,mecozzi2018kkpam}. The signals add a large offset to bipolar symbols and exhibit a high carrier-to-signal power ratio (CSPR); see \cite[Figs. 1, 6, and 9]{mecozzi2016necessary}. One loses $\approx\SI{6.5}{dB}$ in power at 5 bits per channel use (\SI{}{bpcu}) compared to a coherent receiver, which corresponds to losing $\approx\SI{2}{bpcu}$ \cite[Fig.~8]{chou2022phase}. The CSPR increases with the CD level, and the KK receiver loses another $\approx\SI{1.5}{dB}$ for a total loss of $\approx\SI{8}{dB}$; see \cite[Fig.~5c and 7]{chou2022phase}. Deep-learning-based phase retrieval also loses over $\SI{6}{dB}$ \cite[Fig.~8]{orsuti2023deep}.

The KK scheme has more limitations, e.g., it requires a coherent I/Q transmitter for real-valued bipolar signals \cite{mecozzi2018kkpam}. Alternatively, one may generate a double-sideband passband signal and optically filter it to obtain a single-sideband signal that is minimum-phase \cite{mecozzi2018kkpam}. However, the optical filter must have a sharp cutoff.
Also, KK detectors apply a square-root and logarithm to the entries of $\mathbf{Y}$, which can be negative with electrical noise $\mathbf{N}_2$. Practical receivers use clipping \cite{lowery2019clipping}, but this reduces rates \cite[Sec.~2.8]{cover1991elementsofIT} unless the CSPR increases.

\subsection{Oversampling at the Receiver}
One can recover phase information if the receiver sampling rate exceeds the baud rate. This allows the use of bipolar or even complex modulation alphabets. In fact, the paper \cite{mecozzi2018information} showed that DD can operate within $\SI{1}{bit}$ of the coherent capacity if the optical noise $\mathbf{N}_1$ plays the dominant role.
The papers \cite{mecozzi2018information,secondini2020direct,tasbihi2020capacity,tasbihi2021direct,plabst2022achievable,prinz2023successive,plabst2024neural} exploit ISI to detect bipolar real or complex modulations. The power gains of bipolar ASK over unipolar ASK are approximately \SI{6}{dB} and \SI{3}{dB} with and without optical amplification, respectively \cite{secondini2020direct,prinz2023successive,plabst2024neural}. However, it is unclear if one can achieve high rates with competitive complexity. For example, precoding to remove CD requires a coherent transmitter \cite{tasbihi2021direct}, and the neural networks (NNs) of \cite{plabst2024neural} require thousands of multiplications per symbol.

\subsection{Contributions and Organization}
We study bipolar signaling over short-reach optical fiber with DD receivers. The transmitter has a single Mach-Zehnder modulator (MZM) for the in-phase signal component and cannot pre-compensate the complex-valued ISI. We derive a GVAMP detector \cite{rangan2019VAMP,schniter2016VAMPGLM} for links with and without optical amplification. Multi-level coding (MLC) and successive interference cancellation (SIC) increase information rates and permit the use of codes designed for memoryless channels. We compute EXIT functions to predict performance.

We simulate $\SI{300}{\giga Bd}$ transmission over $\SI{4}{\kilo\meter}$ of standard single-mode fiber (SSMF) in the C-band. For optically amplified links, optimized bipolar ASK achieves within 0.26 and \SI{0.1}{bpcu} of the real-alphabet coherent capacity and equiprobable ASK capacity, respectively. Moreover, bipolar ASK gains roughly $6\,\text{dB}$ and $3\,\text{dB}$ in power over unipolar modulation with and without optical amplification, respectively.
The detector complexity is $\mathcal{O}(n_\text{it} \log_2 n + M)$ multiplications per transmit symbol, where $n_\text{it}$ is the number of GVAMP iterations, $n$ is the transmit block length, and $M$ is the modulation alphabet size. For example, for optically amplified SSMF, separate detection and decoding (SDD) requires $n_\text{it} \approx 37$ and $93$ multiplications per information bit (mpib) to achieve $\SI{5}{bpcu}$.

This paper is organized as follows.
Sec.~\ref{sec:preliminaries} reviews notation,  
Sec.~\ref{sec:system-model} describes the system model, and
Sec.~\ref{sec:MLC-SIC} reviews generalized mutual information (GMI) for SDD and MLC-SIC. 
Sec.~\ref{sec:vamp} develops the GVAMP algorithm with linear and non-linear denoisers, and Sec.~\ref{sec:state_evolution} derives EXIT functions. Sec.~\ref{sec:simulation-parameters} and Sec.~\ref{sec:simulation-results} present our simulation setup and results.
Sec.~\ref{sec:conclusions} concludes the paper.
The Appendices provide supporting material on how to update the EP messages in general (Appendix~\ref{appendix-EP}) before specializing to the linear denoiser (Appendix \ref{appendix-LMMSE-denoiser}) and non-linear denoiser (Appendices \ref{appendix-chi2}, \ref{appendix-denoising-w1}, \ref{appendix-saddle_point}).

\section{Preliminaries}
\label{sec:preliminaries}

\subsection{Notation}
\label{subsec:notation}

We write $j=\sqrt{-1}$. Column vectors, their transpose, their complex-conjugate, and their complex-conjugate transpose are written as bold symbols $\mathbf{x}$, $\mathbf{x}\tran$, $\mathbf{x}^*$, and $\mathbf{x}\herm$, respectively. Matrices are also written as uppercase bold letters; the determinant of a square matrix $\mathbf{\Sigma}$ is $|\mathbf{\Sigma}|$. The $n$-dimensional all-zeros vector is $\mnull{n}$, the $n\times n$-dimensional all-zeros matrix is $\mnull{n\times n}$, and the $n \times n$-dimensional identity matrix is $\mident{n}$. We sometimes discard the subscripts and write, e.g., $\mathbf{I}$ for $\mident{n}$. We use $\blkdiag(\mathbf{\Sigma}_1, \mathbf{\Sigma}_2)$ to denote a block diagonal matrix with matrices $\mathbf{\Sigma}_1$ and $\mathbf{\Sigma}_2$ placed along the diagonal and zeros elsewhere.

Scalar and vector strings are written as $x_\kappa^n=(x_\kappa,\ldots,x_n)$ and $\mathbf{X}_\kappa^n= (\mathbf{X}_\kappa,\ldots,\mathbf{X}_n)$, respectively; we omit the subscript if $\kappa = 1$.
Let $\idxset N=\{1,2,\ldots,N\}$. For an ordered list of indices $\mathcal{I} = \{i_1, \ldots i_N\}$, we use $\mathcal{I}_t$ to denote the $t^\text{th}$ element of the list. Define the modulo operator over the interval $[-A/2,A/2)$ with $A>0$, as $x\ \mathrm{mod}\ A = x-kA$ where $k$ is the unique integer for which $x-kA\in [-A/2,A/2)$.

The sinc function is $\sinc(t) = \sin(\pi t)/(\pi t)$. The Dirac- and Kronecker-delta functions are written as $\delta(\cdot)$ in the context of continuous and discrete variables, respectively. The convolution and inner product of $g(t)$ and $h(t)$ are
\begin{align}
    g(t)*h(t)
    & = \int_{\mathbb R} g(\tau)\, h(t-\tau)\, \mathrm{d}\tau \\
    \langle g(t) , h(t) \rangle
    & = \int_{\mathbb R} g(t)\, h(t)^*\, \mathrm{d}t .
\end{align}
The energy of $a(t)$ is $\|a(t)\|^2 = \langle a(t), a(t) \rangle$. The real and imaginary parts of $x$ are $\Re\{x\}$ and $\Im\{x\}$, respectively. Fourier transform pairs are written as $a(t)$ \laplace\, $A(f)$.

RVs and vectors are written in uppercase, e.g., $X$ and $\mathbf{X}$, and their realizations in lowercase, e.g., $x$ and $\mathbf{x}$.
The probability mass function (PMF) and probability density function (PDF) of a vector of discrete and continuous RVs $\mathbf{X}$ are written as $P_{\mathbf{X}}$ and $p_{\mathbf{X}}$, respectively. We discard subscripts if the arguments are upper- or lowercase versions of their RVs.  The conditional PMF of $X$ given $Y$ is $P_{X\lvert Y}$. Similarly, we use $P_{X\lvert Y}( \cdot |y)$ for the PMF of $X$ given $Y=y$. The expectation of $\mathbf{X}$ is $\mathbb{E}[\mathbf{X}]$. Entropy, conditional entropy, the informational divergence, and mutual information are written as in \cite{cover1991elementsofIT}, and we measure the quantities in bits.

\subsection{Gaussian Vectors}
\label{subsec:Gaussian-vectors}
The PDF of a real Gaussian $\mathbf{X}$ with mean $\boldsymbol{\mu}=\mathbb{E}[\mathbf{X}]$ and covariance matrix $\mathbf{\Sigma}=\mathbb{E}[(\mathbf{X}-\boldsymbol{\mu})(\mathbf{X}-\boldsymbol{\mu})\tran]$ is
\begin{align}
     \mathcal{N}(\mathbf{x}; \boldsymbol{\mu}, \mathbf{\Sigma}) 
     = \frac{1}{|2\pi \mathbf{\Sigma}|^{1/2}} 
     e^{- (\mathbf{x}-\boldsymbol{\mu})\tran \mathbf{\Sigma}^{-1}(\mathbf{x}-\boldsymbol{\mu})/2}
\end{align}
where $|2\pi \mathbf{\Sigma}|=(2\pi)^n |\mathbf{\Sigma}|$ if $\mathbf{X}$ has dimension $n$. We write $\mathbf{X} \sim \mathcal{N}(\boldsymbol{\mu}, \mathbf{\Sigma})$.

A complex $\mathbf{X}$ is circularly-symmetric if $e^{j\phi}\mathbf{X}$ has the same PDF as $\mathbf{X}$ for all $\phi$, implying $\mathbb{E}[\mathbf{X}]=\mnull{}$. We study translations of CSCG vectors that we refer to as CSCG-$\boldsymbol{\mu}$. The PDF for mean $\boldsymbol{\mu}$ and covariance matrix $\mathbf{\Sigma}=\mathbb{E}[(\mathbf{X}-\boldsymbol{\mu})(\mathbf{X}-\boldsymbol{\mu})\herm]$ is
\begin{align}
     & \mathcal{CN}
     (\mathbf{x}; \boldsymbol{\mu},\mathbf{\Sigma})
     = \frac{1}{| \pi \mathbf{\Sigma} |}
     e^{-(\mathbf{x}-\boldsymbol{\mu})\herm
         \mathbf{\Sigma}^{-1}
         (\mathbf{x}-\boldsymbol{\mu})}.
\end{align}
We write $\mathbf{X} \sim \mathcal{CN}(\boldsymbol{\mu}, \mathbf{\Sigma})$.

\subsection{Divergence Projections}
We project onto Gaussian densities.
Suppose $g(\cdot)$ is a non-negative function and $c = \int g(\mathbf{x}) \,\mathrm{d}\mathbf{x}$. For real and complex domains, define the respective projections
\begin{subequations}
\begin{align}
    \Proj[g(\cdot)] 
    & := \argmin_{\mathcal{N}(\boldsymbol{\mu},\sigma^2 \mident{})} D\left( g(\cdot)/c \,\big\|\, 
    \mathcal{N}(\boldsymbol{\mu}, \sigma^2 \mident{} ) \right)  \label{eq:min_div_proj} \\
    \CProj[g(\cdot)] 
    & := \argmin_{\mathcal{CN}(\boldsymbol{\mu},\sigma^2 \mident{})} D\left( g(\cdot)/c \,\big\|\, 
    \mathcal{CN}(\boldsymbol{\mu}, \sigma^2 \mident{} ) \right) .
    \label{eq:min_div_cproj}
\end{align}
\end{subequations}
The problems \eqref{eq:min_div_proj} and \eqref{eq:min_div_cproj} are solved by computing the mean and the average variance of $g(\cdot)/c$ (see \cite[Ch.~10.7]{bishop2006pattern}):
\begin{align}
    & \boldsymbol{\mu} = \int \mathbf{x}\,
    \frac{g(\mathbf{x})}{c} \,\mathrm{d}\mathbf{x}, &&
    \sigma^2 = \frac{1}{n} \int  \|\mathbf{x} - \boldsymbol{\mu}\|^2 \frac{g(\mathbf{x})}{c}  \,\mathrm{d}\mathbf{x} .
\end{align}

\section{System Model}
\label{sec:system-model}

\subsection{Continuous-Time Baseband Model}
\label{sec:time-continuous-model}
We study the model shown in Fig.~\ref{fig:continuous_detailed_system_model}; see \cite{plabst2022achievable,prinz2023successive}.  
\begin{figure*}[t!]
    \centering
    \usetikzlibrary{decorations.markings}
\tikzset{node distance=2.3cm}

\pgfdeclarelayer{background}
\pgfdeclarelayer{foreground}
\pgfsetlayers{background,main,foreground}
\tikzset{boxlines/.style = {draw=black!20!white,}}
\tikzset{boxlinesred/.style = {densely dashed,draw=red!50!white,thick}}

\pgfmathsetmacro{\samplerwidth}{30}

\tikzset{midnodes/.style = {midway,above,text width=1.5cm,align=center,yshift=-0.2em}}
\tikzset{midnodesRP/.style = {midway,above,text width=1.5cm,align=center,yshift=-1.4em}}

\begin{tikzpicture}[]
    \node [input, name=input] {Input};
    \node [comblock,right of=input,node distance=1.4cm] (txfilter) {$g_\text{tx}(t)$};
    \node [comblock,right of=txfilter,node distance=2.1cm] (cir) {$h(t)$};
    \node [sum,right of=cir,node distance=1.7cm] (sumnode2) {$+$};
    \node [comblock,right of=sumnode2,node distance=1.4cm] (sld_bw_const) {$g_\mathrm{D}(t)$};
    \node [comblock,right of=sld_bw_const,node distance=2.3cm] (sld) {$\left\lvert \,\cdot\, \right\rvert^2$};
    \node [sum,right of=sld,node distance=1.7cm] (sumnode) {$+$};
    \node [input, name=noise,above of=sumnode,node distance=1cm] {Input};
    \node [input, name=noise2,above of=sumnode2,node distance=1cm] {Input};
    \node [comblock,right of=sumnode,node distance=1.6cm] (rxfilter) {$g_\text{rx}(t)$};
    \node [comblock,right of=rxfilter,minimum width=\samplerwidth pt,node distance=2.1cm,font=\footnotesize] (sampler) {};
    \node [comblock,right of=sampler,node distance=1.9cm] (dsp) {RX};
    \node [input, name=output, right of=dsp,node distance=1.5cm] {Output};
    \draw[thick] (sampler.west) -- ++(\samplerwidth/4 pt,0) --++(\samplerwidth/2.7 pt,\samplerwidth/4.5 pt );
    \draw[thick] (sampler.east) -- ++(-\samplerwidth/3pt,0);

    \draw ($(sampler.west) + (\samplerwidth/4.5 pt,0.25)$)edge[out=0,in=100,-latex,thick] ($(sampler.east) + (-\samplerwidth/2.5 pt,-0.2)$);

    \draw[-latex,thick] (input) -- node[midnodes](s_ti){$X_\kappa$  } (txfilter);

    \draw[-latex,thick] (txfilter) -- node[midnodes](){$X(t)$\\[0.2em] } (cir);
    \draw[-latex,thick] (cir) -- node[midnodes](){$Z'(t)$\\[0.2em] } (sumnode2);
    \draw[-latex,thick] (sumnode2) --  (sld_bw_const);
    \draw[-latex,thick] (sld_bw_const) -- node[midnodes](){$Z''(t)$\\[0.2em] }   (sld);
    \draw[-latex,thick] (sld) -- node[midnodes](){$Y^\prime(t)$\\[0.2em] } (sumnode);
    \draw[-latex,thick] (sumnode) --node[midnodes](){$Y''(t)$\\[0.2em] }  (rxfilter);
    \draw[-latex,thick] (rxfilter) -- node[midnodes](){$Y(t)$\\[0.2em] } (sampler);
    \draw[-latex,thick] (sampler) --  node[midnodes](){$Y_k$\\ [0.2em] } (dsp)  ;

    \draw[-latex,thick] (noise) -- (sumnode);
    \draw[-latex,thick] (noise2) -- (sumnode2);
    \node[above] () at (noise) {$N_2'(t)$};
    \node[above] () at (noise2) {$N_1'(t)$};
    \node[below,yshift=-1em,font=\footnotesize] () at (txfilter) {DAC};
    \node[below,xshift=-2em,yshift=-1em,font=\footnotesize] () at (sampler) {Bandlimited ADC};

    \node[below,yshift=-1em,font=\footnotesize] () at (cir) {Fiber};
    \node[below,xshift=-1.2cm,yshift=-1em,text width=5cm,align=center,font=\footnotesize] () at (sld) {PD with an optical bandwidth constraint};

\end{tikzpicture}
    \caption{Model of a short-reach system with DD \cite{plabst2022achievable}.}
    \label{fig:continuous_detailed_system_model}
\end{figure*}
The source outputs iid real symbols $\left(...,X_1, X_2, X_3,\ldots \right)$ where each $X_\kappa$ has the scaled alphabet $c\cdot \mathcal{M}$ where
\begin{align}
    \mathcal{M} = 
    \frac{1}{M-1} \, \big\{\pm1,\pm3,\ldots \pm (M-1)\big\} + o
    \label{eq:alphabet_A}
\end{align}
and $o \in [0,1]$ is an offset. We refer to $c\cdot \mathcal{M}$ as an $M$-ASK-$o$ alphabet. For example, $o=1$ and $o=0$ give classic unipolar and bipolar ASK, respectively. We use uniformly-distributed symbols, except for Sec.~\ref{subsec:shaping}. A digital-to-analog converter (DAC) generates the baseband waveform
\begin{align}
    X(t)  = \sum\nolimits_{\kappa \in \mathbb Z} X_\kappa \cdot g_\text{tx}(t-\kappa T_\text{s})
    .
    \label{eq:xt_pulseshaping}
\end{align}
where $g_\text{tx}(t)$ is a real pulse and $T_\text{s}$ is the symbol time. The baud rate is $B = 1/T_\text{s}$;
see Table~\ref{tab:Bparams} for the bandwidths of the various devices in Fig.~\ref{fig:continuous_detailed_system_model}.

\begin{table}
\centering
    {\small
      \caption{Bandwidth Parameters}
      {%
        \begin{tabular}{ll}
          \toprule
          Symbol rate & $\B=1/T_\text{s}$\\
          FD-RRC transmit pulse $g_\text{tx}(t)$ & $B(1+\atx)$\\
          Optical fiber $h(t)$ & $\infty$\\
          Photodiode filter  $g_\text{D}(t)$ & $\Bd=\B(1+\atx)$ \\
          Intensity signal $Y'(t)$ & $B'=2 B_\mathrm{D}$ \\
          Receiver electrical filter  $g_\text{rx}(t)$ & $\Brx$ \\
          ADC sampling rate & $\Bsam$\\
          ADC oversampling factor &  $\Nos = \Bsam/\B$ \\
          \bottomrule
        \end{tabular}
        \label{tab:Bparams}
      }
    }
\end{table}

In simulations, we transmit a large number $n$ of symbols, along with a cyclic prefix, to permit the use of fast Fourier transforms (FFTs) for convolutions. We choose the constellation scaling $c$ so that 
\begin{align}
    P_\text{tx} = \frac{1}{n T_\mathrm{s}} \mathbb{E}\left[ \lVert X(t) \rVert^2 \right]
    \label{eq:ptx}
\end{align}
for a prescribed average transmit power $P_\text{tx}$. 

We use frequency-domain root raised cosine (FD-RRC) pulses with roll-off factor $\alpha_\text{tx}$, $0\le \alpha_\text{tx}\le 1$, i.e., we use $g_\text{tx}(t) \,\laplace\, G_\text{RRC}(f)$ where $G_\text{RRC}(f)$ is (see \cite[Sec.~6]{gallager2008principles})
\begin{align}
\begin{cases}
    T_\text{s}, & |f| \!\leq\! \frac{1 - \alpha_\text{tx}}{2T_\text{s}} \\
    T_\text{s}\cos\!\left(\frac{\pi T_\text{s}}{2\alpha_\text{tx}}\big(|f| - \frac{1 - \alpha_\text{tx}}{2T_\text{s}}\big)\right), 
    & \frac{1 - \alpha_\text{tx}}{2T_\text{s}} \!<\! |f| \!\leq \!\frac{1 + \alpha_\text{tx}}{2T_\text{s}} \\
 0,
    & \text{otherwise .}
\end{cases}
\label{eq:fdrrc_fourier}
\end{align}
The double-sided bandwidth is $B(1+\alpha_\text{tx})$.
The choice $\alpha_\text{tx}= 0$ gives $g_\text{tx}(t) = \sinc(Bt) \,\laplace\, T_\text{s}\rect{\left(f/B\right)}$ where
\begin{align}
    \rect(x) :=
    \left\{\begin{array}{rl}
        1, & \text{if } |x| < 1/2 \\
        \frac12, & \text{if } |x| = 1/2 \\
        0, & \text{if } |x| > 1/2.
    \end{array}\right.
\end{align}
The passband signal $\sqrt{2}\Re\{X(t) e^{\mathrm{j} 2\pi f_0 t}\}$ has the same power as \eqref{eq:xt_pulseshaping} for a carrier frequency $f_0 \gg B$ \cite[Sec.~II-A]{wiener_filter_plabst2020}. We assume an ideal MZM for electrical-to-optical conversion.

We study single-polarization transmission over SSMF. This channel exhibits CD with frequency response \cite[Sec.~II-B]{wiener_filter_plabst2020}
\begin{align}
    H(f) = \exp{\left(- \mathrm{j} \omega^2 L_\text{fib} \beta_2/2 \right)}
    \label{eq:cd_response_freq}
\end{align}
where $\beta_2$ is the CD parameter, $\omega=2\pi f$ is the angular frequency and $L_\text{fib}$ is the fiber length. The noise-free signal at the output of the fiber is $Z'(t) = X(t) * h(t)$, and we account for fiber loss in the definition of the SNR; see Sec.~\ref{sec:simulation-results}.

The noise $N_1'(t)$ includes driver amplifier noise, laser phase noise, and optical noise caused by amplified spontaneous emission of an erbium-doped fiber amplifier. We model $N_1'(t)$ as a white CSCG process with a two-sided power spectral density (PSD) $\nu_1/2$ Watts per Hertz per real dimension. 
The noisy optical signal is filtered by a brickwall filter $g_\mathrm{D}(t) = B_\mathrm{D} \sinc(B_\mathrm{D}t) \,\laplace\, \rect(f/B_\mathrm{D})$. We set $B_\mathrm{D} = B(1+\alpha_\text{tx})$, i.e., the bandwidth of $g_\text{D}(t)$ is matched to that of $g_\text{tx}(t)$. 

A PD performs optical-to-electrical conversion of $Z''(t)$ and puts out the intensity signal
\begin{align} 
    Y'(t) = |Z''(t)|^2
    = \big| g_\mathrm{D}(t) * \big( Z'(t) + N_1'(t) \big) \big|^2
    \label{eq:squaring_time_cont}
\end{align}
which occupies twice the bandwidth of $Z''(t)$. 
The signal $Y'(t)$ is corrupted by additive white Gaussian noise (AWGN) $N_2'(t)$ from the electrical receiver components, e.g., the thermal noise of the PD transimpedance amplifier and the analog-to-digital converter (ADC). $N_2'(t)$ is modeled as a real Gaussian process with a two-sided PSD $\nu_2$ Watts per Hertz. Other noise, such as signal-dependent PD shot noise \cite[Sec.~4.4.1]{AgrawalThirdEdFiberOptics}, can be modeled via $N'_1(t)$. Expanding \eqref{eq:squaring_time_cont}, we have
\begin{align}
    Y''(t) &=  |Z(t)|^2  + |N_1(t)|^2 + 2 N_\text{sh}(t)+  N'_2(t)
    \label{eq:squaring_time_cont_expanded}
\end{align}
where the filtered quantities are
\begin{subequations}
\begin{align}
    Z(t) &:= g_\mathrm{D}(t) * Z'(t) \label{eq:filtered_zpt}\\
    N_1(t) &:= g_\mathrm{D}(t) * N_1'(t) \label{eq:filtered_cont_n1}\\
    N_\text{sh}(t) &:= 
    \Re\{ N_1(t) \cdot Z^*(t) \}
    \label{eq:shot_noise_n1}
    .
\end{align}
\end{subequations}
The term $N_\text{sh}(t)$ can approximate the PD shot-noise. The term $|N_1(t)|^2$ can be neglected at intermediate to high signal powers or can be included in $N'_2(t)$. Finally, $Y''(t)$ is filtered by $g_\text{rx}(t)$ and digitized by an ADC.  

\subsection{Discrete-Time Baseband Model}
\subsubsection{Sufficient Statistics}
The PD doubles the bandwidth to $\Bprime = 2 B_\mathrm{D}$. The set of functions bandlimited to $\Bprime$ Hz has a complete orthonormal basis \cite[Sec.~4.6.2]{gallager2008principles}
\begin{align}
    \{\phi_k(t)\}_{k \in \mathbb{Z}} \;\text{ where }\; \phi_k(t) := \sqrt{\Bprime} \sinc(\Bprime t-k) .
    \label{eq:sinc_basis}
\end{align}
Note that $Y''(t) = Y'(t) + N_2'(t)$, where $Y'(t)$ is bandlimited to $\Bprime$ and the noise $N_2'(t)$ has the decomposition 
\begin{align}
    N'_2(t) = \sum\nolimits_{k\in\mathbb{Z}} N_{2k}  \phi_k(t) + N_2^\perp(t)
    \label{eq:n2_decomp}
\end{align}
where the iid $N_{2k} \sim \mathcal{N}(0,\nu_2)$ represent noise inside the bandwidth $\Bprime$, and $N_2^\perp(t)$ represents independent noise outside the bandwidth $\Bprime$.
Thus, collecting the projection values
\begin{align}
\big\langle Y''(t)  \,,\, \phi_k(t) \big\rangle = \big\langle Y'(t) \,,\, \phi_k(t) \big\rangle + \big\langle N_2'(t) \,,\, \phi_k(t) \big\rangle
\label{eq:projection_Yk}
\end{align}
for $k\in \mathbb{Z}$ gives sufficient statistics. Using~\eqref{eq:squaring_time_cont}, along with~\eqref{eq:filtered_zpt} and~\eqref{eq:filtered_cont_n1}, we rewrite the first term of~\eqref{eq:projection_Yk} as
\begin{align}
\big\langle Y'(t) \,,\, \phi_k(t) \big\rangle  %
=  \frac{1}{\sqrt{\Bprime}} \cdot \big|Z(k/\Bprime) + N_1(k/\Bprime)\big|^2
.
\end{align}
The second term of~\eqref{eq:projection_Yk} simplifies to $\langle N_2'(t) \,,\, \phi_k(t) \rangle = N_{2k}$.

\subsubsection{Bandlimited ADC}
\label{sec:bandlimited_adc}
Consider the ADC in Fig.~\ref{fig:continuous_detailed_system_model} with the unit frequency-gain ADC filter
\begin{align}
\grx{t} =  \Brx  \sinc(\Brx t)
\label{eq:rx_filter}
\end{align}
with double-sided bandwidth $\Brx$. The filter puts out
\begin{align}
Y(t) = \grx{t} * \left( Y'(t) + N_2'(t) \right)
\label{eq:yt_filtered_adc}
\end{align}
which is sampled at the rate $\Bsam$, yielding
\begin{align}
    Y_k = Y(k/\Bsam).
\label{eq:yt_filtered_sampled_adc}
\end{align}
Define the oversampling factor $\Nos := \Bsam/\B$. The samples~\eqref{eq:yt_filtered_sampled_adc} are thus sufficient statistics if the ADC bandwidth and sampling rate are chosen as $\Brx = \Bsam = \Bprime$; see~\eqref{eq:projection_Yk}. 

We discuss the parameters $\Brx, \Bsam$ for sinc and FD-RRC pulses. The sinc pulse bandwidth is $\B$, and the intensity signal $Y'(t)$ has bandwidth $\Bprime = 2\B$. The choice $\Brx = 2\B$ with $\Nos = 2$ makes~\eqref{eq:yt_filtered_sampled_adc} sufficient statistics. Instead, an FD-RRC pulse with $\atx > 0$ has an intensity signal bandwidth $\Bprime > 2 \B$. To obtain sufficient statistics, we require $\Brx > 2\B$ and $\Nos > 2$. In practice, ADC samples are costly, so we consider only $\Nos=2$ and $\Bsam = 2 \B$.
Moreover, we consider two practical choices for $\Brx$ that do not provide sufficient statistics:
\begin{itemize}
\item $\Brx = 2 \B$, which models an ADC with an anti-aliasing filter that removes high-frequency noise;
\item $\Brx = \Bprime$, which models an ADC with an aliasing filter that passes some high-frequency noise.
\end{itemize}
For example, Sec.~\ref{sec:practical_tx_excess_bw} studies an FD-RRC pulse with $\alpha_\text{tx}=99\%$ and an ADC with bandwidth $\Brx = 4B$ and $N_\text{os} = 2$, even though one requires $N_\text{os} \approx 4$ for sufficient statistics.

\subsubsection{Discrete-Time Model for Sinc Pulses}
\label{sec:discrete_model}
We derive~\eqref{eq:yt_filtered_sampled_adc} for sinc pulses ($\Bd = \B$) and the ADC parameters $\Brx = 2B$, $\Bsam = 2\B$, $\Nos = 2$; see Table~\ref{tab:Bparams}. We have sufficient statistics, and the ADC filter leaves the transmit signal components unaffected.

Let $a(t) := g_\text{tx}(t) * h(t)$ be the combined DAC and fiber response with taps $a_k := a(k/\Bsam)$ for $k\in\mathbb Z$. We choose $K$ so that the $a_k$ are negligible for $k\not\in [-K,K]$. The channel memory in input and output symbols is  $K$ and $2K$, respectively. Collect the channel taps in the vector
\begin{align}
    \mathbf{a} := [a_{-K}, \ldots,  a_{K}  ]\tran \in \mathbb{C}^{2K+1 \times 1} .
    \label{eq:avec}
\end{align}

Consider the receiver samples $(Y_k)_{k=1}^{m}$, where $m = \Nos n$ for $n$ inputs $(X_\kappa)_{\kappa = 1}^n$. The discrete-time channel output is
\begin{align}
    \mathbf{Y} = \left| \mathbf{A}_{\text{Toep}}\cdot \mathbf{\Xi}_{n+K} [\mathbf{S}_0\tran , \mathbf{X}\tran ]\tran  +  \mathbf{N}_1 \right|^{2} + \mathbf{N}_2
    \label{eq:io_vec_matrix}
\end{align}
where
\begin{itemize}
    \item $\mathbf{A}_{\text{Toep}} \dimC{m \times (m + 2K)}$ is a Toeplitz matrix with shifted copies of the time-reversal of $\mathbf{a}\tran$ as its rows;
    \item $\mathbf{\Xi}_n = \mident{n} \otimes [1,0]\tran$ is a $2n \times n$-dimensional upsampling matrix;
    \item $\mathbf{S}_0 = [X_{-K+1},\ldots,X_0]\tran $ is the initial state, which we can control to act as a cyclic prefix; 
    \item the input and output vectors are
\begin{subequations}
\begin{alignat}{7}
    &\mathbf{X} = && \big[X_1, && && \! X_2, \quad \dots \quad && X_{ n} && &&\big]\tran   \dimR{n} \\
    &\mathbf{Y} = && \big[Y_1, && Y_2, && Y_3, \quad \dots && Y_{m-1}, && Y_{ m} &&\big]\tran   \dimR{m}
    \label{eq:yvec}
\end{alignat}
\end{subequations}
    where $m=2n$ and we redefined the indexing so that $Y_k := Y((k-K+1)/\Bsam)$; see~\eqref{eq:projection_Yk}. 
\end{itemize}

The noise vectors
\begin{align}
    & \mathbf{N}_1 \sim \mathcal{CN}(\mnull{m}, \mathbf{\Sigma}_{\mathbf{N}_1}), 
    &&
    \mathbf{N}_2 \sim \mathcal{N}(\mnull{m}, \mathbf{\Sigma}_{\mathbf{N}_2} )
    \label{eq:discrete_n1_n2}
\end{align}
have covariance matrices with entries
\begin{subequations}
\begin{align}
    [\mathbf{\Sigma}_{\mathbf{N}_1}]_{i,j} = \varphi_{N_1}([i-j]/\Bsam)
     \label{eq:n1_cov}
    \\   
    [\mathbf{\Sigma}_{\mathbf{N}_2}]_{i,j} = \varphi_{N_2}([i-j]/\Bsam)
    \label{eq:n2_cov}
\end{align}
\end{subequations}
based on the autocorrelation functions (ACFs)
\begin{subequations}
\begin{align}
    \varphi_{N_1}(\tau) &:= \nu_{1} B_\mathrm{D} \cdot \sinc(B_\mathrm{D} \tau ) 
    \label{eq:autocorr_n1}
    \\
    \varphi_{N_2}(\tau) &:= \nu_{2} \Brx \cdot  \sinc(\Brx \tau ) 
    \label{eq:autocorr_n2}
    .
\end{align}
\end{subequations}
The average pre-PD noise power is $\nu_{N_1} := \nu_{1} B_\mathrm{D}$ and \eqref{eq:n2_cov} simplifies to $\mathbf{\Sigma}_{\mathbf{N}_2} = \nu_{N_2}  \mident{m}$, where $\nu_{N_2} := \nu_{2} \Brx$ is the average post-PD noise power.

\subsubsection{Surrogate Model for FD-RRC Pulses}
\label{sec:discrete_model_fdrrc}
Consider FD-RRC transmit pulses with $\atx > 0$. The ADC signal after front-end filtering with bandwidth $\Brx$ is 
\begin{align}
Y(t) = \underbrace{\grx{t} * Y'(t)}_{\displaystyle\text{ Filtered intensity signal}} +  \underbrace{\grx{t} * N_2'(t)}_{\displaystyle\text{Filtered electrical noise}\quad}
.
\label{eq:yt_filtered_adc_repeat}
\end{align}
We sample~\eqref{eq:yt_filtered_adc_repeat} with $\Bsam = 2\B$ and $\Nos=2$ to obtain the samples~\eqref{eq:yt_filtered_sampled_adc} $\vY \in \mathbb{R}^m$ as in~\eqref{eq:yvec}. However, for receiver processing, we use the surrogate channel model $q_{\vY|\vX}$ that corresponds to~\eqref{eq:io_vec_matrix} with PD bandwidth $\Bd = (1+\atx)\B$ and ADC bandwidth $\Brx$. 

We discuss the impact of the surrogate model for different choices of $\atx$ and $\Brx$. We focus on FD-RRC pulses with small $\atx = 1\%$, where $\Bd$ is only slightly larger than $\B$. Thus, the bandwidth $\Bprime$ of $Y'(t)$ is only slightly larger than $2\B$ and the ADC filter with bandwidth $\Brx = 2 \B$ filters out very little signal energy. The surrogate $q_{\vY|\vX}$ thus accurately models the electrical noise (see~\eqref{eq:discrete_n1_n2}), but neglects the effect of $\grx{t}$ on the intensity $Y'(t)$. Fig.~\ref{fig:aa_receiver} shows the spectrum of $|\gtx{t}|^2$ together with the bandwidth of $\grx{t}$. The relative energy removed by the ADC filter is below $\SI{1e-7}{}$.

When optical noise dominates, we choose $\Brx = \Bprime$. Thus, the ADC filter has no effect and the surrogate $q_{\vY|\vX} = p_{\vY|\vX}$ correctly models the aliased samples~\eqref{eq:yt_filtered_sampled_adc}. 

\begin{figure}
    \centering
    \input{Fig/anti_aliasing_receiver.tex}
    \caption{Spectrum of $|\gtx{t}|^2$ for a FD-RRC pulse with $\atx=1\%$. The dotted lines mark the bandwidth $\Brx = 2\B$ of $\grx{t}$.}
    \label{fig:aa_receiver}
\end{figure}

\subsection{Cyclic Convolution}
\subsubsection{Channel Matrix}
To simplify computation of~\eqref{eq:io_vec_matrix}, we use a cyclic prefix with block length $n > K$, i.e., we set $\mathbf{S}_0 = [X_{n-K+1},\ldots, X_{n}]$ so that 
\begin{align}
    &\mathbf{A}_{\text{Toep}}\cdot \mathbf{\Xi}_{n+K} [\mathbf{S}_0\tran , \mathbf{X}\tran ]\tran
    = \mathbf{A} \mathbf{X}, 
    && \mathbf{A} :=\Ac \mathbf{\Xi}_n 
    \label{eq:A_cp_simple}
\end{align}
where $\Ac$ is a circulant matrix whose columns are cyclically shifted copies of $\mathbf{a}$ in~\eqref{eq:avec}. 
DFTs decompose the convolution into independent parallel channels, and FFTs speed up the simulations. 
The following steps are for $N_\text{os}=2$. The steps for $N_\text{os} > 2$ differ slightly and require adjustments.

We have $\Ac = \mathbf{F}_m\herm \mathbf{\Lambda} \mathbf{F}_m $, where $\mathbf{F}_m \in \mathbb{C}^{m \times m}$ is the unitary DFT matrix and $\mathbf{\Lambda} := \diag(\lambda_0, \ldots \lambda_{m-1})$ where the $\lambda_u$ 
are DFT values of the zero-padded ${\mathbf{a}}$ in \eqref{eq:avec} and $u = 0, 1, \ldots, m-1$.
For large $m$, the $\lambda_u$ become samples of the discrete-time Fourier transform (DTFT) of $\mathbf{a}$  on the frequency interval $[0, \Bsam)$, i.e., we have
\begin{align}
    \lambda_u &= \Bsam \cdot e^{-\mathrm{j} 2 \pi K \frac{u}{m}} \cdot A\left(\Bsam \left(\frac{u}{m}\ \mathrm{mod}\ 1 \right)\right) 
    \label{eq:eigs_channel}
\end{align}
with $A(f) = G_\text{RRC}(f) H(f)$, where the scaling factor $\Bsam$ results from the Poisson summation formula. Now observe that~\eqref{eq:cd_response_freq} gives $|H(f)|=1$ and \eqref{eq:fdrrc_fourier} gives
\begin{align}
    G_\text{RRC}(f)^2 + G_\text{RRC}(f-B)^2
    = T_\text{s}^2
\end{align}
for $0 \le f \le B$. With $\Nos = 2$ and $\Bsam = \Nos \B$, we have 
\begin{align}
    |\lambda_u|^2 + |\lambda_{u+n}|^2 = \Bsam^2 T_\text{s}^2
    = \Nos^2
    \label{eq:lambda_u}
\end{align}
for $u=0,1,\dots,n-1$; see Fig.~\ref{fig:evd_svd} below.

Multiplying $\Ac$ by $\mathbf{\Xi}_n$ (see \eqref{eq:A_cp_simple}) gives 
\begin{align}
    \mathbf{A} = \mathbf{F}_m\herm \mathbf{\Lambda} (\mathbf{F}_m \mathbf{\Xi}_n)
    = \mathbf{F}_m\herm \mathbf{\Lambda} \cdot 
    \frac{1}{\sqrt{N_\text{os}}} 
    \begin{bmatrix}
        \mathbf{F}_{n} \\ \mathbf{F}_{n}
    \end{bmatrix} .
    \label{eq:A_eff_chan}
\end{align}
The singular values of \eqref{eq:A_eff_chan} are computed via 
\begin{align}
    \mathbf{A}\herm \mathbf{A} 
    = \begin{bmatrix}
        \mathbf{F}_{n}\herm & \mathbf{F}_{n}\herm
    \end{bmatrix} 
    \frac{\mathbf{\Lambda}\herm\mathbf{\Lambda}}{N_\text{os}}
        \begin{bmatrix}
        \mathbf{F}_{n} \\ \mathbf{F}_{n}
    \end{bmatrix} 
    = \mathbf{F}_n\herm \mathbf{\Sigma}_\mathrm{s}^2 \mathbf{F}_n 
\end{align}
with
\begin{align}
    \mathbf{\Sigma}_\mathrm{s}^2 = \tfrac{1}{N_\text{os}}  |\mathbf{\Lambda}_1|^2  +  \tfrac{1}{N_\text{os}}|\mathbf{\Lambda}_2|^2
    \label{eq:sigma_mat_sq}
\end{align}
and $\mathbf{\Lambda} = \blkdiag(\mathbf{\Lambda}_{1}, \mathbf{\Lambda}_{2})$ with diagonal  $\mathbf{\Lambda}_{1}, \mathbf{\Lambda}_{2} \in \mathbb{C}^n$ that each contain $n = m/2$ eigenvalues of $\Ac$. 
The expression \eqref{eq:sigma_mat_sq} is a scaled-identity matrix for FD-RRC pulses; see \eqref{eq:lambda_u} and Fig.~\ref{fig:evd_svd} below. The SVD thus gives 
\begin{align}
    \mathbf{A} = \underbrace{\tfrac{1}{N_\text{os}}\mathbf{F}_m\herm  
    \begin{bmatrix}
        \mathbf{\Lambda}_1\\  \mathbf{\Lambda}_2
    \end{bmatrix} 
    }_{\displaystyle \mathbf{U}_\mathrm{s}} \cdot \underbrace{\sqrt{N_\text{os}}\, \mident{n}}_{\displaystyle \mathbf{\Sigma}_\mathrm{s}} \cdot \underbrace{\mathbf{F}_n}_{\displaystyle \mathbf{V}_\mathrm{s}\herm}.
    \label{eq:A_svd}
\end{align}
with semi-unitary $\mathbf{U}_\mathrm{s}$, scaled-identity $\mathbf{\Sigma}_\mathrm{s}$ and unitary $\mathbf{V}_\mathrm{s}$. Furthermore, we have $\rank(\mathbf{A}) = n$.

\subsubsection{Optical Noise}
The covariance matrix $\mathbf{\Sigma}_{\mathbf{N}_1}$ \eqref{eq:n1_cov} of the optical noise $\mathbf{N}_1$ is symmetric Toeplitz. We approximate $\mathbf{\Sigma}_{\mathbf{N}_1}$ by a circulant matrix
\begin{align}
    \mathbf{\Sigma}_{\mathbf{N}_1}
    \approx  \mathbf{F}_m\herm \mathbf{\Lambda}_\mathrm{D}  \mathbf{F}_m
    \label{eq:noise_cov1_approx}
\end{align}
where $\mathbf{\Lambda}_\mathrm{D} = \diag(\lambda_{\mathrm{D},0}, \ldots \lambda_{\mathrm{D},m-1})$ with eigenvalues 
\begin{align}
   \lambda_{\mathrm{D},u} = \Bsam \nu_1 \rect\left( \left(\modone{\frac{u}{m}}\right) \frac{\Bsam}{B_\mathrm{D}}\right)
   \label{eq:psd_opt_noise}
\end{align}
for $u=0,1,\dots,m-1$. 
One obtains the spectrum \eqref{eq:psd_opt_noise} by sampling the DTFT of $\varphi_{N_1}(\cdot)$ in \eqref{eq:n1_cov} at $f=\frac{u}{m}\Bsam$; see \eqref{eq:eigs_channel} and \cite[Sec.~1.2]{gray2006toeplitz}.

The approximation \eqref{eq:noise_cov1_approx} is accurate for large $m$ and asymptotically equivalent to~\eqref{eq:n1_cov} if $g_\text{D}(t)$ is given a small positive roll-off factor \cite{gray2006toeplitz}. Moreover, \eqref{eq:noise_cov1_approx} reduces the complexity of the LMMSE denoiser \eqref{eq:tab_sigma_n1_tild_def}-\eqref{eq:w2_den_var}, since the channel matrices~\eqref{eq:A_svd}-\eqref{eq:noise_cov1_approx} diagonalize with DFTs; see~\eqref{eq:sigma_u_giv_y_simp}. Such diagonalization is possible only when the EP messages have the same bases as the channel matrices, which, for DFTs, includes scaled identity covariance matrices but not general diagonal covariance matrices. Also, FFTs accelerate the DFTs.

\subsubsection{Precoding}
\label{sec:precoding}
GVAMP is not necessarily replica Bayes-optimal for our channels. However, we observed that inference improved by inserting a $n \times n$ precoder $\mathbf{P}$ to approximate \emph{right}-rotational invariance, i.e., we replaced \eqref{eq:A_svd} with
\begin{align}
    \mathbf{A} &:= \mathbf{U}_\mathrm{s} \mathbf{\Sigma}_\mathrm{s}
    \mathbf{V}_\mathrm{s}\herm \mathbf{P} .
    \label{eq:A_svd2}
\end{align}
To simplify simulations, we use FFT-based precoders with
\begin{align}
    \mathbf{P} = \mathbf{F}\herm_n \mathbf{\Lambda}_\mathrm{P} \mathbf{F}_n
    \label{eq:fft_precoder}
\end{align}
where $\mathbf{\Lambda}_\mathrm{P} := \diag(\lambda_{\mathrm{P},0}, \ldots \lambda_{\mathrm{P},n-1})$. Inserting \eqref{eq:fft_precoder} into \eqref{eq:A_svd2} slightly modifies \eqref{eq:A_svd} to become
\begin{align}
    \mathbf{A} = \underbrace{\tfrac{1}{N_\text{os}}\mathbf{F}_m\herm 
    \begin{bmatrix}
        \mathbf{\Lambda}_1\\  \mathbf{\Lambda}_2
    \end{bmatrix} 
    }_{\displaystyle \mathbf{U}_\mathrm{s}} \cdot \underbrace{\sqrt{N_\text{os}}\, \mident{n}}_{\displaystyle \mathbf{\Sigma}_\mathrm{s}} \cdot \underbrace{\mathbf{\Lambda}_\mathrm{P} \mathbf{F}_n}_{\displaystyle  \mathbf{\Lambda}_\mathrm{P} \mathbf{V}_\mathrm{s}\herm}.
    \label{eq:A_svd3}
\end{align}

We use real precoding with unit-modulus eigenvalues, i.e., $\lambda_{\mathrm{P},u} = e^{\mathrm{j} \psi_u}$ and $\lambda_{\mathrm{P},n-u} = \lambda^*_{\mathrm{P},u}$ for $u = 1,\ldots,n/2-1$ where $n$ is even. We choose the $\psi_u$ iid from a uniform distribution:
\begin{align}
    & \psi_u \sim 
    \, \mathcal{U}[0,2 \pi), 
    && u \in \{ 1, \ldots, n/2-1 \} 
    \label{eq:precoder_diag}
\end{align}
and $\psi_0$ and $\psi_{n/2}$ are drawn independently and uniformly at random from $\{0, \pi\}$. Note that complex precoding can zero-force the unitary fiber response; however, this requires two MZMs. Precoding can also compensate for the bandwidth limitations of the transmitter or receiver components.

\section{Multilevel Coding and Successive Interference Cancellation}
\label{sec:MLC-SIC}

\subsection{Achievable Information Rates}
\label{subsec:AIRs}

Let $\mathbf{X}$ have dimension $n$ and $\mathbf{Y}$ have dimension proportional to $n$. The mutual information and GMI rates are
\begin{subequations}
\begin{align}
    I_{n}(\mathbf{X}; \mathbf{Y}) & = \frac1n I(\mathbf{X}; \mathbf{Y}) \\
    I_{n,\text{GMI}}(\mathbf{X}; \mathbf{Y}) & = \sup_{s \geq 0} 
    \frac1n \mathbb{E}\!\left[
    \log_2 \frac{q(\mathbf{X},\mathbf{Y})^s}{ \mathbb{E}_{\widetilde{\mathbf{X}}} \big[ q( \widetilde{\mathbf{X}}, \mathbf{Y})^s  \big] }
    \right]
    \label{eq:gmi}
\end{align}
\end{subequations}
where $q(\cdot)$ is a non-negative function called the decoding metric, and $\mathbb{E}_{\widetilde{\mathbf{X}}}[\cdot]$ denotes taking expectation with respect to $\widetilde{\mathbf{X}}$ only, where $\widetilde{\mathbf{X}}$ has the same distribution as $\mathbf{X}$. We have 
\begin{equation}
    I_{n,\text{GMI}}(\mathbf{X}; \mathbf{Y})
    \le I_{n}(\mathbf{X}; \mathbf{Y})
\end{equation}
so the GMI rate is an achievable information rate (AIR); see \cite{gallager1968,Kaplan:93:Mismatched,Scarlett20,kramer2023adaptive}. We define the limiting rate
\begin{align}
    I(\mathcal{X};\mathcal{Y}) & := \lim_{n\rightarrow \infty} I_{n}(\mathbf{X}; \mathbf{Y}).
\end{align}

\subsection{Separate Detection and Decoding}
\label{subsec:SDD}
Joint detection and decoding (JDD) can approach the rate $I(\mathcal{X};\mathcal{Y})$, but JDD is usually too complex to implement \cite{muller_capacity_separate2004}. Two practical alternatives use SDD in a turbo loop \cite{Douillard:95:iterative,alexander-ETT98,wang1999iterative} or with MLC-SIC \cite{imai1977new,wachsmann1999multi,Guess:00:ISI,pfister2001achievable,soriaga2007}.
AMP detectors can perform SDD, e.g., they have been used for turbo detection and decoding (TDD) \cite{liu2021capacity,liu2024capacity,chi2024gampgvamp}. For linear systems and iid Gaussian $\mathbf{A}$, the receiver can approach the rate $I(\mathcal{X};\mathcal{Y})$ for $\mathbf{X}$ with iid entries \cite{liu2021capacity}. The results extend to right-rotationally invariant $\mathbf{A}$ \cite{liu2024capacity}. The paper \cite{chi2024gampgvamp} compares achievable rates of TDD with GAMP and GVAMP detectors for non-linear $f(\cdot)$ corresponding to clipping and quantization, and with iid Gaussian $\mathbf{A}$ and partial unitary $\mathbf{A}$. However, error control codes for TDD should be carefully designed to match the decoder and detector EXIT functions, which depend on channel memory, signal-to-noise ratio (SNR), and modulation format \cite{ashikhmin04}. In contrast, MLC with SIC can use codes designed for memoryless channels \cite{pfister2001achievable,soriaga2007}.

\subsection{MLC-SIC AIRs}
\label{subsec:SDD-AIRs}

We combine SDD with MLC-SIC as in \cite{Guess:00:ISI,pfister2001achievable,soriaga2007,prinz2023successive}. 
Consider the APPs $M_\kappa=P_{X_\kappa|\mathbf{Y}}(\cdot |\mathbf{y})$ for $\kappa \in \idxset n$ that are functions of $\mathbf{y}$. We have
\begin{align}
    I_n(\mathbf{X};\mathbf{Y}) 
    &\overset{(a)}\ge \frac{1}{n} \sum\nolimits_{\kappa=1}^n H(X_\kappa) -  H(X_\kappa|\mathbf{Y})  
    \nonumber %
    \\
    &\overset{(b)}=   \frac{1}{n} \sum\nolimits_{\kappa=1}^n I(X_\kappa ; M_\kappa) := I_\text{$n$,SDD} \label{eq:separate_detection_decoding_rate}
\end{align}
where step $(a)$ follows because conditioning cannot increase entropy, and step $(b)$ because $M_\kappa$ is a sufficient statistic \cite[Ch.~2.9]{cover1991elementsofIT}. 
The rate \eqref{eq:separate_detection_decoding_rate} corresponds to treating $(M_\kappa)_{\kappa\in\mathbb Z}$ as memoryless: inserting $q(\mathbf{x},\mathbf{y}) = \prod_\kappa M_\kappa(x_{\kappa}) /P(x_{\kappa})$ in \eqref{eq:gmi} gives $I_{n,\text{GMI}}(\mathbf{X}; \mathbf{Y}) = I_{n,\text{SDD}}$.

SIC creates $S$ strings of length $\siclength = n/S$ (assume $\siclength \in \mathbb{Z}$)
\begin{align}
    \mathbf{V}_{\sics}
    = \big( V_{\sics,\sicindex} \big)_{\sicindex=1}^{\siclength}
    = \big(X_{\mathcal{I}^\ell_1}, X_{\mathcal{I}^\ell_2}, \ldots, X_{\mathcal{I}^\ell_N}   \big) 
    \label{eq:subsampling}
\end{align}
where $\mathcal{I}^\ell := \{\sics + (t-1)  S \mid 1 \leq t \leq N \}$.  Stacking the $\mathbf{V}_{\sics}$ in \eqref{eq:subsampling} forms the vector $\mathbf{V} = ( \mathbf{V}_\sics )_{\sics=1}^S$. 
We have $I_n(\mathbf{X};\mathbf{Y}) = I_n(\mathbf{V};\mathbf{Y})$ and (see \cite[Eq.~(11)-(13)]{prinz2023successive})
\begin{align}
    I_n(\mathbf{V};\mathbf{Y}) %
    \geq \underbrace{\frac{1}{S} \sum_{\sics=1}^S  \frac{1}{\siclength} \sum_{\sicindex=1}^{\siclength} I\left(V_{\sics,\sicindex} ; \mathbf{Y},\mathbf{V}^{\sics-1}\right)}_{\displaystyle := I_\text{$n$,SIC}}
    \geq I_{n,\text{SDD}}
    \label{eq:sic_inequality_alt}
    .
\end{align}
Define $M_{\ell,t} := P_{V_{\ell,t}|\mathbf{Y}, \mathbf{V}^{\ell-1}}(\cdot |\mathbf{y}, \mathbf{v}^{\ell-1})$ so the normalized inner sum of \eqref{eq:sic_inequality_alt} is
\begin{align}
    I^{\ell}_{N\!,\text{SIC}} = \frac{1}{\siclength} \sum\nolimits_{t = 1}^N I(V_{\ell,t};  M_{\ell,t})
    \label{eq:per_stage_rate}
\end{align}
which is non-decreasing in $\ell$ and  at most $H(V_{\ell,t})$. The rate $I_\text{$n$,SIC}$ corresponds to using $q(\mathbf{v}, \mathbf{y}) = \prod_{\ell, t} M_{\ell,t}(v_{\ell,t})/P(v_{\ell,t})$ in \eqref{eq:gmi} and we obtain $I_{n,\text{GMI}}(\mathbf{V}; \mathbf{Y}) = I_{n,\text{SIC}}$. 

\begin{figure}
    \centering
    \usetikzlibrary{decorations.markings}
\tikzset{node distance=1cm}

\pgfdeclarelayer{background}
\pgfdeclarelayer{foreground}
\pgfsetlayers{background,main,foreground}

\pgfmathsetmacro{\samplerwidth}{30}

\tikzset{redbox/.style = {draw,minimum height=1.9em,minimum width=3.6em,fill=white}}
\tikzset{bluebox/.style = {draw,minimum height=1.9em,minimum width=3.6em,fill=white}}
\tikzset{greenbox/.style = {draw,minimum height=1.9em,minimum width=3.6em,, opacity=1,fill=white},
mydash/.style={rectangle,fill=white}}
\tikzset{dot/.style = {anchor=base,fill,circle,inner sep=1pt}}
\tikzstyle{point}=[fill,shape=circle,minimum size=3pt,inner sep=0pt]

\begin{tikzpicture}[]
\footnotesize

\node[] (y) {};
\node[redbox,right of=y,align=center,node distance=1.2cm,font=\footnotesize,label={[xshift=+0.5cm,yshift=-0.05cm]above:{}}] (app1) {APP 1};
\node[redbox,right of=app1,align=center,node distance=1.20cm] (dec1) {Dec. 1};

\node[bluebox,below of=app1,yshift=-0.10cm,xshift=1.55cm,align=center,label={[xshift=0.72cm,yshift=-0.05cm]above left:{}}] (app2) {APP 2};
\node[bluebox,right of=app2,node distance=1.20cm,align=center] (dec2) {Dec. 2};

\node[below of=app2,node distance=0.70cm,xshift=-0.2cm,yshift=0.1cm] (recDots) {$\ddots$};
\node[greenbox,below of=app2,yshift=-0.15cm,xshift=1.9cm,align=center,label={[xshift=0.7cm,yshift=-0.05cm]above left:{}}] (appM) {APP $S$};
\node[greenbox,right of=appM,node distance=1.2cm,align=center] (decM) {Dec. $S$};

\draw[-latex] (app1 -|  y) -- (app1);
\draw (app1) --(dec1);

\draw[-latex] (app1 -| y) |- (appM);
\draw[-latex] (app1 -| y) |- (app2);

\node[xshift=-0.3cm,yshift=-0.2cm](y_desc) at(y |- app2) {$\mathbf{Y}$}; 

\draw[-latex] (dec1) -| node[mydash,pos=0.87]{$\rvdots$} (appM.30) node[midway,above](v1label) {$\hat{\mathbf{V}}_1$};
\draw[-latex] (dec1) -| (app2.40);
\draw[-latex] (app2) --  (dec2) -| node[mydash,pos=0.70]{$\rvdots$} (appM.47) node[midway,above] {$\hat{\mathbf{V}}_2$};

\node[xshift=1.2cm,font=\footnotesize](out1) at(dec1 -| decM) {$\hat{\mathbf{V}}_1$}; 
\node[xshift=1.2cm,font=\footnotesize](out2) at(dec2 -| decM) {$\hat{\mathbf{V}}_2$};
\node[xshift=1.2cm,font=\footnotesize](outM) at(decM -| decM) {$\hat{\mathbf{V}}_S$};

\draw (appM) -- (decM); 
\path[] (app1) -- (appM);

\draw[-latex,] (dec1 -| appM.21) -- (out1);
\draw[-latex,] (dec2) -- (out2);
\draw[-latex,] (decM) -- (outM);

\draw [decorate, decoration = {mirror,brace}] ($(app1.west) + (0,-0.35)$) -- node[midway,below,font=\footnotesize] {Stage 1} ($(dec1.east) + (0,-0.35)$);
\draw [decorate, decoration = {mirror,brace}] ($(app2.west) + (0,-0.35)$) -- node[midway,below,font=\footnotesize] {Stage 2} ($(dec2.east) + (0,-0.35)$);
\draw [decorate, decoration = {mirror,brace}] ($(appM.west) + (0,-0.35)$) -- node[midway,below,font=\footnotesize] {Stage $S$} ($(decM.east) + (0,-0.35)$);

\end{tikzpicture}
    \caption{SIC receiver with $S$ stages.}
    \label{fig:sic_receiver}
\end{figure}

Fig.~\ref{fig:sic_receiver} shows a SIC receiver with $S$ stages \cite{plabst2024neural}. The first stage performs SDD and computes $M_{1,t}=P_{V_{1,t} | \mathbf{Y}}(\cdot | \mathbf{y})$, $t \in \idxset N$, and the estimate $\hat{\mathbf{V}}_1$. The second stage computes $\hat{M}_{2,t} := P_{V_{2,t} | \mathbf{V}_1,\mathbf{Y}}(\cdot |  \hat{\mathbf{v}}_1,\mathbf{y})$, $t \in \idxset N$, and $\hat{\mathbf{V}}_2$. The other stages work similarly. 
Define the limiting rates
\begin{align}
    & I^{\ell}_{\text{SIC}} = \lim_{N \rightarrow \infty } I^{\ell}_{N\!,\text{SIC}}, 
    &&
    I_{\text{SIC}} = \lim_{n \rightarrow \infty } I_{n,\text{SIC}}.
\end{align}
MLC-SIC achieves $I_{\text{SIC}}$ if $\hat{\mathbf{V}}_\ell=\mathbf{V}_\ell$ for all $\ell$.
This can be accomplished by encoding at rates less than $I^{\ell}_{\text{SIC}}$ for all $\ell$.
SIC approaches $I(\mathcal{X};\mathcal{Y})$ by increasing $S$ \cite[Fig.~6]{prinz2023successive}. In general, $S$ should grow with the total memory $K$, modulation alphabet size, and ISI magnitude; see \cite[Sec.~IV]{prinz2023successive}.

\section{EP Messages}
\label{sec:vamp}
Consider SIC stage $\ell$ and the discrete-time model based on writing \eqref{eq:io_vec_matrix} with \eqref{eq:A_cp_simple} as
\begin{align}
    \mathbf{Y} %
    &=  \big| %
     \underbrace{
     \Ad \mathbf{U} +  \mathbf{S}
    + \mathbf{N}_1 }_{\displaystyle := \mathbf{W}}
    \big|^{2} + \mathbf{N_2}
    \label{eq:dp_vector_model}
\end{align}
where
\begin{align}
    \label{eq:sic_interference}
    & \Ad \mathbf{U} = \sum\nolimits_{i \geq \ell } \mathbf{A}_i \mathbf{V}_i,  
    &&
    \Sint= \sum\nolimits_{i < \ell } \mathbf{A}_i \mathbf{V}_i
\end{align}
and 
\begin{subequations}
\begin{alignat}{2}
    \Ad & = [\mathbf{A}_\ell, \mathbf{A}_{\ell+1}, \ldots \mathbf{A}_S] && \in \mathbb{C}^{m \times n'}
    \label{eq:ad_construction}\\
    \Usym & = [\mathbf{V}_\ell\tran, \mathbf{V}_{\ell+1}\tran, \ldots \mathbf{V}_S\tran]\tran && \in \mathbb{R}^{n'}
    \label{eq:usym_remaining_constr}
\end{alignat}
\end{subequations}
where $\mathbf{A}_\ell := [\mathbf{a}_{\mathcal{I}^\ell_1}, \ldots \mathbf{a}_{\mathcal{I}^\ell_N}]$, $\mathbf{a}_\kappa$ is the $\kappa^\text{th}$ column of $\mathbf{A}$, $\mathbf{V}_\ell = [X_{\mathcal{I}^\ell_1}, \ldots X_{\mathcal{I}^\ell_N}]\tran$ and $n' = (S-\ell-1)N$. The vector $\mathbf{S} \in \mathbb{C}^{m}$ is known interference from the symbols $(\mathbf{V}_i)_{i<\ell}$ of the previous SIC stages. $\mathbf{W}$ models the signal before the PD, including the pre-intensity noise.

\subsection{GVAMP Messages via EP}
We wish to compute $M_{\ell,t} = P_{V_{\ell,t} | \mathbf{S},\mathbf{Y}}(\cdot | \mathbf{s},\mathbf{y})$ for $t \in \idxset N$; see \eqref{eq:per_stage_rate}. Belief propagation cannot be implemented for general densities; we instead use EP with Gaussian messages. 

Consider the conditional probability factorization
\begin{align}
    \underbrace{P(\usym)}_{\displaystyle t_1(\usym)} \,
    \underbrace{p(\z|\usym,\sint)}_{\displaystyle t_2(\usym,\z)} \, \underbrace{p(\mathbf{y}|\z)}_{\displaystyle t_3(\z)}
    \label{eq:joint-pdf}
\end{align}
which gives the factor graph in Fig.~\ref{fig:factor_graph} where
\begin{itemize}[leftmargin=*] \itemsep 0pt
    \item factor node $t_1 = P_{\Usym}$ represents the real-valued iid prior;
    \item factor node $t_2(\mathbf{u},\mathbf{w}) = \mathcal{CN}(\mathbf{w}; \Ad \mathbf{u} + \mathbf{s}, \mathbf{\Sigma}_{\mathbf{N}_1})$ represents the linear equation relating $\mathbf{U}$ and the complex-valued $\Z$ under optical noise $\mathbf{N}_1$;
    \item factor node $t_3 =\mathcal{N}(|\z|^2, \nu_{N_2} \mident{})$ represents the DD squaring function and electrical noise;
    \item we discarded $\sint$ and $\mathbf{y}$ in $t_1$ and $t_3$ because they are constants. 
\end{itemize}    

\begin{figure}
    \centering
    \resizebox{
    \IfOneCol{0.45\columnwidth}
    \IfTwoCol{0.8\columnwidth}
    }
    {!}{\begin{tikzpicture}[scale = 1.58]
    \newcommand{\vertex}{\node[vertex]}
    \tikzset{vertex/.style = {circle, draw, inner sep = 0pt, minimum size = 10pt}}
    \vertex[label = $\Usym$](up) at (-1,0) {};
    \vertex[label = $\Z$](wp) at (1,0) {};
    \tikzset{vertex/.style = {rectangle, fill = none, draw=black, inner sep = 0pt, minimum size = 13pt}};
    \tikzset{vertex/.style = {rectangle, fill = none, draw=black, inner sep = 0pt, minimum size = 18pt}}
    \vertex[label = above: %
    ](a) at (-2,0) {$t_1$};
    \vertex[label = above: 
    ](b) at (0,0) {$t_2$};
    \vertex[](c) at (2,0) {$t_3$};

    \tikzset{myarR/.style={
        -{latex}, %
        red,
        shorten <=0.20cm,
        shorten >=0.20cm
    }}; 

    \tikzset{myarC/.style={
        -{latex}, %
        blue,
        shorten <=0.20cm,
        shorten >=0.20cm
    }};

    \draw (a)--(up);
    \draw (up)--(b);
    \draw (b)--(wp);
    \draw (wp)--(c);
    
\end{tikzpicture}}
    \caption{Factor graph of GVAMP. Circles and squares represent variable and factor nodes, respectively.}
    \label{fig:factor_graph}
\end{figure}

The EP functions $\widetilde{t}_1,\widetilde{t}_2,\widetilde{t}_3$ approximate the respective factor nodes $t_1,t_2,t_3$:
\begin{subequations} 
\label{eq:post_approx_fact}
\begin{align}
    \label{eq:post_approx_fact_a}
    t_1(\mathbf{u}) &\;\leftrightarrow\; \widetilde{t}_1(\mathbf{u}) \\
    t_2(\mathbf{u},\mathbf{w}) &\;\leftrightarrow\; \widetilde{t}_2(\mathbf{u},\mathbf{w}) = \widetilde{t}_{21}(\mathbf{u}) \cdot \widetilde{t}_{23}(\mathbf{w}) \label{eq:factor_c_approx}
    \\
    t_3(\mathbf{w}) &\;\leftrightarrow \;\widetilde{t}_3(\mathbf{w})
    \label{eq:post_approx_fact_e}
\end{align}
\end{subequations}
where \eqref{eq:factor_c_approx} models $\widetilde{t}_{21}$ and $\widetilde{t}_{23}$ as independent priors. We choose
\begin{subequations} 
\begin{align}
    \widetilde{t}_{1}  = \mathcal{N}(\mathbf{r}_2, \nu_{U_2} \mathbf{I})  \label{eq:t1til} && 
    \widetilde{t}_{21} = \mathcal{N}(\mathbf{r}_1, \nu_{U_1} \mathbf{I}) \\
    \widetilde{t}_{23} =\mathcal{CN}(\mathbf{p}_1, \nu_{W_1} \mathbf{I})  && 
    \widetilde{t}_{3}  =\mathcal{CN}(\mathbf{p}_2, \nu_{W_2})  \label{eq:t3til}
\end{align}
\end{subequations} 
and Appendix~\ref{appendix-EP} shows how the parameters of~\eqref{eq:t1til}-\eqref{eq:t3til} are updated; see~\eqref{eq:ext_input}-\eqref{eq:ext_output} and~\eqref{eq:ext_lmmse_u}-\eqref{eq:ext_lmmse_w}. 

Fig.~\ref{fig:block-diagram-GVAMP} depicts the EP operation, and Table~\ref{tab:explicit_denoisers} gives the input and LMMSE denoiser expressions. The LMMSE denoiser is derived in Appendix \ref{appendix-LMMSE-denoiser}, and the non-linear output denoiser is derived in Appendices \ref{appendix-chi2}, \ref{appendix-denoising-w1}, and \ref{appendix-saddle_point}. The ``ext'' blocks generate extrinsic messages.

\begin{figure*}
\centering
\resizebox{
    \IfOneCol{0.95\columnwidth}
    \IfTwoCol{1.9\columnwidth}
    }
    {!}{\begin{tikzpicture}
  \node[block, fill=gray!15] (p_u) {\normalsize $\Usym$};
  \node[block, fill=gray!15, minimum width=3.2cm,,text width=3.0cm, right= 5cm of p_u] (z_uv) {\normalsize  $\Z = \Ad \Usym + \mathbf{s} + \mathbf{N}_1$};
  \node[block, fill=gray!15, right= 5cm of z_uv,minimum width=1.5cm,text width=2.8cm] (phi_z) {\normalsize
  $\bm{Y} = |\Z|^{2} + \mathbf{N}_2$};

  \node[above of=p_u,node distance=1.8cm]{\textbf{Input denoiser}}; 
  \node[above of=z_uv,node distance=1.8cm]{\textbf{LMMSE denoiser}}; 
  \node[above of=phi_z,node distance=1.8cm]{\textbf{Output denoiser}}; 
  
  \node[blockExt,right=of p_u, xshift=0.4cm, yshift=-1cm] (ext_pu_to_z_uv) {$\mathrm{\textbf{ext}}$};
  \node[blockExt,left=of z_uv, xshift=-0.4cm, yshift=1cm] (ext_z_uv_to_p_u) {$\mathrm{\textbf{ext}}$};
  \node[blockExt,right=of z_uv, xshift=0.4cm, yshift=-1cm] (ext_z_uv_to_y) {$\mathrm{\textbf{ext}}$};
  \node[blockExt,left=of phi_z, xshift=-0.4cm, yshift=1cm] (ext_y_to_z_uv) {$\mathrm{\textbf{ext}}$};

  \draw [-latex,very thick] (p_u.east |- ext_pu_to_z_uv) -- 
  node [midway,below=0em,align=center ] { $\hat{\usym}_1$}
  node [midway,below=1.3em,align=center ] {${\alpha}_1$}
  (ext_pu_to_z_uv.west);
  \draw [-latex,very thick] (ext_pu_to_z_uv) --
  node [midway,below=0em,align=center ] { $\mathbf{r}_2$}
  node [midway,below=1.3em,align=center ] {$\nu_{U_2}$}
  (z_uv.west |- ext_pu_to_z_uv)
  node [pos=0.25](ext_between_pu_z_uv){};
  \draw [-latex,very thick] (z_uv.east |- ext_z_uv_to_y.west) --
  node [midway,below=0em,align=center ] { $\hat{\z}_2$}
  node [midway,below=1.3em,align=center ] {${\beta}_2$}
  (ext_z_uv_to_y.west);
  \draw [-latex,very thick] (ext_z_uv_to_y.east) --
  node [midway,below=0em,align=center ] { $\mathbf{p}_1$}
  node [midway,below=1.3em,align=center ] {$\nu_{W_1}$}
  (phi_z.west |- ext_z_uv_to_y)
  node [pos=0.25](ext_between_z_uv_y){};
  \draw [-latex,very thick] (phi_z.west |- ext_y_to_z_uv.east) --
  node [midway,above=0em,align=center ] { $\hat{\z}_1$}
  node [midway,above=1.3em,align=center ] {${\beta}_1$}
  (ext_y_to_z_uv.east);
  \draw [-latex,very thick] (ext_y_to_z_uv) --
  node [midway,above=0em,align=center ] { $\mathbf{p}_2$}
  node [midway,above=1.1em,align=center ] {$\nu_{W_2}$}
  (z_uv.east |- ext_y_to_z_uv)
  node [pos=0.25](ext_between_y_z_uv){};
  \draw [-latex,very thick] (z_uv.west |- ext_z_uv_to_p_u.east) --
  node [midway,above=0em,align=center ] { $\hat{\usym}_2$}
  node [midway,above=1.3em,align=center ] {${\alpha}_2$}
  (ext_z_uv_to_p_u.east);
  \draw [-latex,very thick] (ext_z_uv_to_p_u) --
  node [midway,above=0em,align=center ] { $\mathbf{r}_1$}
  node [midway,above=1.1em,align=center ] {$\nu_{U_1}$}
  (p_u.east |- ext_z_uv_to_p_u)
  node [pos=0.25](ext_between_z_uv_pu){};
  
  \draw [-latex,very thick] (ext_between_pu_z_uv.center) --
  (ext_z_uv_to_p_u.south);
  \draw [-latex,very thick] (ext_between_z_uv_y.center) --
  (ext_y_to_z_uv.south);
  \draw [-latex,very thick] (ext_between_y_z_uv.center) --
  (ext_z_uv_to_y.north);
  \draw [-latex,very thick] (ext_between_z_uv_pu.center) --
  (ext_pu_to_z_uv.north);
\end{tikzpicture}}
\caption{GVAMP with an input denoiser (left), LMMSE denoiser (center), and output denoiser (right). The ``ext'' blocks generate extrinsic messages.}
\label{fig:block-diagram-GVAMP}
\end{figure*}

\begin{table*}
\renewcommand{\arraystretch}{1.0}
\newcommand{\stepequation}{\refstepcounter{equation}(\theequation)} %
\setlength{\tabcolsep}{0pt} %

\caption{Input and LMMSE denoiser expressions.}
\label{tab:explicit_denoisers}
\begin{tabular*}{\textwidth}{
  @{\extracolsep{\fill}}
  l %
  >{$\textstyle{}}r<{$} %
  @{\extracolsep{0pt}}
  >{$\textstyle{}}l<{$} %
  @{\extracolsep{\fill}}
  >{\stepequation}r
}
\midrule
\textit{definition} & 
c &= \sum_{a \in \mathcal{U}} P_U(a)\,  \mathcal{N}(a; r_{1,\kappa}, \nu_{U_1}) 
\\[4pt]
mean \textit{(component)}& 
\hat{u}_{1,\kappa} 
&= \frac{1}{c} \sum\nolimits_{a \in \mathcal{U}} a\, P_U(a)\,  \mathcal{N}(a; r_{1,\kappa}, \nu_{U_1}) &
\label{eq:x_den_mean}
\\[4pt]
variance \textit{(component)} & 
\alpha_{1,\kappa}  
&= 
\frac{1}{c} \sum\nolimits_{a \in \mathcal{U}} |a|^2\, P_U(a)\,  \mathcal{N}(a; r_{1,\kappa}, \nu_{U_1}) - |\hat{u}_{1,\kappa}|^2 
&
\label{eq:x_den_var}
\\[4pt]
\midrule
\textit{definitions} &   \mathbf{Q}_{\mathbf{U}} &= \big( \Re\big\{\Ad\herm \big(\widetilde{\mathbf{\Sigma}}_{\mathbf{N}_1} \big/ 2 \big)^{-1}  \Ad\big\} + \tfrac{1}{\nu_{U_2}}\mident{} \big)^{-1}  \;
\text{ and } \;
\mathbf{\widetilde{\Sigma}}_{\mathbf{N}_1}
= \mathbf{\Sigma}_{\mathbf{N}_1} + \nu_{W_2} \mident{} \;
&
\label{eq:tab_sigma_n1_tild_def}
\\[4pt]
mean & \hat{\usym}_2 
&= 
 \mathbf{Q}_{\mathbf{U}}
 \big( \Re\big\{\Ad\herm \big( \widetilde{\mathbf{\Sigma}}_{\mathbf{N}_1} \big/ 2 \big)^{-1}   (\mathbf{p}_2-\mathbf{s})\big\} + \tfrac{1}{\nu_{U_2}} \mathbf{r}_2  \big) 
& 
\label{eq:u2_den_mean}
\\[4pt]
mean & \hat{\z}_2  
&= 
\nu_{W_2} \widetilde{\mathbf{\Sigma}}_{\mathbf{N}_1}^{-1}  
(\Ad \hat{\mathbf{u}}_2  +  \mathbf{s}  - \mathbf{p}_2 ) + \mathbf{p}_2
&
\label{eq:w2_den_mean}
\\[4pt]
variance \textit{(average)}  &
\alpha_2
&= 
(1/n') \, \tr\left\{\mathbf{Q}_{\mathbf{U}} \right\} 
& 
\label{eq:u2_den_var}
\\[4pt]
variance \textit{(average)}
 & \beta_2 
&= 
(1/m) \,   \trace\big\{  \nu_{W_2} \mident{m} -  \nu_{W_2}^2
\big( \widetilde{\mathbf{\Sigma}}_{\mathbf{N}_1}^{-1}  -  
\widetilde{\mathbf{\Sigma}}_{\mathbf{N}_1}^{-1} \Re\left\{\Ad \,
\mathbf{Q}_{\mathbf{U}} \,
\Ad\herm \right\}  \widetilde{\mathbf{\Sigma}}_{\mathbf{N}_1}^{-1} \big) 
\big\}
& 
\label{eq:w2_den_var}
\\[4pt]
\bottomrule
\end{tabular*}

\end{table*}

\subsection{Decoding Metrics}
We wish to estimate $P(v_{\ell,t} | \mathbf{y}, \mathbf{v}^{\ell-1})$, $t \in \idxset N$. Recall that $\Usym = [\mathbf{V}_\ell\tran, \mathbf{V}_{\ell+1}\tran, \ldots, \mathbf{V}_S\tran]\tran$ has the symbols of stage $\ell$ and the remaining stages. 
After several GVAMP iterations, we compute a posteriori probability (APP) estimates using the surrogate model \eqref{eq:surrogate_r_u}. GVAMP thus acts as a detector with input $(\mathbf{y},\mathbf{v}^{\ell-1})$ and output $(\mathbf{r}_1, \nu_{U_1})$, namely the estimates $\mathbf{r}_1$ and their reliabilities $\nu_{U_1}$.
The GMI \eqref{eq:gmi} gives the lower-bound 
\begin{align}
    I_\text{$n$,SIC} \geq 
    \underbrace{\frac{1}{S}  \sum\nolimits_{\sics=1}^S  \frac{1}{\siclength} \sum\nolimits_{\sicindex=1}^{\siclength}
    H(V_{\ell,t}) - \mathbb{E}\!\left[-
    \log_2 Q_{\ell,t} 
    \right]}_{\displaystyle := I_{q,n,\text{SIC}}}
    \label{eq:sic_gmi}
\end{align}
where 
\begin{align}
    Q_{\ell,t}(v_{\ell,t})
    = P_U(v_{\ell,t}) \cdot \mathcal{N}(v_{\ell,t}; r_{1,t}, \hat{\nu}_{U_1}) / c
    \label{eq:qel_demapper}
\end{align}
and the constant $c$ normalizes to a PMF.
The GMI $s$-parameter is absorbed by $\hat{\nu}_{U_1}$ that we optimize by a line search.

\section{GVAMP and EXIT Charts}
\label{sec:state_evolution}
We refine the expressions in Table \ref{tab:explicit_denoisers} to describe the GVAMP algorithm and EXIT functions.

\subsection{GVAMP with Approximate Variances}
We derive the following approximations for \eqref{eq:u2_den_var} and \eqref{eq:w2_den_var}:
\begin{align}
    \alpha_2 & = \frac{\nu_{U_2}( \nu_{N_1}/\gamma + \nu_{W_2})}{2 N_\text{os}\nu_{U_2} +\nu_{N_1}/\gamma + \nu_{W_2}}
    \label{eq:alpha2_simp} \\
    \beta_2 & = \frac{\nu_{W_2}\, \nu_{N_1}}{\nu_{W_2}+ \nu_{N_1}/\gamma}  
     +  \left(1 - \frac{\ell-1}{S}\right) \frac{\alpha_2 \, \nu_{W_2}^2}{(\nu_{W_2}+ \nu_{N_1}/\gamma)^2}
    \label{eq:beta2_simp}
\end{align}
where $\nu_{N_1} = B_\mathrm{D} \nu_1$, see below~\eqref{eq:autocorr_n2}, and $\gamma:=B_\mathrm{D}/\Bsam$. 

To begin, observe that the channel of SIC stage $\ell$ is
\begin{equation}
    \Ad = \mathbf{A} \mathbf{\Gamma}
    \label{eq:ad_fast1}
\end{equation}
where the SIC stage matrix $\mathbf{\Gamma} \in \mathbb{R}^{n \times n'}$ is obtained from $\mident{n}$ by deleting the $N(\ell-1)$ columns with indices $\bigcup_{k=1}^{\ell-1} \mathcal{I}^k$; see \eqref{eq:ad_construction}. We thus have $\mathbf{\Gamma}\tran \mathbf{\Gamma}=\mident{n'}$.

Next, the PSD of the filtered $N_1'(t)$ is flat within the bandwidth $B_\mathrm{D}=(1+\alpha_\text{tx}) B$ of $g_\text{tx}(t)$, and we thus have $\nu_{N_1} = B_\mathrm{D} \nu_1 = \gamma \Bsam\nu_1$. 
Fig.~\ref{fig:evd_svd} shows the singular values and eigenvalues of $\mathbf{A}$. Fig.~\ref{fig:evd_noise_cov} plots $\mathbf{\Lambda}_\mathrm{D} = \blkdiag(\mathbf{\Lambda}_{\mathrm{D},1}, \mathbf{\Lambda}_{\mathrm{D},2})$ with diagonal  $\mathbf{\Lambda}_{\mathrm{D},1}, \mathbf{\Lambda}_{\mathrm{D},2} \in \mathbb{C}^n$ that each contain $n = m/2$ eigenvalues of the circulant approximation to $\mathbf{\Sigma}_{\mathbf{N}_1}$ \eqref{eq:noise_cov1_approx}. 
\begin{figure}[!t]
    \centering
    \input{Fig/EVD_SVD_channel}
    \caption{Singular values and modulus-squared eigenvalues of $\mathbf{A}$ for a FD-RRC pulse with $\alpha_\text{tx} = 0.2$. }
    \label{fig:evd_svd}
\end{figure}
\begin{figure}[!t]
    \centering
    \definecolor{mycolor1}{rgb}{0.00000,0.44700,0.74100}%
\definecolor{mycolor2}{rgb}{0.85000,0.32500,0.09800}%
\definecolor{mycolor3}{rgb}{0.92900,0.69400,0.12500}%

\begin{tikzpicture}
\begin{axis}[%
width=0.68*\figwidth,
height=0.32*\figheight,
scale only axis,
xmin=1,
xmax=128,
ymin=-0.0,
ymax=1.0,
enlarge y limits={abs=0.01}, %
axis background/.style={fill=white},
xmajorgrids,
xtick={1,64+1,128},
xticklabels={1,$n/2$,$n$},
xlabel={index},
ylabel style={align=center},
ytick={0,1,2/sqrt(2),2},
yticklabels={0,1,$\sqrt{N_\text{os}}$,$N_\text{os}$},
legend pos=south east,
ymajorgrids,
grid=both,
legend style={legend cell align=left, align=left, draw=white!15!black, legend pos=outer north east,
},
]
\addplot [color=mycolor1,solid,line width=1] table[row sep=crcr,y expr=\thisrowno{1}*1]{%
1	1\\
77	1\\
77  0\\
128	0\\
};
\addlegendentry{$\mathbf{\Lambda}_{\mathrm{D},1}$}; 
\addplot [color=black!50!mycolor1,dashed,line width=1,mark=*,mark size=1.6,mark options={solid,line width=0.7pt,fill=white}] table[row sep=crcr,y expr=\thisrowno{1}*1]{%
1 0\\
53 0\\
53  1\\
128 1\\
};
\addlegendentry{$\mathbf{\Lambda}_{\mathrm{D},2}$}; 
\end{axis}
\end{tikzpicture}%
    \caption{Eigenvalues of the optical noise covariance \eqref{eq:noise_cov1_approx} for $g_\mathrm{D}(t)$ with  $B_\mathrm{D} = (1+\alpha_\text{tx})B$, $\alpha_\text{tx} = 0.2$ and $\Bsam \nu_1 := 1$.}
    \label{fig:evd_noise_cov}
\end{figure}

To derive \eqref{eq:alpha2_simp}, use \eqref{eq:noise_cov1_approx} and $\widetilde{\mathbf{\Sigma}}_{\mathbf{N}_1}$ in
\eqref{eq:tab_sigma_n1_tild_def} to approximate
\begin{align}
    \Ad\herm (\widetilde{\mathbf{\Sigma}}_{\mathbf{N}_1} \big/ 2)^{-1}\Ad 
    &\approx 2 \, \Ad\herm \mathbf{F}_m\herm (\mathbf{\Lambda}_\mathrm{D} + \nu_{W_2} \mident{m})^{-1} \mathbf{F}_m  \Ad .
    \label{eq:Q_part_simplified}
\end{align}
For the final two terms, insert \eqref{eq:A_svd3} to write
\begin{align}
    \mathbf{F}_m \Ad = \mathbf{F}_m \mathbf{A} \mathbf{\Gamma} = \frac{1}{\sqrt{N_\text{os}}} 
    \begin{bmatrix}
        \mathbf{\Lambda}_1 \\ \mathbf{\Lambda}_2
    \end{bmatrix} \mathbf{\Lambda}_\mathrm{P}\mathbf{F}_n \mathbf{\Gamma}
     \label{eq:Q_part_simplified2}
\end{align}
and (see~Fig.~\ref{fig:evd_svd}-\ref{fig:evd_noise_cov})
\begin{align}
    \tfrac{2}{N_\text{os}} \begin{bmatrix}
        \mathbf{\Lambda}_1\herm & \mathbf{\Lambda}_2\herm
    \end{bmatrix}
    (\mathbf{\Lambda}_\mathrm{D} + \nu_{W_2} \mident{m})^{-1}
    \begin{bmatrix}
        \mathbf{\Lambda}_1 \\ \mathbf{\Lambda}_2
    \end{bmatrix} \approx \frac{2 N_\text{os}}{\Bsam\nu_1+\nu_{W_2}}\, \mident{n} .
    \label{eq:Q_part_simplified3}
\end{align}
We may thus approximate
\begin{align}
    2 \Re\{\Ad\herm \widetilde{\mathbf{\Sigma}}_{\mathbf{N}_1}^{-1} \Ad \} + \tfrac{1}{\nu_{U_2}} \mident{n'} 
    \approx
    \left(\frac{2 N_\text{os}}{\nu_{N_1}/\gamma+\nu_{W_2}} + \frac{1}{\nu_{U_2}} \right) \mident{n'}.
     \label{eq:alpha2_simp_inv}
\end{align}
The expression \eqref{eq:tab_sigma_n1_tild_def} gives
\begin{align}
    \mathbf{Q}_{\mathbf{U}} \approx \frac{\nu_{U_2}( \nu_{N_1}/\gamma + \nu_{W_2})}{2 N_\text{os}\nu_{U_2} +\nu_{N_1}/\gamma + \nu_{W_2}} \mident{n'}
    \label{eq:sigma_u_giv_y_simp}
\end{align}
and \eqref{eq:alpha2_simp} follows by \eqref{eq:u2_den_var}. Observe that $\alpha_2<\nu_{U_2}$ for $\gamma \in (0,1]$ and $(1 - \tfrac{\ell-1}{S}) \in (0,1]$, and that $0 < \nu_{W_2}$ and $0 < \nu_{U_2}$. 

To derive \eqref{eq:beta2_simp}, consider the argument of the trace of \eqref{eq:w2_den_var}. We approximate
\begin{align}
   &\trace\big\{ \nu_{W_2}\mident{} - \nu_{W_2} ^2 \widetilde{\mathbf{\Sigma}}_{\mathbf{N}_1}^{-1} \big\} \IfTwoCol{\nonumber\\
   & \qquad} \approx m \nu_{W_2} -  \nu_{W_2}^2  \trace\big\{  (\mathbf{\Lambda}_\mathrm{D} + \nu_{W_2} \mathbf{I}_m)^{-1}   \big\} .
   \label{eq:beta2-part1}
\end{align}
From \eqref{eq:psd_opt_noise}, the $u^\text{th}$ diagonal entry of $\mathbf{\Lambda}_\mathrm{D} + \nu_{W_2} \mathbf{I}_m$ is
\begin{align}
    \begin{cases}
        \Bsam \nu_1 + \nu_{W_2}, & \modone{\tfrac{u}{m}} <  \frac{\gamma}{2}\\
        \frac12 (\Bsam \nu_1 + \nu_{W_2}), & \modone{\tfrac{u}{m}} = \frac{\gamma}{2} \\
        \nu_{W_2},  &  \modone{\tfrac{u}{m}} > \frac{\gamma}{2} .
    \end{cases}
\end{align}
For large $m$, \eqref{eq:beta2-part1} is thus approximately
\begin{align}
   & m\,\nu_{W_2} -  m\,\nu_{W_2}^2 \left( \gamma \frac{1}{\Bsam \nu_1 + \nu_{W_2}} + (1-\gamma) \frac{1}{\nu_{W_2}} \right) \IfTwoCol{\nonumber \\
   &} = m \frac{\nu_{W_2}  \nu_{N_1}}{\Bsam \nu_1 + \nu_{W_2}}
\end{align}
where we used $\nu_{N_1}=\gamma \Bsam \nu_1$. 

For the final term in the trace of \eqref{eq:w2_den_var}, we use \eqref{eq:sigma_u_giv_y_simp} and substitute $\Re\{\Ad\Ad\herm\} = \tfrac{1}{2}(\Ad\Ad\herm + (\Ad\Ad\herm)^*)$ to write
\begin{align}
    \frac{\alpha_2 \nu_{W_2}^2}{2} \trace\Big\{ 
    \, \widetilde{\mathbf{\Sigma}}_{\mathbf{N}_1}^{-2} (\Ad\Ad\herm +  (\Ad\Ad\herm)^*)
    \Big\}.
    \label{eq:first_term_trace_beta2}
\end{align}
Consider the first term in \eqref{eq:first_term_trace_beta2} and apply the same steps as for simplifying \eqref{eq:Q_part_simplified} to obtain the real-valued
\begin{align}
    \Ad\herm\widetilde{\mathbf{\Sigma}}_{\mathbf{N}_1}^{-2}\Ad 
    \approx
    \frac{N_\text{os}}{(\nu_{N_1}/\gamma  + \nu_{W_2})^2}\, \mident{n'} .
    \label{eq:Q_part_simplified_2}
\end{align}
Now substitute  $n'=(S-\ell+1)N$, $N=n/S$, and $m=N_\text{os} n$. Observe that $\beta_2<\nu_{W_2}$  for $\gamma \in (0,1]$ and $(1 - \tfrac{\ell-1}{S}) \in (0,1]$, and $0 < \nu_{W_2}$ and $0 < \alpha_2$. Note that \eqref{eq:Q_part_simplified} and \eqref{eq:sigma_u_giv_y_simp} simplify the expressions for the means \eqref{eq:u2_den_mean} and \eqref{eq:w2_den_mean}.

Finally, inserting the approximation~\eqref{eq:alpha2_simp} into \eqref{eq:ext_lmmse_u} gives the variance 
\begin{align}
    \nu_{U_1} = \frac{\nu_{U_2} \alpha_2 }{\nu_{U_2} - \alpha_2} = \frac{\nu_{N_1}/\gamma + \nu_{W_2}}{2 N_\text{os}} 
    \label{eq:ext_simpl_nuU1}
\end{align}
and mean
\begin{align}
    \mathbf{r}_1 &= 
    \displaystyle\frac{\nu_{U_2}\vdIIh - \alpha_2 \mathbf{r}_2}{\nu_{U_2}-\alpha_2}
    =  \nu_{U_1} \Re\big\{\Ad\herm \big( \widetilde{\mathbf{\Sigma}}_{\mathbf{N}_1} \big/ 2 \big)^{-1}   (\mathbf{p}_2-\mathbf{s})\big\}
    \label{eq:ext_simpl_r1}
\end{align}
which does not depend on the message $(\mathbf{r}_2, \nu_{U_2})$. This is because $\mathbf{Q}_{\mathbf{U}}$ is a scaled identity matrix, indicating that the components of the Gaussian posterior in \eqref{eq:surrogate_y_ad_u_appdx} are iid.
We summarize the GVAMP algorithm in Algorithm~\ref{algo:GVAMP}.

{
\begin{center}
\centering
\begin{algorithm}[t]
\caption{GVAMP for SIC stage $\ell$}\label{algo:GVAMP}
\usetikzlibrary{decorations.pathreplacing,calc}
\newcommand{\tikzmark}[1]{\tikz[overlay,remember picture] \node (#1) {};}

\newcommand*{\AddNote}[4]{%
    \begin{tikzpicture}[overlay, remember picture]
        \draw [decoration={brace,amplitude=0.5em},decorate, thick]
            ($(#3)!(#1.north)!($(#3)-(0,1)$)$) --  
            ($(#3)!(#2.south)!($(#3)-(0,1)$)$)
                node [align=center, text width=2.5cm, pos=0.5, anchor=west, font=\footnotesize] {#4};
    \end{tikzpicture}
}%

\begin{algorithmic}[1]
\IfTwoCol{\small}
\Statex \textbf{Initialization:} extrinsic $(\mathbf{p}_1, \nu_{W_1})$.
\Repeat
\Statex \LeftComment{1}{\textit{Output Denoiser:}}
    \tikzmark{top}
\State Calculate $(\hat{\mathbf{w}}_1, 
\beta_1)$~\eqref{eq:z_den_mean_final}-\eqref{eq:z_den_var_final}. 
\State Convert to extrinsic~\eqref{eq:ext_output}.  
\tikzmark{bottom}
\vspace{3pt}
\Statex \LeftComment{1}{\textit{LMMSE Denoiser:}} \tikzmark{toptwo}
\State  Calculate extrinsic $(\mathbf{r}_1, \nu_{U_1})$~\eqref{eq:ext_simpl_nuU1}-\eqref{eq:ext_simpl_r1}.  \tikzmark{right}\tikzmark{righttwo}
\Statex \LeftComment{1}{\textit{Input Denoiser:}}
\State Calculate $(\hat{\mathbf{u}}_1, \alpha_1)$~\eqref{eq:x_den_mean}-\eqref{eq:x_den_var}. 
\State Convert to extrinsic~\eqref{eq:ext_input}. 
\Statex \LeftComment{1}{\textit{LMMSE Denoiser:}}
\State Calculate $(\hat{\mathbf{w}}_2, \beta_2)$~\eqref{eq:w2_den_mean}-\eqref{eq:beta2_simp}. %
\State Convert to extrinsic~\eqref{eq:ext_lmmse_w}. \tikzmark{bottomtwo}
\vspace{3pt}
\Until{convergence}
\State Compute decoding metrics $Q_{\ell,t}(v_{\ell,t})$, $t \in \idxset N$~\eqref{eq:qel_demapper}.
\State \textbf{Return} $Q_{\ell,t}(v_{\ell,t})$.
\AddNote{top}{bottom}{right}{$1^\text{st}$ half-iteration.}
\AddNote{toptwo}{bottomtwo}{righttwo}{$2^\text{nd}$ half-iteration.}
\end{algorithmic}

\end{algorithm}
\end{center}
}

\subsection{Variance-Based EXIT Functions}
\label{subsec:vEXIT-functions}
We analyze GVAMP convergence by combining the input and LMMSE denoisers in Fig.~\ref{fig:block-diagram-GVAMP} into one module; see Fig.~\ref{fig:variance_exit}. The output denoiser corresponds to the $1^\text{st}$ half-iteration, while the combined denoiser corresponds to the $2^\text{nd}$ half-iteration in Algorithm~\ref{algo:GVAMP}. 
\begin{figure}[!t]
    \centering
    \begin{tikzpicture}[node distance=5.5cm, auto]

      \node[block, minimum width=2.3cm, minimum height=1.0cm] (prior) {$T_2$};
      
      \node[draw, block, minimum width=2.3cm, minimum height=1.0cm, right of=prior] (output) {$T_1$};
      
      \draw[latex-,thick] ([yshift=8pt]prior.east) -- node[above,text width=1cm,align=center]{$\nu_{W_2}$} ([yshift=8pt]output.west);
    
      \draw[-latex,thick] ([yshift=-8pt]prior.east) -- node[below,text width=1cm,align=center]{$\nu_{W_1}$} ([yshift=-8pt]output.west);

      \node[above,yshift=0.5cm, text width=3.3cm,font=\footnotesize] at(input){Input \& LMMSE denoiser}; 
      \node[above,yshift=0.5cm,font=\footnotesize] at(output){Output denoiser};

\end{tikzpicture}
    \caption{Iterations between two modules.}
    \label{fig:variance_exit}
\end{figure}
Define the EXIT functions
\begin{align}
   \nu_{W_2} = T_1(\nu_{W_1}), \quad
   \nu_{W_1} = T_2(\nu_{W_2})
   \label{eq:variance_tfs}
\end{align}
that depend on the variances $(\nu_{W_1},\nu_{W_2})$. To compute \eqref{eq:variance_tfs}, we use the surrogates (see \cite[Sec.~IV]{rangan2019VAMP})
\begin{subequations}
\label{eq:awgn_perturbed_extrinsics_WU}
\begin{align}
    \mathbf{P}_i &= \mathbf{W} + \mathbf{N}'', && \mathbf{N}'' \sim \mathcal{CN}(\mnull{}, \nu_{W_i} \mident{})
    \label{eq:awgn_perturbed_extrinsics_W} \\
    \mathbf{R}_i &= \mathbf{U} + \mathbf{N}', && \mathbf{N}' \sim \mathcal{N}(\mnull{}, \nu_{U_i} \mident{}) 
     \label{eq:awgn_perturbed_extrinsics_U}
\end{align}
\end{subequations}
for $i=1,2$. We calculate $T_1(\nu_{W_1})$ with $\mathbf{P}_1$ in \eqref{eq:awgn_perturbed_extrinsics_W}; see the $1^\text{st}$ half-iteration of Algorithm~\ref{algo:GVAMP}. We calculate $T_2(\nu_{W_2})$ by activating the LMMSE denoiser to compute $\nu_{U_1}$ via \eqref{eq:ext_simpl_nuU1}, running the input denoiser for $\mathbf{R}_1$ in \eqref{eq:awgn_perturbed_extrinsics_U}, and finally running the LMMSE denoiser to compute \eqref{eq:beta2_simp} and $\nu_{W_1}$ via \eqref{eq:ext_lmmse_w}; see the $2^\text{nd}$ half-iteration of Algorithm~\ref{algo:GVAMP}.

Iterating between the two modules in Fig.~\ref{fig:variance_exit} leads to a fixed point corresponding to the intersection of the EXIT functions $T_1$ and $T_2$; see
Sec.~\ref{subsec:EXIT-Charts} below. The EXIT fixed point agrees with the actual GVAMP fixed point under certain conditions, i.e., $n \rightarrow \infty$, bi-rotationally invariant $\mathbf{A}$, and others; see \cite[Thm.~1]{rangan2019VAMP} and \cite[Thm.~1]{fletcher2018inference}. We again remark that $\mathbf{A}$ in \eqref{eq:A_svd2} is not drawn from a bi-rotationally invariant ensemble. Nevertheless, the EXIT predictions were accurate for large CD and random precoding.

\section{Simulation Parameters}
\label{sec:simulation-parameters}
We study short-reach fiber-optic links with the parameters in Table~\ref{tab:simparams}. We consider CD \eqref{eq:cd_response_freq} and optical and electrical noise  \eqref{eq:discrete_n1_n2} as the link impairments, and discard the Kerr effect. The program code is available at \cite{ddsicamp2025plabst}. 

\begin{table}[!t]
\caption{Short-reach fiber-optic system parameters}
\centering
{\renewcommand{\arraystretch}{0.95}
\IfTwoCol{\footnotesize}
\begin{tabular}{ll} 
\toprule
Carrier wavelength & \SI{1550}{\nano\meter} (C-band)\\
Symbol rate & $B = \SI{300}{\giga Bd}$ \\
Fiber length & \SI{4}{\kilo\meter} \\
Attenuation factor & \SI{0.2}{dB \per\kilo\meter} (or \SI{0.046}{\per\kilo\meter})  \\
Group velocity dispersion & $\beta_2=\SI{-2.168e-23}{\second^2\per\kilo\meter}$\\
\midrule
DAC & FD-RRC with $\alpha_\text{tx} = 1\%$ \\
TX bandwidth & $(1 + \alpha_\text{tx})B$ \\
SSMF response &$H(f) = \exp{(-\mathrm{j}\, \beta_2/2\, \omega^2 L_\text{fib})}$\\
Pre-DD complex AWGN & See \eqref{eq:discrete_n1_n2}, ACF \eqref{eq:n1_cov}\\
Post-DD real AWGN & See \eqref{eq:discrete_n1_n2}, ACF \eqref{eq:n2_cov}\\
ADC oversampling factor & $N_\text{os} = 2$\\
Receive filter &$g_\text{rx}(t) = \Brx \,\mathrm{sinc}(\Brx  t)$\\
Precoder & Random orthogonal (FFT-based) \eqref{eq:fft_precoder}--\eqref{eq:precoder_diag} \\
\bottomrule
\end{tabular}
}
\label{tab:simparams}
\end{table}

\subsection{Channel and Receiver Parameters}
We study $L_\text{fib} = \SI{4}{\kilo\meter}$ of SSMF in the C-band,  which is common for campus data centers. The FD-RRC pulse and CD introduce long ISI. We approximate the combined response $a(t)=g_\text{tx}(t)*h(t)$ by a filter with $K' = K \cdot N_\text{os} +1$ taps where $K = 250$; see the text below \eqref{eq:yvec}. This choice includes over $99.9\%$ of the energy of $a(t)$. We add a cyclic prefix with $K$ symbols to simplify simulation. Note that, for the same CD, one may reduce $L_\text{fib}$ and increase $B$ using \eqref{eq:cd_response_freq}, or vice versa.

Optical noise is modeled as complex AWGN; see Sec.~\ref{sec:time-continuous-model}. The optical filter $g_\mathrm{D}(t)$ before DD has bandwidth $B_\mathrm{D}=(1+\alpha_\text{tx})B$ and removes out-of-band noise. The DD doubles the bandwidth of $Z''(t)$ to slightly more than $2 B$; see Fig.~\ref{fig:continuous_detailed_system_model}. Electrical noise due to the PD and a low-noise amplifier (LNA) is modeled as real baseband AWGN. The ADC applies the unit frequency-gain sinc filter $g_\text{rx}(t)$ with bandwidth $\Brx$ and samples with $N_\text{os} = 2$. We choose $\Brx$ based on whether optical or electrical noise dominates. In general, the receiver samples do not provide sufficient statistics. The GVAMP detector uses the surrogate channel model $q_{\vY|\vX}$ from Sec.~\ref{sec:discrete_model_fdrrc}.

Define the average optical receive power as 
\begin{align}
    P_\text{rx} = P_\text{tx} \cdot \e^{- \alpha L_\text{fib}}
    \label{eq:prx}
\end{align}
where $P_\text{tx}$ is the average transmit power~\eqref{eq:ptx} and the exponential term accounts for fiber loss with the attenuation factor $\alpha$ listed in Table~\ref{tab:simparams}. 
\subsection{Transmitter Parameters}
We transmit $N_\text{b} = 128$ blocks each having $n=2048$ $M$-ASK-$o$ symbols; see \eqref{eq:alphabet_A}. A random real orthogonal precoder $\mathbf{P}$ randomizes $\mathbf{A}$; see \eqref{eq:A_svd2}--\eqref{eq:precoder_diag}. The DAC uses FD-RRC pulses with $\alpha_\text{tx} = 1\%$ at the baud rate $B = \SI{300}{GBd}$, generating a real baseband signal $X(t)$ with two-sided bandwidth $(1+\alpha_\text{tx})B$; see \eqref{eq:xt_pulseshaping}-\eqref{eq:fdrrc_fourier}. 

\subsection{Detector Tuning}
\label{subsec:adaptive-denoising}
\subsubsection{Adaptive Damping}
Define the product $Q(\cdot) := \medint\prod_{\;\;\ell >1,t} Q_{\ell,t}(\cdot|\mathbf{y},\mathbf{v}^{\ell-1})$. 
We measure improvements of the APPs \eqref{eq:qel_demapper} in stage $\ell$ via the divergence cost (see \cite[Sec.~IV-B]{parkerBIGAMP2014}) 
\begin{equation}
\begin{aligned}
    &D\big( Q(\cdot) \,\big|\big|\, P_{\mathbf{U}|\mathbf{Y},\mathbf{V}^{\ell-1}}(\cdot|\mathbf{y},\mathbf{v}^{\ell-1} ) \big) \IfTwoCol{\\
    &} \propto 
    D\big(Q(\cdot) \,\big|\big|\,  P_{\Usym}(\cdot)  \big)
    - \sum\nolimits_\mathbf{u} Q(\mathbf{u})  \log_2 p(\mathbf{y}|\mathbf{v}^{\ell-1}, \mathbf{u}) .
    \label{eq:variational_bayes}
\end{aligned}
\end{equation}
The first term in \eqref{eq:variational_bayes} is easy to compute analytically. To approximate the second term in~\eqref{eq:variational_bayes} we first calculate 
\begin{align}
    &\CProj \left[\medint\int \mathcal{N}(\mathbf{u}; \vdIh, {\alpha}_1 \mathbf{I}) \, \mathcal{CN}(\mathbf{w}; \Ad \mathbf{u} + \mathbf{s}, \mathbf{\Sigma}_{\mathbf{N}_1}) \mathrm{d}\mathbf{u} \right] \IfTwoCol{\nonumber \\
    &} =\mathcal{CN}(\mathbf{w}; \Ad \hat{\mathbf{u}}_1 + \mathbf{s}, (\alpha_1 (1 - \tfrac{\ell-1}{S}) + \nu_{N_1}) \mident{} ) 
    \label{eq:gaussian_post_out_approx}
\end{align}
where we used \eqref{eq:ad_fast1} and \eqref{eq:A_svd}. The approximation of the second term in \eqref{eq:variational_bayes} is 
\begin{align}
    \E\left[ \log_2 q(\mathbf{y}|\mathbf{W}) \right]
    \label{eq:div2_approx}
\end{align}
where the expectation is with respect to \eqref{eq:gaussian_post_out_approx} and $q_{\vY|\mathbf{W}}$ corresponds to the surrogate model from Sec.~\ref{sec:discrete_model_fdrrc} with $\mathbf{W}$ defined in~\eqref{eq:dp_vector_model}. We approximate \eqref{eq:div2_approx} via Monte-Carlo sampling.

We use the adaptive damping strategy from \cite[Sec.~IV-B]{parkerBIGAMP2014} for the extrinsic messages 
$(\mathbf{p}_2, \nu_{W_2})$; see \cite[Sec.~III-D]{schniter2012phase}. At each iteration $n_\text{it}$, check if \eqref{eq:variational_bayes} is less than the maximum cost of the previous $n_\mathrm{W}+1$  iterations. If not, the step is considered unsuccessful, and the damping factor is halved. GVAMP retries the step until the cost decreases, the damping factor falls below \SI{1e-2}{}, or the maximum iteration count of 250 is exceeded. For each successful step, the damping factor is multiplied by 1.1 until it reaches a maximum of 1. We choose $n_\mathrm{W} = 10$ and initialize the damping factor to 1. After at least 10 iterations, we record the iteration count at which the damping factor is next reduced. This value $n_\text{trig}$ toggles the variance annealing described next.

\subsubsection{Variance Annealing}
we anneal until iteration $n_\text{trig}-1$ to avoid local minima. The LMMSE denoiser first calculates $(\hat{\mathbf{u}}_2, \alpha_2)$ in \eqref{eq:u2_den_mean} and \eqref{eq:alpha2_simp} for white optical noise ($\gamma = 1$) and 
\begin{align}
    &\nu_{N_1} = \nu_{N_1}^{(\mathbf{u})},  && \mathbf{\Sigma}_{\mathbf{N}_1} = \nu_{N_1}\mident{} 
\end{align}
with the noise variances listed in Table~\ref{tab:variance_annealing}. 
Next, calculate $(\hat{\mathbf{w}}_2, \beta_2)$ via \eqref{eq:w2_den_mean} and \eqref{eq:beta2_simp}, but now with 
\begin{align}
    &\nu_{N_1} = c \, \nu_{N_1}^{(\mathbf{u})}, &&  \mathbf{\Sigma}_{\mathbf{N}_1} = \nu_{N_1}\mident{} 
\end{align}
where $ 0 < c \leq 1$. To see the effect, rewrite \eqref{eq:w2_den_mean} as the mixture 
\begin{align}
    \hat{\mathbf{w}}_2 = \xi \cdot (\mathbf{A} \hat{\mathbf{u}}_2 + \mathbf{s}) + \left(1 - \xi\right) \cdot \mathbf{p}_2
    \label{eq:convex_comb_lmmse}
\end{align}
where $\xi:= \nu_{W_2} / (\nu_{W_2} + c \, \nu_{N_1}^{(\mathbf{u})})$. For $c < 1$, $\hat{\mathbf{w}}_2$ shifts focus to $\mathbf{A} \hat{\mathbf{u}}_2 + \mathbf{s}$ with the denoised $\hat{\mathbf{u}}_2$. We chose $c = 1/4$ to dampen the output denoiser's influence during the first few iterations. This heuristic stabilized the algorithm for short fibers, and was less important for long fibers and large $o$.

The noise variances in Table~\ref{tab:variance_annealing} are set depending on $P_\text{rx,dB} = 10 \log_{10}P_\text{rx}$~\eqref{eq:prx}; the relationship was coarsely optimized for one low and high-SNR point by fitting an exponential function. We switched off annealing and used the actual LMMSE denoiser at iteration count $n_\text{trig}$.

We initialized Algorithm~\ref{algo:GVAMP} by drawing $\mathbf{p}_1$ from $\mathcal{CN}(\mnull{}, \nu_{W_1} \mident{})$ and setting $\nu_{W_1} = 10 \cdot P_\text{rx} \trace(\Ad \Ad\herm)/m$.

\begin{table}[t!]
\centering
\setlength{\tabcolsep}{4pt} %
\renewcommand{\arraystretch}{1.7} %
\caption{Variance Annealing for the LMMSE Denoiser}
\label{tab:variance_annealing}
\IfTwoCol{\footnotesize}
\begin{tabular}{cl}
\toprule
Noise &  Parameter Settings  \\
\midrule
Optical
 & 
\multirow{2}{*}{
$
\begin{aligned}
\nu_{N_1}^{(\mathbf{u})} &= \max \big\lbrace 1,\; 0.36 \cdot e^{0.19 P_\text{rx,dB}} \big\rbrace&& n_\text{it} = 1 , \ldots , n_\text{trig} -1 \\
\nu_{N_1}^{(\mathbf{u})} &= 0.47 \cdot e^{0.2 P_\text{rx,dB}}  && n_\text{it} = 1 , \ldots , n_\text{trig} -1 \\
\end{aligned}
$
}
\\
 Electrical  &     \\
\hline
\end{tabular}
\end{table}

\section{Simulation Results}
\label{sec:simulation-results}

\subsection{Optically Amplified Link}
\label{sec:results_opt_noise}
Suppose the optical noise dominates. We model the channel as having no electrical noise ($\nu_{N_2}=0$) and set the ADC filter bandwidth to $\Brx = \B'$. This gives $q_{\mathbf{Y}|\mathbf{X}} = p_{\mathbf{Y}|\mathbf{X}}$ with aliased samples $\mathbf{Y}$ that are not quite sufficient statistics; recall that $\alpha_\text{tx} = 1\%$.
We have $p(\mathbf{y}|\mathbf{w}) = \delta(\mathbf{y}-|\mathbf{w}|^2)$ but instead use the surrogate $q_{\mathbf{Y}|\mathbf{W}}$ in~\eqref{eq:div2_approx} to model $\mathbf{Y} = |\mathbf{W} + \mathbf{N}_1'|^2$ with CSCG $\mathbf{N}_1'$ with a small $\nu_{N_1'} = \SI{1e-6}{}$. The reason for adding a small amount of noise is to obtain a smooth and closed-form $q(\mathbf{y}|\mathbf{w})$ that is a generalized chi-square density with two degrees of freedom; see Appendix~\ref{appendix-chi2}.
\subsubsection{SNR Definition and Capacity Bounds}
\label{sec:snr_def_capacity_bounds}
A coherent detector has access to the optical signal $Z''(t)$ before the PD. It puts out sufficient statistics $\mathbf{Z}'' \in \mathbb{C}^n$ by projecting $Z''(t)$ onto the set of orthogonal functions $\{a(t- \kappa T_\text{s})\}_{\kappa=1}^n$, where $a(t)$ is the combined impulse response of the transmit pulse and CD channel; see above~\eqref{eq:avec}. Projecting is equivalent to matched-filtering $Z''(t)$ with $a(t)^*$ and sampling the filter output at rate $B$; see e.g., \cite[Sec.~III-B]{essiambre2010capacity}.

Define the SNR of the coherent receiver as 
\begin{align}
     P_\text{rx} \big/ (B \nu_1/2)
    \label{eq:snr_opt}
\end{align}
where $P_\text{rx}$ is the received power and $B \nu_{1}/2$ is the power of the real optical noise component, i.e., the in-phase noise component, after matched-filtered sampling. 

The coherent capacity for real-valued signaling is
\begin{align}
      C_{\mathbb{R},\text{coh}}
      & = \frac{1}{2} \log_2\left(1 + \frac{\Prx}{B\nu_{1}/2}\right)
    \label{eq:coh_capacity}
\end{align}
where Gaussian inputs achieve capacity.
We can upper-bound the DD information rates by
\begin{align}
    I_n(\mathbf{X}; \mathbf{Y}) \le 
    I_n( \mathbf{X}; \mathbf{Z}'' )\le C_{\mathbb{R},\text{coh}}.
    \label{eq:dpi_coherent}
\end{align}
Moreover, the same steps as in \cite{mecozzi2018information} give
\begin{align}
    C_{\text{DD},\mathbb{R}} \ge  C_{\mathbb{R},\text{coh}}  - 1
    \label{eq:dd_capacity_lb}
\end{align}
for $\atx = 0$. That is, the real-signaling DD capacity is within \SI{1}{bpcu} of the real-signaling coherent capacity.

\subsubsection{GVAMP}
Let $ \Bsam \nu_1 = 1 $ and $\Bsam = 2B$; see Table~\ref{tab:Bparams}. The colored optical noise~\eqref{eq:autocorr_n1} after the PD filter $g_\mathrm{D}(t)$ has $\nu_{N_1} = \gamma = (1+\alpha_\text{tx})/2$, i.e., $\nu_{N_1} \approx 1/2$ for $\alpha_\text{tx} = 1\%$. Compared to \eqref{eq:snr_opt}, the GVAMP noise power per sample is approximately twice that of the coherent receiver. Intuitively, coherent receivers for real-valued signals ignore the noise's imaginary component. At the same time, GVAMP sees only the intensity signal, which mixes the transmitted signal with the real and imaginary noise components.

We run GVAMP with adaptive damping and compare over a window of $n_\mathrm{W} = 10$ iterations.
Fig.~\ref{fig:vamp_opt_rates} plots the AIRs for $S=4$ SIC stages and $M$-ASK-$o$ for $M \in\{4,8,16,32,64\}$ and $o = 0.2$. The figure also shows AIRs for $M=64$ and $o=1$, but omits smaller modulation orders because the rates are similar. The offset $o=0.2$ is significantly more power efficient than $o=1$, gaining $\approx\SI{5.6}{dB}$ at intermediate to high SNRs. The power gap to the real coherent capacity \eqref{eq:coh_capacity} is $\SI{2.4}{dB}$. Equivalently, $o=0.2$ gains $\approx\SI{0.9}{bpcu}$ over $o=1$ for a wide SNR range and operates within $\SI{0.2}{bpcu}$ of the equiprobable ASK capacity (the maximal rates of equiprobable-ASK with a coherent receiver \cite[Fig.~1]{forney1998modulation}) and $\SI{0.4}{bpcu}$ of the coherent capacity. 
The curves even improve upon the lower bound \eqref{eq:dd_capacity_lb}.

Fig.~\ref{fig:vamp_opt_rates} also shows EXIT predictions for $M=64$. The SE curves accurately predict the AIRs for intermediate to high SNRs, with an estimation error of $\approx\SI{0.4}{dB}$ for $o=0.2$ and $\approx\SI{0.6}{dB}$ for $o=1$. 
\begin{figure}[!t]
    \centering
    \pgfkeys{
    /pgf/number format/.cd,
    precision=1,
}

\pgfdeclarelayer{background}
\pgfdeclarelayer{foreground}
\pgfsetlayers{background,main,foreground}

\begin{tikzpicture}[spy using outlines={magnification=4, connect spies}]

\begin{axis}[%
yminorticks=true,
xmajorgrids,
ymajorgrids,
yminorgrids,
minor x tick num=9,
minor y tick num=4,
ytick={0,1,2,3,4,5,6},
grid=both,
title style={yshift=-5pt,},
legend style={legend cell align=left,  draw=white!15!black, %
legend pos=south east
},
xlabel style={font=\color{white!15!black}},
ylabel style={font=\color{white!15!black}},
axis background/.style={fill=white},
scale only axis,
width=\figwidth,
height=\figheight,
xmin=-10,
xmax=50,
Rate_VS_PRX_NuN1_HALVE,
ymin=0,
ymax=6.02,
minor y tick num=4,
legend cell align=left,
legend style={font=\small,cells={align=left},%
},
]

\addplot[draw=none,line width = 1.2, black,domain=-10:40,name path=Cr,forget plot] (x,{1/2*log2(1 + 10^(x/10))});
\addlegendentry{$C_{\mathbb{R},\text{coh}}$};

\addplot[line width = 0.7, black,domain=-10:40,decoration={ 
text align={left, left indent=0.6cm},
text along path,
raise=2pt,
text={|\small|Coherent Capacity},
},
mark options={},
postaction={decorate},] (x,{1/2*log2(1 + 10^(x/10))});

\addplot[line width = 0.6, black,densely dotted,name path global=Rr,] table [x=SNR, y=IqYX, col sep=comma,x expr= \thisrow{SNR}]{Fig/256-ASK-AWGN.txt};
\addlegendentry{$R_\text{ASK}$};

\pgfplotsinvokeforeach{4,8,16,32}
{
\addplot[ASK,name path global=#1-ASK-0.2,forget plot] table [x=SNR, y=IqYX, col sep=comma,x expr= \thisrow{SNR}+6]{plots/vamp_v11_r1/PRE/r1,#1-ASK-0.2,S=4,C=0,Rb=1,n1=1,n2=0,n1m=0,n2m=0,P=cO,ps=RRC,a=0.01,Nsp=250,Npi=0,n=2048,L=4,Rs=300,Nr=1,Nb=128,I=250.txt};
}

\addplot[ASK,name path global=64-ASK-0.2,] table [x=SNR, y=IqYX, col sep=comma,x expr= \thisrow{SNR}+6]{plots/vamp_v11_r1/PRE/r1,64-ASK-0.2,S=4,C=0,Rb=1,n1=1,n2=0,n1m=0,n2m=0,P=cO,ps=RRC,a=0.01,Nsp=250,Npi=0,n=2048,L=4,Rs=300,Nr=1,Nb=128,I=250.txt};
\addlegendentry{$M$-ASK-0.2};

\addplot[PAM,,name path global=64-ASK-1] table [x=SNR, y=IqYX, col sep=comma,x expr= \thisrow{SNR}+6]{plots/vamp_v11_r1/PRE/r1,64-ASK-1.0,S=4,C=0,Rb=1,n1=1,n2=0,n1m=0,n2m=0,P=cO,ps=RRC,a=0.01,Nsp=250,Npi=0,n=2048,L=4,Rs=300,Nr=1,Nb=128,I=250.txt};
\addlegendentry{64-ASK-1};

\addplot[ASK,EXIT,name path global=64-ASK-0.2-SE,forget plot] table [x=SNR, y=EXIT, col sep=comma,x expr= \thisrow{SNR}+6]{plots/vamp_v11_r1/PRE/r1,64-ASK-0.2,S=4,C=0,Rb=1,n1=1,n2=0,n1m=0,n2m=0,P=cO,ps=RRC,a=0.01,Nsp=250,Npi=0,n=2048,L=4,Rs=300,Nr=1,Nb=128,I=250.txt};

\addplot[PAM,EXIT,name path global=64-ASK-1-SE,forget plot] table [x=SNR, y=EXIT, col sep=comma,x expr= \thisrow{SNR}+6]{plots/vamp_v11_r1/PRE/r1,64-ASK-1.0,S=4,C=0,Rb=1,n1=1,n2=0,n1m=0,n2m=0,P=cO,ps=RRC,a=0.01,Nsp=250,Npi=0,n=2048,L=4,Rs=300,Nr=1,Nb=128,I=250.txt};

\begin{pgfonlayer}{foreground}
  \pgfplotsinvokeforeach{8,16,32,64}
  {

    \MeasureXDistance{{0.85*log2(#1)}}{Cr}{#1-ASK-0.2}{anchor=east,xshift=-0.6cm,font=\normalsize};
    \MeasureXDistance{{0.85*log2(#1)}}{#1-ASK-0.2}{64-ASK-1}{anchor=west,font=\normalsize};

  }
\end{pgfonlayer}

\coordinate (spypoint) at (axis cs:29,4.61);
\coordinate (magnifyglass) at (axis cs:0,4.8);

\end{axis}

\spy [size=2cm] on (spypoint) in node[fill=white] at (magnifyglass);

\end{tikzpicture}%
    \caption{AIRs for optically amplified links with $L_\text{fib} = \SI{4}{\kilo\meter}$, $B = \SI{300}{\giga Bd}$, $S=4$, and $M=4,8,16,32,64$. The curve labeled $R_\text{ASK}$ shows the equiprobable-ASK capacity with a coherent receiver \cite[Fig.~1]{forney1998modulation}. The two dash-dotted curves are EXIT predictions for $M=64$.}
    \label{fig:vamp_opt_rates}
\end{figure}

Fig.~\ref{fig:vamp_opt_rates_stage} compares SIC rates for $64$-ASK-$0.2$ and their EXIT predictions. SDD $(\ell=1)$ is almost optimal at high SNR, and the EXIT predictions are accurate for all stages above $5\,\text{bpcu}$. The EXIT predictions are less accurate at low SNR, especially for $\ell = 1$. This gap might be because GVAMP gets stuck in a local maximum due to the structured matrix $\mathbf{A}$; note that phase retrieval does not always work for structured unitary matrices such as the DFT; see \cite[Table~I]{waldspurger2015phase} and \cite[p.~11]{dong2023phasetutorial}.
At higher SIC levels, $\mathbf{A}$ is subsampled based on known prior symbols, see \eqref{eq:ad_construction}, which reduces structure and appears to avoid local minima. We observed that GVAMP agrees better with the EXIT predictions for $\ell = 3,4$.

\begin{figure}[!t]
    \centering
    \pgfkeys{
    /pgf/number format/.cd,
    precision=1,
}

\begin{tikzpicture}[spy using outlines={magnification=2, connect spies}]

\begin{axis}[%
yminorticks=true,
xmajorgrids,
ymajorgrids,
yminorgrids,
minor x tick num=9,
minor y tick num=4,
ytick={0,1,2,3,4,5,6},
grid=both,
title style={yshift=-5pt,},
legend style={legend cell align=left,  draw=white!15!black,  %
legend pos=south east
},
xlabel style={font=\color{white!15!black}},
ylabel style={font=\color{white!15!black}},
axis background/.style={fill=white},
scale only axis,
width=\figwidth,
height=\figheight,
xmin=-10,
xmax=50,
Rate_VS_PRX_NuN1_HALVE,
ymin=0,
ymax=6.02,
minor y tick num=4,
legend pos= south east,
legend cell align=left,
legend style={font=\small,cells={align=left},%
},
,
]

\addplot[draw=none,line width = 0.7, black,domain=-10:40,name path=Cr,forget plot] (x,{1/2*log2(1 + 10^(x/10))});
\addlegendentry{$C_{\mathbb{R},\text{coh}}$};

\addplot[line width = 0.7, black,domain=-10:40,decoration={ 
text align={left, left indent=4.7cm},
text along path,
raise=2pt,
text={|\small|Coherent Capacity},
},
mark options={},
postaction={decorate},] (x,{1/2*log2(1 + 10^(x/10))});

\addplot[ASK,mark=*,mark options={mark size=1pt,solid},densely dashed,name path global=ASK-0.2-s1] table [x=SNR, y=IqYXs_1, col sep=comma,x expr= \thisrow{SNR}+6]{plots/vamp_v11_r1/PRE/r1,64-ASK-0.2,S=4,C=0,Rb=1,n1=1,n2=0,n1m=0,n2m=0,P=cO,ps=RRC,a=0.01,Nsp=250,Npi=0,n=2048,L=4,Rs=300,Nr=1,Nb=128,I=250.txt};
\addlegendentry{$\ell = 1$};

\addplot[ASK,mark=*,mark options={mark size=1pt,solid},densely dashdotted,name path global=ASK-0.2-s2] table [x=SNR, y=IqYXs_2, col sep=comma,x expr= \thisrow{SNR}+6]{plots/vamp_v11_r1/PRE/r1,64-ASK-0.2,S=4,C=0,Rb=1,n1=1,n2=0,n1m=0,n2m=0,P=cO,ps=RRC,a=0.01,Nsp=250,Npi=0,n=2048,L=4,Rs=300,Nr=1,Nb=128,I=250.txt};
\addlegendentry{$\ell = 2$};

\addplot[ASK,mark=*,mark options={mark size=1pt,solid},densely dashdotdotted,name path global=ASK-0.2-s3] table [x=SNR, y=IqYXs_3, col sep=comma,x expr= \thisrow{SNR}+6]{plots/vamp_v11_r1/PRE/r1,64-ASK-0.2,S=4,C=0,Rb=1,n1=1,n2=0,n1m=0,n2m=0,P=cO,ps=RRC,a=0.01,Nsp=250,Npi=0,n=2048,L=4,Rs=300,Nr=1,Nb=128,I=250.txt};
\addlegendentry{$\ell = 3$};

\addplot[ASK,mark=*,densely dotted,mark options={mark size=0.8pt,solid},name path global=ASK-0.2-s4] table [x=SNR, y=IqYXs_4, col sep=comma,x expr= \thisrow{SNR}+6]{plots/vamp_v11_r1/PRE/r1,64-ASK-0.2,S=4,C=0,Rb=1,n1=1,n2=0,n1m=0,n2m=0,P=cO,ps=RRC,a=0.01,Nsp=250,Npi=0,n=2048,L=4,Rs=300,Nr=1,Nb=128,I=250.txt};
\addlegendentry{$\ell = 4$};

\addplot[EXIT,densely dashed,name path global=EXIT-0.2-s1] table [x=SNR, y=EXITs_1, col sep=comma,x expr= \thisrow{SNR}+6]{plots/vamp_v11_r1/PRE/r1,64-ASK-0.2,S=4,C=0,Rb=1,n1=1,n2=0,n1m=0,n2m=0,P=cO,ps=RRC,a=0.01,Nsp=250,Npi=0,n=2048,L=4,Rs=300,Nr=1,Nb=128,I=250.txt};
\addlegendentry{EXIT};

\addplot[EXIT,densely dashdotted,name path global=EXIT-0.2-s2,forget plot] table [x=SNR, y=EXITs_2, col sep=comma,x expr= \thisrow{SNR}+6]{plots/vamp_v11_r1/PRE/r1,64-ASK-0.2,S=4,C=0,Rb=1,n1=1,n2=0,n1m=0,n2m=0,P=cO,ps=RRC,a=0.01,Nsp=250,Npi=0,n=2048,L=4,Rs=300,Nr=1,Nb=128,I=250.txt};

\addplot[EXIT,densely dashdotdotted,name path global=EXIT-0.2-s3,forget plot] table [x=SNR, y=EXITs_3, col sep=comma,x expr= \thisrow{SNR}+6]{plots/vamp_v11_r1/PRE/r1,64-ASK-0.2,S=4,C=0,Rb=1,n1=1,n2=0,n1m=0,n2m=0,P=cO,ps=RRC,a=0.01,Nsp=250,Npi=0,n=2048,L=4,Rs=300,Nr=1,Nb=128,I=250.txt};

\addplot[EXIT,densely dotted,,name path global=EXIT-0.2-s4,forget plot] table [x=SNR, y=EXITs_4, col sep=comma,x expr= \thisrow{SNR}+6]{plots/vamp_v11_r1/PRE/r1,64-ASK-0.2,S=4,C=0,Rb=1,n1=1,n2=0,n1m=0,n2m=0,P=cO,ps=RRC,a=0.01,Nsp=250,Npi=0,n=2048,L=4,Rs=300,Nr=1,Nb=128,I=250.txt};

\coordinate (spypoint) at (axis cs:11,1.6);
\coordinate (magnifyglass) at (axis cs:0,4.8);

\end{axis}

\spy [size=2cm] on (spypoint) in node[fill=white] at (magnifyglass);

\end{tikzpicture}%
    \caption{SIC AIRs for optically amplified links with $L_\text{fib} = \SI{4}{\kilo\meter}$, $64$-ASK-$0.2$, $S=4$ SIC stages  and $B = \SI{300}{\giga Bd}$. Black curves without markers, with the same linestyle, are the EXIT predictions of the stage AIRs.}
    \label{fig:vamp_opt_rates_stage}
\end{figure}

Fig.~\ref{subfig:TRACE_PRE_I} plots SDD rates for 64-ASK-0.2 and $L_\text{fib}=\SI{4}{\kilo\meter}$ against the number of iterations. We chose an SNR of \SI{33.5}{dB} to operate around $85\%$ of the maximum rate $\log_2 M$.
The area marks the $1^\text{st}$ and $99^\text{th}$ percentile rates at which the algorithm converged to a value around the mean across 256 transmitted blocks. The annealing strategy is apparent where the rate saturates at a plateau. Upon deactivating annealing, as described in Sec.~\ref{subsec:adaptive-denoising}, the rate increases again and saturates close to the value predicted by the EXIT analysis. 

\begin{figure}
\centering
\subfloat[\IfTwoCol{\footnotesize} Initial damping; Table~\ref{tab:variance_annealing}.\label{subfig:TRACE_PRE_I}]%
{\centering
\begin{minipage}[t]{0.48\columnwidth}%
\centering
\pgfplotsset{
  log x ticks with fixed point/.style={
      xticklabel={
        \pgfkeys{/pgf/fpu=true}
        \pgfmathparse{exp(\tick)}%
        \pgfmathprintnumber[fixed relative, precision=3]{\pgfmathresult}
        \pgfkeys{/pgf/fpu=false}
      }
  }
}
\begin{tikzpicture}
\begin{axis}[%
width=0.43*\figwidth,
height=0.8*\figheight,
scale only axis,
xmode=log,
xmin=1,
xmax=100,
xminorticks=true,
ymin=-0.01,
ymax=6,
ytick={0,1,2,3,4,5,6},
axis background/.style={fill=white},
title style={yshift=-5pt,},
xmajorgrids,
xminorgrids,
ymajorgrids,
legend style={font=\small,legend cell align=left, align=left, draw=white!15!black},
ylabel={Rate [bpcu]},
xlabel={Iterations},
legend pos= north west,
log x ticks with fixed point
]

\addplot [color=mycolor6, line width=1.5pt]
table[x=Iter, y=Mean, col sep=comma]{Fig/v11_P1_TRACE_PRE_64ASK-0.2_SNR=33.5dB.txt};
\addlegendentry{Mean};

\addplot [draw=none,color=black,  name path=A, densely dashed, line width=0.1pt,]
table[x=Iter, y=UP, col sep=comma]{Fig/v11_P1_TRACE_PRE_64ASK-0.2_SNR=33.5dB.txt};

\addplot [draw=none,color=black,  name path=B, densely dashed, line width=0.1pt,]
table[x=Iter, y=DOWN, col sep=comma]{Fig/v11_P1_TRACE_PRE_64ASK-0.2_SNR=33.5dB.txt};

\addplot[mycolor6!30,forget plot] fill between[of=A and B];

\addplot[black, densely dashed,thick] coordinates {
    (65, 0)
    (65, 6)
 } node[left,pos=0.45,rotate=90,opacitylabel,font=\normalsize, fill opacity=1, xshift=0.2cm, text=black, inner sep=2pt]{$65$ iterations};

\node at(axis cs: 5, 2)[opacitylabel,rotate=60,text=black,thick,left,font=\normalsize]{Annealing}; 

\end{axis}
\end{tikzpicture}%
%
\end{minipage}
}%
\hfill%
\subfloat[\IfTwoCol{\footnotesize} Optimized damping; Table~\ref{tab:mpib_optimized}.\label{subfig:TRACE_PRE_II}]%
{\centering
\begin{minipage}[t]{0.48\columnwidth}%
\centering
\pgfplotsset{
  log x ticks with fixed point/.style={
      xticklabel={
        \pgfkeys{/pgf/fpu=true}
        \pgfmathparse{exp(\tick)}%
        \pgfmathprintnumber[fixed relative, precision=3]{\pgfmathresult}
        \pgfkeys{/pgf/fpu=false}
      }
  }
}
\begin{tikzpicture}
\begin{axis}[%
width=0.43*\figwidth,
height=0.8*\figheight,
scale only axis,
xmode=log,
xmin=1,
xmax=100,
xminorticks=true,
ymin=-0.01,
ymax=6,
ytick={0,1,2,3,4,5,6},
axis background/.style={fill=white},
title style={yshift=-5pt,},
xmajorgrids,
xminorgrids,
ymajorgrids,
legend style={font=\small,legend cell align=left, align=left, draw=white!15!black},
xlabel={Iterations},
legend pos= north west,
log x ticks with fixed point
]

\addplot [color=mycolor6, line width=1.5pt]
table[x=Iter, y=Mean, col sep=comma]{Fig/v11_P1_TRACE_OPTIMPRE_64ASK-0.2_SNR=33.5dB.txt};
\addlegendentry{Mean};

\addplot [draw=none,color=black,  name path=A, densely dashed, line width=0.1pt,]
table[x=Iter, y=UP, col sep=comma]{Fig/v11_P1_TRACE_OPTIMPRE_64ASK-0.2_SNR=33.5dB.txt};

\addplot [draw=none,color=black,  name path=B, densely dashed, line width=0.1pt,]
table[x=Iter, y=DOWN, col sep=comma]{Fig/v11_P1_TRACE_OPTIMPRE_64ASK-0.2_SNR=33.5dB.txt};

\addplot[mycolor6!30,forget plot] fill between[of=A and B];

\addplot[black,dashed,thick] coordinates {
    (38, 0)
    (38, 6)
} node[left,pos=0.45,rotate=90,opacitylabel,font=\normalsize,fill opacity=1,text=black,  xshift=0.2cm,inner sep=2pt]{$38$ iterations}; 

\node at(axis cs: 5, 2)[opacitylabel,rotate=60,text=black,thick,left,font=\normalsize]{Annealing}; 

\end{axis}
\end{tikzpicture}%
\end{minipage}
}
\caption{SDD rates for optically amplified links with $L_\text{fib}=\SI{4}{\kilo\meter}$, $64$-ASK-$0.2$, $B=\SI{300}{\giga Bd}$ and a SNR of \SI{33.5}{dB}. Plot $(a)$ uses the damping parameters in Table~\ref{tab:variance_annealing}, while $(b)$  uses the optimized damping of Table~\ref{tab:mpib_optimized}.}
\label{fig:TRACE_PRE}
\end{figure}

We next plot GVAMP rates for different fiber lengths and offsets $o$ for fixed SNRs and $32$-ASK. Fig.~\ref{fig:cd_plot} shows that the memory of the combined response $a(t)$ significantly increases from $\SI{0.5}{\kilo\meter}$ to $\SI{10}{\kilo\meter}$. 
Fig.~\ref{fig:subfig_contour_opt_noise} shows that GVAMP benefits from a small offset $o \approx 0.15-0.2$, while large $o$ wastes transmit power. The rates of GVAMP increase until $L_\text{fib} \approx \SI{2}{\kilo\meter}$ and remain for constant for $L_\text{fib}$ between $\SI{2}{\kilo\meter}$ and $\SI{10}{\kilo\meter}$. This suggests that CD benefits the detector by ``randomizing'' the channel \eqref{eq:A_eff_chan}; see Sec.~\ref{sec:precoding}. One can introduce dispersion at the transmitter, for example, by using additional fiber or chirped Bragg gratings. One can omit precoding for points above the red dotted line, as the rate loss is less than $\SI{0.1}{bpcu}$. 
\begin{figure*}[!t]
    \centering
    \input{Fig/cd_plot}
    \caption{Magnitude $|a(t)|$ for different fiber lengths $L_\text{fib}$. Solid and dashed curves are for FD-RRC pulses with $\alpha_\text{tx} = 1\%$ and $\alpha_\text{tx} = 99\%$, respectively.}
    \label{fig:cd_plot}
\end{figure*}

\begin{figure*}[!t]
    \centering
    \subfloat[{\small Optical Noise}\label{fig:subfig_contour_opt_noise}]%
    {\centering
    \begin{minipage}[t]{0.242\textwidth}%
\begin{tikzpicture}
\pgfplotsset{
        colormap={parula}{
            rgb255=(53,42,135)
            rgb255=(15,92,221)
            rgb255=(18,125,216)
            rgb255=(7,156,207)
            rgb255=(21,177,180)
            rgb255=(89,189,140)
            rgb255=(165,190,107)
            rgb255=(225,185,82)
            rgb255=(252,206,46)
            rgb255=(249,251,14)
        },
        }
    \begin{axis}[
        xlabel={$L_\text{fib}$ [\SI{}{\kilo\meter}]},
        ylabel={Offset $o$},
        title={$\text{SNR}=\SI{22}{dB}$ },
        small,
        xtick distance=2,
        ytick distance=0.2, 
        view={0}{90},
        colorbar,         
        colorbar style={
        tickwidth=2pt,
        xshift=-13pt,
        x tick style={color=black},
        y tick style={color=black},
        ytick distance=0.2,
        scaled y ticks = false,
        samples=6,    
        },
        colorbar/width=2.5mm,
        x tick style={color=black},
        y tick style={color=black},
        tickwidth=2pt,
        grid=none,
        width=0.92\textwidth,
        height=4.0cm,
        title style={yshift=-6.5pt,},
    ]
        \addplot3 [
        surf,
        shader=interp, 
        mesh/cols=15,  
        ] table[ col sep=comma] {Fig/r1_surf_PRE_22dB.txt}; 

        \addplot3 [
        mycolor6,densely dotted,thick,
        ] table[ col sep=comma] {Fig/r1_surf_PRE_22dB_noPrecoder_threshold.txt}; 
        
    \end{axis}
\end{tikzpicture}
\end{minipage}%
\hspace{5pt}
\begin{minipage}[t]{0.242\textwidth}
\begin{tikzpicture}
\pgfplotsset{
        colormap={parula}{
            rgb255=(53,42,135)
            rgb255=(15,92,221)
            rgb255=(18,125,216)
            rgb255=(7,156,207)
            rgb255=(21,177,180)
            rgb255=(89,189,140)
            rgb255=(165,190,107)
            rgb255=(225,185,82)
            rgb255=(252,206,46)
            rgb255=(249,251,14)
        },
        }
    \begin{axis}[
        xlabel={$L_\text{fib}$ [\SI{}{\kilo\meter}]},
        title={$\text{SNR}=\SI{30}{dB}$},
        small,
        view={0}{90},
        colorbar,       
        xtick distance=2,
        ytick distance=0.2, 
        colorbar style={
        xshift=-13pt,
        tickwidth=2pt,
        x tick style={color=black},
        y tick style={color=black},
        ytick distance=0.3,
        scaled y ticks = false,
        samples=5,    
        },
        colorbar/width=2.5mm,
        x tick style={color=black},
        y tick style={color=black},
        tickwidth=2pt,
        grid=none,
        width=0.92\textwidth,
        height=4.0cm,
        title style={yshift=-6.5pt,},
    ]
        \addplot3 [
        surf,
        shader=interp, 
        mesh/cols=15,  
        ] table[ col sep=comma] {Fig/r1_surf_PRE_30dB.txt}; 

         \addplot3 [
        mycolor6,densely dotted,thick,
        ] table[ col sep=comma] {Fig/r1_surf_PRE_30dB_noPrecoder_threshold.txt}; 
        
    \end{axis}
\end{tikzpicture}
\end{minipage}
    }
    \subfloat[{\small Electrical Noise}\label{fig:subfig_contour_el_noise}]%
    {\centering
    \begin{minipage}[t]{0.242\textwidth}%
\begin{tikzpicture}

\pgfplotsset{
        colormap={parula}{
            rgb255=(53,42,135)
            rgb255=(15,92,221)
            rgb255=(18,125,216)
            rgb255=(7,156,207)
            rgb255=(21,177,180)
            rgb255=(89,189,140)
            rgb255=(165,190,107)
            rgb255=(225,185,82)
            rgb255=(252,206,46)
            rgb255=(249,251,14)
        },
        }
    \begin{axis}[
        xlabel={$L_\text{fib}$ [\SI{}{\kilo\meter}]},
        ylabel={Offset $o$},
        title={$\text{SNR}=\SI{8}{dB}$ },
        small,
        xtick distance=2,
        ytick distance=0.2, 
        view={0}{90},
        colorbar,  
        colorbar style={
        tickwidth=2pt,
        xshift=-13pt,
        x tick style={color=black},
        y tick style={color=black},
        scaled y ticks = false,
        samples=8,    
        },
        colorbar/width=2.5mm,
        x tick style={color=black},
        y tick style={color=black},
        tickwidth=2pt,
        grid=none,
        width=0.92\textwidth,
        height=4.0cm,
        title style={yshift=-6.5pt,},
    ]
        \addplot3 [
        surf,
        shader=interp, 
        mesh/cols=15,  
        ] table[ col sep=comma] {Fig/r1_surf_POST_8dB.txt}; 

        \addplot3 [
        mycolor6,densely dotted,thick,
        ] table[ col sep=comma] {Fig/r1_surf_POST_8dB_noPrecoder_threshold.txt};

    \end{axis}
\end{tikzpicture}
\end{minipage}%
\hspace{5pt}
\begin{minipage}[t]{0.242\textwidth}
\begin{tikzpicture}

\pgfplotsset{
        colormap={parula}{
            rgb255=(53,42,135)
            rgb255=(15,92,221)
            rgb255=(18,125,216)
            rgb255=(7,156,207)
            rgb255=(21,177,180)
            rgb255=(89,189,140)
            rgb255=(165,190,107)
            rgb255=(225,185,82)
            rgb255=(252,206,46)
            rgb255=(249,251,14)
        },
        }
    \begin{axis}[
        xlabel={$L_\text{fib}$ [\SI{}{\kilo\meter}]},
        title={$\text{SNR}=\SI{14}{dB}$},
        small,
        view={0}{90},
        colorbar,       
        xtick distance=2,
        ytick distance=0.2, 
        colorbar style={
        xshift=-13pt,
        tickwidth=2pt,
        x tick style={color=black},
        y tick style={color=black},
        ytick distance=0.3,
        scaled y ticks = false,
        samples=6,    
        },
        colorbar/width=2.5mm,
        x tick style={color=black},
        y tick style={color=black},
        tickwidth=2pt,
        grid=none,
        width=0.92\textwidth,
        height=4.0cm,
        title style={yshift=-6.5pt,},
    ]
        \addplot3 [
        surf,
        shader=interp, 
        mesh/cols=15,  
        ] table[ col sep=comma] {Fig/r1_surf_POST_14dB.txt}; 

        \addplot3 [
        mycolor6,densely dotted,thick,
        ] table[ col sep=comma] {Fig/r1_surf_POST_14dB_noPrecoder_threshold.txt};

    \end{axis}
\end{tikzpicture}
\end{minipage}
    }
    \caption{$32$-ASK rates for SIC with $S=4$ vs.\ offset $o$ and fiber lengths $L_\text{fib}$ from \SI{100}{\meter} to \SI{10}{\kilo\meter} for $(a)$ optical noise only and $(b)$ electrical noise only. The rates with and without a precoder differ by less than $ \SI{0.1}{bpcu}$ above the dotted red curve.}
    \label{fig:contour}
\end{figure*}

\subsection{Unamplified Link}
Now suppose there is no optical amplification and $\nu_{N_1}=0$. We set the ADC filter bandwidth to $\Brx = 2B$ and define the electrical SNR as 
\begin{align}
    P_\text{rx} \big/ \nu_{N_2}
    \label{eq:snr_el}
\end{align}
with $P_\text{rx}$ defined in~\eqref{eq:prx} and $\nu_{N_2} = \nu_2 \Brx$ is the power of the electrical AWGN within the ADC filter bandwidth $\Brx$; see~\eqref{eq:autocorr_n2}. The ADC filter removes a small portion of $Y'(t)$, and the samples are thus not sufficient statistics. To simplify, we choose $\nu_{N_2} = 1$.

We run GVAMP with adaptive damping and variance annealing, see Sec.~\ref{subsec:adaptive-denoising}. In the initial iterations, the LMMSE denoiser assumes the presence of intermediate optical noise. At iteration $n_\text{trig}$, the optical noise variance is fixed to $\nu_{N_1} = 1.5\%$ to support convergence. This is especially useful at high SNR. For the output denoiser expressions for electrical noise \eqref{eq:z_den_mean_final}-\eqref{eq:z_den_var_final}, we use the saddle point approximation in Appendix~\ref{appendix-saddle_point}.

Fig.~\ref{fig:vamp_el_rates} shows AIRs for $S=4$ SIC stages and $M$-ASK, $M \in\{4,8,16,32,64\}$, $L_\text{fib}=\SI{4}{\kilo\meter}$ and $o=0.2$. We plot the rates for $M=64$ and state-of-the-art systems using $o=1$, but omit smaller modulation orders due to similar behavior. Using $o=0.2$ gains up to $\approx\SI{3}{dB}$ over $o = 1$ with $64$-ASK when operating at $\approx 85$\% of the maximum rate. Equivalently, $o=0.2$ gains $\approx\SI{0.9}{bpcu}$ over $o=1$ at an SNR of \SI{17}{dB}. For $o=0.2$, the EXIT predictions are within a gap of $\SI{0.4}{dB}$ over the entire SNR range. For $o=1$, the predictions are less accurate and show a gap of $\SI{0.7}{dB}$ at high SNR. Fig.~\ref{fig:vamp_el_rates_stage} compares the SIC rates for $\ell = 1,2,3,4$ for $64$-ASK-$0.2$ and their EXIT predictions. SIC improves the rates substantially, as higher SIC stages gain over SDD ($\ell = 1$). The EXIT predictions for $\ell \geq 2$ agree well with the simulations.

Fig.~\ref{fig:subfig_contour_el_noise} shows $S=4$ rates for different offsets $o$ and fiber lengths $L_\text{fib}$ for an SNR of $\SI{8}{dB}$ and $\SI{14}{dB}$. A small offset $o$ between $0.15$ and $0.2$ achieves the maximum rate for fibers longer than $\approx\SI{2}{\kilo\meter}$ and assists GVAMP. The plot suggests that beyond $\SI{2}{\kilo\meter}$, the rates are roughly invariant to fiber length and change only with $o$. One should choose a small offset $o\approx 0.2$ to operate efficiently. One can omit precoding for points above the red dotted line, as the rate loss is less than $\SI{0.1}{bpcu}$. 

\begin{figure}[!t]
    \centering
    \pgfdeclarelayer{background}
\pgfdeclarelayer{foreground}
\pgfsetlayers{background,main,foreground}

\begin{tikzpicture}[]

\begin{axis}[%
yminorticks=true,
width=\figwidth,
height=\figheight,
xmajorgrids,
ymajorgrids,
yminorgrids,
minor x tick num=4,
minor y tick num=4,
grid=both,
legend style={legend cell align=left,  draw=white!15!black, font=\small,  %
legend pos=south east
},
xlabel style={font=\color{white!15!black}},
ylabel style={font=\color{white!15!black}},
axis background/.style={fill=white},
scale only axis,
xmin=-5,
xmax=22,
ytick={0,1,...,7},
Rate_VS_PRX_NuN2,
ymin=0,
ymax=6.0,
ylabel shift = 0pt,legend pos=outer north east,
legend cell align=left,
legend style={cells={align=left}},
legend pos= north west,
title style={yshift=-5pt,},
]

\pgfplotsinvokeforeach{4,8,16,32}
{
\addplot[ASK,name path global=#1-ASK-0.2,forget plot] table [x=SNR, y=IqYX, col sep=comma,x expr= \thisrow{SNR}]{plots/vamp_v11_r1/POST/r1,#1-ASK-0.2,S=4,C=0,Rb=0,n1=0,n2=1,n1m=0,n2m=1,P=cO,ps=RRC,a=0.01,Nsp=250,Npi=0,n=2048,L=4,Rs=300,Nr=1,Nb=128,I=250.txt};
}
\addplot[ASK,name path global=64-ASK-0.2] table [x=SNR, y=IqYX, col sep=comma,x expr= \thisrow{SNR}]{plots/vamp_v11_r1/POST/r1,64-ASK-0.2,S=4,C=0,Rb=0,n1=0,n2=1,n1m=0,n2m=1,P=cO,ps=RRC,a=0.01,Nsp=250,Npi=0,n=2048,L=4,Rs=300,Nr=1,Nb=128,I=250.txt};
\addlegendentry{$M$-ASK-$0.2$};

\addplot[ASK,densely dashdotted,name path global=64-ASK-0.2-EXIT,forget plot] table [x=SNR, y=EXIT, col sep=comma,x expr= \thisrow{SNR}]{plots/vamp_v11_r1/POST/r1,64-ASK-0.2,S=4,C=0,Rb=0,n1=0,n2=1,n1m=0,n2m=1,P=cO,ps=RRC,a=0.01,Nsp=250,Npi=0,n=2048,L=4,Rs=300,Nr=1,Nb=128,I=250.txt};

\addplot[PAM,name path global=64-ASK-1] table [x=SNR, y=IqYX, col sep=comma,x expr= \thisrow{SNR}]{plots/vamp_v11_r1/POST/r1,64-ASK-1.0,S=4,C=0,Rb=0,n1=0,n2=1,n1m=0,n2m=1,P=cO,ps=RRC,a=0.01,Nsp=250,Npi=0,n=2048,L=4,Rs=300,Nr=1,Nb=128,I=250.txt};
\addlegendentry{64-ASK-$1.0$};

\addplot[PAM,densely dashdotted,name path global=64-ASK-1-EXIT] table [x=SNR, y=EXIT, col sep=comma,x expr= \thisrow{SNR}]{plots/vamp_v11_r1/POST/r1,64-ASK-1.0,S=4,C=0,Rb=0,n1=0,n2=1,n1m=0,n2m=1,P=cO,ps=RRC,a=0.01,Nsp=250,Npi=0,n=2048,L=4,Rs=300,Nr=1,Nb=128,I=250.txt};

\pgfplotsinvokeforeach{4, 8,16,32,64}
{
\MeasureXDistance{{0.85*log2(#1)}}{#1-ASK-0.2}{64-ASK-1}{anchor=east,xshift=-1.1cm,font=\normalsize};
}

\end{axis}
\end{tikzpicture}%
    \caption{AIRs for unamplified links with $L_\text{fib} = \SI{4}{\kilo\meter}$, $B = \SI{300}{\giga Bd}$, $S=4$. The two dash-dotted curves are EXIT predictions for $M=64$. }
    \label{fig:vamp_el_rates}
\end{figure}

\begin{figure}[!t]
    \centering 
    \pgfkeys{
    /pgf/number format/.cd,
    precision=1,
}

\begin{tikzpicture}[spy using outlines={magnification=2, connect spies}]

\begin{axis}[%
yminorticks=true,
xmajorgrids,
ymajorgrids,
yminorgrids,
minor x tick num=4,
minor y tick num=4,
ytick={0,1,2,3,4,5,6},
grid=both,
title style={yshift=-5pt,},
legend style={legend cell align=left,  draw=white!15!black,  %
legend pos=south east
},
xlabel style={font=\color{white!15!black}},
ylabel style={font=\color{white!15!black}},
axis background/.style={fill=white},
scale only axis,
width=\figwidth,
height=\figheight,
xmin=-5,
xmax=22,
ymin=0,
ymax=6.02,
minor x tick num=4,
minor y tick num=4,
Rate_VS_PRX_NuN2,
legend pos= south east,
legend cell align=left,
legend style={font=\small,cells={align=left}}
,
]

\addplot[ASK,mark=*,densely dashed,mark options={mark size=1pt,solid},name path global=ASK-0.2-s1] table [x=SNR, y=IqYXs_1, col sep=comma,x expr= \thisrow{SNR}]{plots/vamp_v11_r1/POST/r1,64-ASK-0.2,S=4,C=0,Rb=0,n1=0,n2=1,n1m=0,n2m=1,P=cO,ps=RRC,a=0.01,Nsp=250,Npi=0,n=2048,L=4,Rs=300,Nr=1,Nb=128,I=250.txt};
\addlegendentry{$\ell = 1$};

\addplot[ASK,mark=*,mark options={mark size=1pt,solid},densely dashdotted,name path global=ASK-0.2-s2] table [x=SNR, y=IqYXs_2, col sep=comma,x expr= \thisrow{SNR}]{plots/vamp_v11_r1/POST/r1,64-ASK-0.2,S=4,C=0,Rb=0,n1=0,n2=1,n1m=0,n2m=1,P=cO,ps=RRC,a=0.01,Nsp=250,Npi=0,n=2048,L=4,Rs=300,Nr=1,Nb=128,I=250.txt};
\addlegendentry{$\ell = 2$};

\addplot[ASK,mark=*,mark options={mark size=1pt,solid},densely dashdotdotted,name path global=ASK-0.2-s3] table [x=SNR, y=IqYXs_3, col sep=comma,x expr= \thisrow{SNR}]{plots/vamp_v11_r1/POST/r1,64-ASK-0.2,S=4,C=0,Rb=0,n1=0,n2=1,n1m=0,n2m=1,P=cO,ps=RRC,a=0.01,Nsp=250,Npi=0,n=2048,L=4,Rs=300,Nr=1,Nb=128,I=250.txt};
\addlegendentry{$\ell = 3$};

\addplot[ASK,mark=*,densely dotted,mark options={mark size=0.8pt,solid},name path global=ASK-0.2-s4] table [x=SNR, y=IqYXs_4, col sep=comma,x expr= \thisrow{SNR}]{plots/vamp_v11_r1/POST/r1,64-ASK-0.2,S=4,C=0,Rb=0,n1=0,n2=1,n1m=0,n2m=1,P=cO,ps=RRC,a=0.01,Nsp=250,Npi=0,n=2048,L=4,Rs=300,Nr=1,Nb=128,I=250.txt};
\addlegendentry{$\ell = 4$};

\addplot[EXIT,densely dashed,name path global=SE-0.2-s1] table [x=SNR, y=EXITs_1, col sep=comma,x expr= \thisrow{SNR}]{plots/vamp_v11_r1/POST/r1,64-ASK-0.2,S=4,C=0,Rb=0,n1=0,n2=1,n1m=0,n2m=1,P=cO,ps=RRC,a=0.01,Nsp=250,Npi=0,n=2048,L=4,Rs=300,Nr=1,Nb=128,I=250.txt};
\addlegendentry{EXIT};

\addplot[EXIT,densely dashdotted,,name path global=SE-0.2-s1,forget plot] table [x=SNR, y=EXITs_2, col sep=comma,x expr= \thisrow{SNR}]{plots/vamp_v11_r1/POST/r1,64-ASK-0.2,S=4,C=0,Rb=0,n1=0,n2=1,n1m=0,n2m=1,P=cO,ps=RRC,a=0.01,Nsp=250,Npi=0,n=2048,L=4,Rs=300,Nr=1,Nb=128,I=250.txt};

\addplot[EXIT,densely dashdotdotted,,name path global=SE-0.2-s1,forget plot] table [x=SNR, y=EXITs_3, col sep=comma,x expr= \thisrow{SNR}]{plots/vamp_v11_r1/POST/r1,64-ASK-0.2,S=4,C=0,Rb=0,n1=0,n2=1,n1m=0,n2m=1,P=cO,ps=RRC,a=0.01,Nsp=250,Npi=0,n=2048,L=4,Rs=300,Nr=1,Nb=128,I=250.txt};

\addplot[EXIT,densely dotted,,name path global=SE-0.2-s1,forget plot] table [x=SNR, y=EXITs_4, col sep=comma,x expr= \thisrow{SNR}]{plots/vamp_v11_r1/POST/r1,64-ASK-0.2,S=4,C=0,Rb=0,n1=0,n2=1,n1m=0,n2m=1,P=cO,ps=RRC,a=0.01,Nsp=250,Npi=0,n=2048,L=4,Rs=300,Nr=1,Nb=128,I=250.txt};

\coordinate (spypoint) at (axis cs:15,5.15);
\coordinate (magnifyglass) at (axis cs:-0.5,4.8);

\end{axis}

\spy [size=2cm] on (spypoint) in node[fill=white] at (magnifyglass);

\end{tikzpicture}%
    \caption{SIC AIRs for unamplified links with $L_\text{fib} = \SI{4}{\kilo\meter}$, $64$-ASK-$0.2$, $B = \SI{300}{\giga Bd}$ and $S=4$. Black curves without markers, with the same linestyle, are the EXIT predictions of the stage AIRs. }
    \label{fig:vamp_el_rates_stage}
\end{figure}

\subsection{Transmitter with Excess Bandwidth}
\label{sec:practical_tx_excess_bw}
Consider optical amplification, and FD-RRC pulses with $\alpha_\text{tx} = 99\%$, resulting in a rapid pulse decay. We set $\mathbf{P} = \mident{n}$ to reduce complexity, latency, and the transmitter peak-to-average power ratio. We further set the optical brickwall filter bandwidth to match $\alpha_\text{tx}$, i.e., $B_\mathrm{D} \approx 2B$. This gives $\gamma \approx 1$ and roughly doubles the optical noise power to $\nu_{N_1} \approx 1$; see Sec.~\ref{sec:results_opt_noise}. The GVAMP receiver noise power is now roughly four times that of the coherent receiver in \eqref{eq:snr_opt}.

The DD output has a two-sided bandwidth of roughly $4B$. We choose the electrical receive filter as a unit-gain brickwall filter with a two-sided bandwidth $\Brx = 4B$, but sample with $N_\text{os} = 2$ to avoid increasing the analog receiver complexity. These aliased samples are not sufficient statistics.

Fig.~\ref{fig:vamp_practical_rates} plots 16-ASK rates with $o=0.25$ and $o=1$ for $L_\text{fib} = \SI{3}{\kilo\meter}$, which corresponds to operating at the rate transition boundary of Fig.~\ref{fig:subfig_contour_opt_noise}. The figure also shows AIRs for $\alpha_\text{tx} = 1\%$, a random precoder \eqref{eq:fft_precoder}, and filter choices from Sec.~\ref{sec:results_opt_noise}. For the same modulation, the power gap between $\alpha_\text{tx} = 1\%$ and $\alpha_\text{tx} = 99\%$ is $\approx\SI{3}{dB}$ at high SNR. This gap is likely because the optical noise power doubled compared to Sec.~\ref{sec:results_opt_noise} and the receiver does not have sufficient statistics. For both $\alpha_\text{tx}$, the SNR gap between ASK with $o=0.25$ and legacy ASK with $o = 1$ is $\approx\SI{5}{dB}$ at $85\%$ of the maximum rate. Interestingly, $\alpha_\text{tx} = 99\%$ improves the EXIT prediction error to less than $\SI{0.2}{dB}$ everywhere. The predictions improve especially at low SNRs compared to Fig.~\ref{fig:vamp_opt_rates}, which uses $\alpha_\text{tx} = 1\%$. This suggests that increasing $\alpha_\text{tx}$ closes the EXIT prediction gap, which is relevant for short-reach links that allow excess bandwidth to relax analog component requirements. 

\begin{figure}[!t]
    \centering
    \pgfkeys{
    /pgf/number format/.cd,
    precision=1,
}

\pgfdeclarelayer{fg}    %
\pgfsetlayers{main,fg}  %

\begin{tikzpicture}[]

\begin{axis}[%
yminorticks=true,
xmajorgrids,
ymajorgrids,
yminorgrids,
minor y tick num=4,
ytick={0,1,2,3,4},
grid=both,
title style={yshift=-5pt,},
legend style={legend cell align=left,  draw=white!15!black, %
legend pos=south east
},
xlabel style={font=\color{white!15!black}},
ylabel style={font=\color{white!15!black}},
axis background/.style={fill=white},
scale only axis,
width=\figwidth,
height=\figheight,
xmin=0,
xmax=40,
Rate_VS_PRX_NuN1_HALVE,
ymin=0,
ymax=4.01,
minor x tick num=4,
minor y tick num=4,
legend pos= south east,
legend cell align=left,
legend style={font=\small,cells={align=left}},
]

\addplot[ASK,mark=*,mark options={mark size=1.4pt},solid,name path global=16-ASK-0.25,] table [x=SNR, y=IqYX, col sep=comma,x expr= \thisrow{SNR}+6]{plots/vamp_v11_r1/PRE/r1,16-ASK-0.25,S=4,C=0,Rb=1,n1=1,n2=0,n1m=0,n2m=0,P=cO,ps=RRC,a=0.01,Nsp=250,Npi=0,n=2048,L=3,Rs=300,Nr=1,Nb=128,I=250.txt};
\addlegendentry{$\alpha_\text{tx} = 1\%$};

\addplot[PAM,mark=triangle*,mark options={mark size=1.8pt},solid,name path global=16-ASK-1,] table [x=SNR, y=IqYX, col sep=comma,x expr= \thisrow{SNR}+6]{plots/vamp_v11_r1/PRE/r1,16-ASK-1.0,S=4,C=0,Rb=1,n1=1,n2=0,n1m=0,n2m=0,P=cO,ps=RRC,a=0.01,Nsp=250,Npi=0,n=2048,L=3,Rs=300,Nr=1,Nb=128,I=250.txt};
\addlegendentry{$\alpha_\text{tx} = 1\%$};

\addplot[ASK,mark=*,mark options={mark size=1.4pt,fill=white,solid},name path global=16-ASK-0.25_R_0.99,] table [x=SNR, y=IqYX, col sep=comma,x expr= \thisrow{SNR}+6]{plots/vamp_v11_r1/PRE/r1,16-ASK-0.25,S=4,C=0,Rb=1,n1=1,n2=0,n1m=0,n2m=0,P=_,ps=RRC,a=0.99,Nsp=250,Npi=0,n=2048,L=3,Rs=300,Nr=1,Nb=128,I=250.txt};
\addplot[ASK,line width=0.7pt,mycolor6!70!black,densely dashdotted,name path global=16-ASK-0.25_R_0.99-EXIT,forget plot] table [x=SNR, y=EXIT, col sep=comma,x expr= \thisrow{SNR}+6]{plots/vamp_v11_r1/PRE/r1,16-ASK-0.25,S=4,C=0,Rb=1,n1=1,n2=0,n1m=0,n2m=0,P=_,ps=RRC,a=0.99,Nsp=250,Npi=0,n=2048,L=3,Rs=300,Nr=1,Nb=128,I=250.txt};
\addlegendentry{$\alpha_\text{tx} = 99\%$};

\addplot[PAM,mark=triangle*,mark options={mark size=1.8pt,fill=white,solid},solid,name path global=16-ASK-1_R_0.99] table [x=SNR, y=IqYX, col sep=comma,x expr= \thisrow{SNR}+6]{plots/vamp_v11_r1/PRE/r1,16-ASK-1.0,S=4,C=0,Rb=1,n1=1,n2=0,n1m=0,n2m=0,P=_,ps=RRC,a=0.99,Nsp=250,Npi=0,n=2048,L=3,Rs=300,Nr=1,Nb=128,I=250.txt};
\addplot[PAM,line width=0.7pt,mycolor1!70!black,densely dashdotted,name path global=16-ASK-1_R_0.99-EXIT,forget plot] table [x=SNR, y=EXIT, col sep=comma,x expr= \thisrow{SNR}+6]{plots/vamp_v11_r1/PRE/r1,16-ASK-1.0,S=4,C=0,Rb=1,n1=1,n2=0,n1m=0,n2m=0,P=_,ps=RRC,a=0.99,Nsp=250,Npi=0,n=2048,L=3,Rs=300,Nr=1,Nb=128,I=250.txt};
\addlegendentry{$\alpha_\text{tx} = 99\%$};

 \begin{pgfonlayer}{fg}  

\MeasureXDistance{{log2{(16)} * 0.850}}{16-ASK-0.25}{16-ASK-1}{anchor=east,xshift=-1.3cm,font=\normalsize};
\MeasureXDistance{{log2{(16)} * 0.830}}{16-ASK-0.25_R_0.99}{16-ASK-1_R_0.99}{anchor=west,font=\normalsize};

\draw[line width=0.5pt] (axis cs:19.2,2.5) ellipse (\OneColMul*0.42cm and 0.07cm) node[left,xshift=-0.6cm,opacitylabel,font=\normalsize]{$o=0.25$};
\draw[line width=0.5pt] (axis cs:24,2.5) ellipse (\OneColMul*0.42cm and 0.07cm) node[right,xshift=+0.6cm,opacitylabel,font=\normalsize]{$o=1$};

\end{pgfonlayer}

\end{axis}
\end{tikzpicture}%
    \caption{AIRs for 16-ASK-$o$ and optically amplified links with $L_\text{fib} = \SI{3}{\kilo\meter}$, $B = \SI{300}{\giga Bd}$, and $S=4$. $\alpha_\text{tx} = 1\%$ uses the precoder \eqref{eq:fft_precoder} and $\alpha_\text{tx} = 99\%$ uses no precoder. Dash-dotted lines are EXIT predictions for $\alpha_\text{tx} = 99\%$.}
    \label{fig:vamp_practical_rates}
\end{figure}

\subsection{Computational Complexity}
We measure complexity by counting multiplications. The FFT-based LMMSE denoiser dominates with $\mathcal{O}(\log_2 n)$ multiplications per symbol and iteration. We neglect the complexities of the input and output denoisers, since both can be precomputed and stored in lookup tables. 
Also, for higher-order modulation with $M \geq 16 $, the input denoiser \eqref{eq:projection_ep_I_II-1} complexity can be reduced by approximating the PMF $P_\mathbf{U}$ with a PDF $p_\mathbf{U}$.
Let $n_{\text{it}}$ be the iteration at which GVAMP reaches $99.5\%$ of its ultimate rate. The asymptotic per-symbol complexity of GVAMP scales as
\begin{align}
    \mathcal{O}( n_\text{it}  \log_2 n + M ).
    \label{eq:apsc}
\end{align}

Fig.~\ref{fig:complexity} plots the GVAMP complexities in mpib for optically amplified SSMF and SDD, i.e., one encoding and decoding stage.
Using mpib means dividing the number of multiplications per symbol by the rate, which is here $0.85\log_2 M$. The figure also shows the complexities of linear frequency-domain detectors \cite{plabst2022achievable}, the forward-backward algorithm (FBA), Gibbs sampling (GS) \cite{prinz2023successive}, and NNs \cite{plabst2024neural}, evaluated for $L_\text{fib}=\SI{0.5}{\kilo\meter}$ where these methods are computationally feasible. The papers \cite{plabst2022achievable,prinz2023successive,plabst2024neural} study unamplified links; the complexities for optically amplified links are similar.

Linear detectors with symbol-wise soft-demapping have the lowest complexity $\mathcal{O}(\log_2 n + M)$, but they lose significant rate with strong dispersion and  DD \cite[Fig.~5a]{wiener_filter_plabst2020}. For example, the linear detector of \cite{wiener_filter_plabst2020} could not exceed AIRs of $\approx\SI{1}{bpcu}$. 

Channel shortening combines a linear filter and FBA with small memory, denoted as L-FBA \cite{rusek2012optimal,wettlin2020dsp}. For instance, using a unit-memory FBA has complexity $\mathcal{O}(\log_2 n + M^2)$ and improves performance for moderate CD and small $M$. We used a cascade of three processing steps: a 201-tap Wiener filter \cite{wiener_filter_plabst2020}, followed by an optimized 9-tap linear filter, followed by the FBA for a 3-tap linear surrogate channel model; see \cite[Eq.~(39)]{plabst2022achievable}. The 9-tap filter and surrogate model parameters were jointly optimized using the approach of \cite[Appendix]{plabst2022achievable}. Even then, the L-FBA could not exceed AIRs of $\approx\SI{1}{bpcu}$.

\begin{figure}
    \centering
    \pgfdeclareplotmark{mystar}{
    \node[star, star points=5, star point ratio=0.5, draw=black, solid, fill=magenta, minimum width=3pt, inner sep=0pt, outer sep=0pt, anchor=center] {};
}

\pgfmathsetmacro{\Rperc}{0.85}

\begin{tikzpicture}
	\begin{axis}[
    legend columns=4, 
    scale only axis,
    width=0.98*\figwidth,
    height=0.8*\figheight,
    xmode=log,
    ymode=log,
    xmajorgrids,
    xminorgrids,
    ymajorgrids,
    yminorgrids,
    yminorticks=true,
    grid=both,
    xmin=3.5,
    xmax=72,
    ymin=1E0,
    ymax=1E7,
    title style={yshift=-5pt,},
    ytick={1E0,1E1,1E2,1E3,1E4,1E5,1E6,1E7,1E8},
    xtick={4,8,16,32,64,128},
    xticklabels={4,8,16,32,64,128},
    xlabel={$M$},
    ylabel={mpib},
    legend style={font=\small,legend cell align=left,draw=white!15!black, row sep=-2.5pt,  
    at={([yshift=7pt]1,1)},
    anchor=south east,
    }, 
    title={},
    ]

    \addplot[FBA,ASK,mark=asterisk,solid,  mark size=2.2pt, mark options={line width=1pt,solid},line width=0.7pt] 
    table[row sep=crcr, y expr= {\thisrow{y} / (\Rperc * log2(\thisrow{x})) }] {
    x y\\
    4 1e6 \\ %
    8 16e6 \\ %
    } node [pos=0.46, above=0.5pt,rotate=21.5,opacitylabel,,font=\footnotesize,] {$\approx\SI{0.5}{\kilo\meter}$}; 
    \addlegendentry{FBA~\cite{plabst2022achievable}};

    \addplot[GS,PAM,mark=square*, solid, mark size=1.6pt, mark options={line width=1pt,solid},line width=0.7pt] table[row sep=crcr, y expr= {\thisrow{y} / (\Rperc * log2(\thisrow{x})) }] {
    x y\\
    4 2E5\\ %
    8 2E5\\ %
    32 2E7\\ %
    } node [pos=0.5, above=1pt,rotate=20,opacitylabel,,font=\footnotesize] {$\approx\SI{0.5}{\kilo\meter}$}; 
    \addlegendentry{GS~\cite{prinz2023successive}};

    \addplot[RNN,mark=triangle*,solid, mark size=2pt, mark options={line width=1pt,solid},line width=0.9pt
    ] table[row sep=crcr, y expr= {\thisrow{y} / (\Rperc * log2(\thisrow{x})) }] {
    x y\\
    4 3E4 \\ %
    8 5E4 \\ %
    16 1E5 \\ %
    32 2E5 \\ %
    64 5E5 \\ %
    } node [pos=0.57, above=1pt,rotate=8,opacitylabel,font=\footnotesize, text=mycolorrnn!80!black] {$\approx\SI{0.5}{\kilo\meter}$} ; 
    \addlegendentry{NN~\cite{plabst2024neural}};

    \addplot[QAM,mark=diamond*,solid, mark size=2.5pt, mark options={fill=white,line width=1pt,solid},line width=0.9pt] table[x=ModSize, col sep=comma, y expr=(\thisrow{nitOpt} * log2(2048) + \thisrow{ModSize}) / (\Rperc * log2(\thisrow{ModSize})),  col sep=comma] 
    {Fig/v11_r1_iterations_Lfib=4_ratePerc=0.85_off=0.2.txt} 
    node [pos=0.37, above=3pt,rotate=-4,opacitylabel,font=\footnotesize,text=mycolor5!70!black] {\SIrange{0.1}{10}{\kilo\meter} (optical/electrical noise)}; 
    \addlegendentry{GVAMP};

    \addplot[QAM,mark=diamond*,solid, mark size=2.5pt, mark options={fill=white,line width=1pt,solid},densely dashed,name path global=vampopt,line width=1.1pt,mark size=2.5pt, ] 
    table[x=M, col sep=comma, y expr=(\thisrow{nitmin} * log2(2048) + \thisrow{M})/ (\Rperc * log2(\thisrow{M}))]{Fig/r1_optim_complexity.txt}; 
    \addlegendentry{opt. GVAMP};

     \addplot[black,solid, name path global=lineq, line width=0.5pt,mark=*, mark size=1.8pt, mark options={fill=white,line width=0.5pt,solid}] table[row sep=crcr, y expr=(log2(2048) + \thisrow{modsize}^2)/ (\Rperc * log2(\thisrow{modsize}))] {
        modsize iter \\
        4  0 \\ %
        8  0 \\ %
        16 0 \\ %
        32 0 \\ %
        64 0 \\ %
        }; 
    \addlegendentry{L-FBA};

    \addplot[black,densely dashed, name path global=lineq, line width=0.5pt,mark=*, mark size=1.8pt, mark options={fill=white,line width=0.5pt,solid}] table[row sep=crcr, y expr=(\thisrow{iter} * log2(2048) + \thisrow{modsize}) / (\Rperc * log2(\thisrow{modsize}))] {
    modsize iter \\
    4  1 \\ %
    8  1 \\ %
    16 1 \\ %
    32 1 \\ %
    64 1 \\ %
    }; %
    \addlegendentry{Linear};

\draw[doubarrsmall, black, line width=0.4pt](axis cs: 8,36) -- node[midway,left, xshift=-0.3cm,opacitylabel,font=\normalsize]{$\times\, 4$}(axis cs: 8, 85);
\draw[doubarrsmall, black, line width=0.4pt](axis cs: 16,12) -- node[midway,right, xshift=+0.3cm,opacitylabel,font=\normalsize]{$\times\, 12$}(axis cs: 16, 60);

\end{axis}
\end{tikzpicture}
    \caption{SDD complexities. The GVAMP complexity is for an optically amplified link and $n=2048$. All curves are computed at 85\% of the maximum rate $\log_2 M$. All detectors except for GVAMP are either computationally infeasible (FBA, GS, NN) or rate-limited (L-FBA, linear) for larger $L_\text{fib}$.}
    \label{fig:complexity}
\end{figure}

The FBA complexity is $M^{K+1}$, where $K$ is the channel memory of the combined sampled response; see Fig.~\ref{fig:cd_plot}. This complexity is prohibitive for long fiber and/or large $M$. The GS and NN receivers reduce complexity, but still require many multiplications \cite[Fig.~8]{plabst2024neural}. Also, the NN receiver must be trained for a specific $M$, fiber lengths, and SNRs.

The FBA, GS, and NN curves in Fig.~\ref{fig:complexity} are taken from \cite{plabst2024neural}, which studied a different fiber length and baud rate $B$. The CD response length is proportional to $B^2  L_\text{fib}$, cf. \eqref{eq:cd_response_freq}, so the results in \cite{plabst2024neural} translate to $B=\SI{300}{\giga Bd}$ and $L_\text{fib}\approx\SI{0.5}{\kilo\meter}$. The mpib of the first SIC stage (SDD) dominate the complexity because no interference is removed.
For $L_\text{fib} = \SI{4}{\kilo\meter}$, the GS is run as in \cite[Fig.~13]{plabst2024neural}, except that the symbol memory was increased to $99$. However, the AIRs for GS did not exceed \SI{2}{bpcu} at high SNR. Similarly, the AIRs of the NN detector remained at $\approx\SI{2}{bpcu}$, even for large NNs with $\approx\SI{1e5}{mpib}$ \cite[Table~III]{plabst2024neural} and extensive training. The FBA must again significantly truncate the channel memory to remain viable, and could not exceed $\approx \SI{0.25}{bpcu}$.

GVAMP's mpib are similar for different $M$, $L_\text{fib}$ ranging from $\SIrange{0.1}{10}{\kilo\meter}$, and optically amplified and unamplified links. We thus show only one curve labeled as ``optical/electrical noise''.
One can reduce complexity by optimizing the damping and annealing for each scenario (modulation format, SNR, fiber length). Table~\ref{tab:mpib_optimized} lists the optimized parameters at 85\% of the maximum rate $\log_2 M$. The parameter $\alpha_\text{anneal}$ replaces the exponential prefactor of Table~\ref{tab:variance_annealing}. A smaller $n_\mathrm{W}$ results in a more dynamic adjustment of the damping factor, which shrinks the plateau in Fig.~\ref{subfig:TRACE_PRE_II} compared to Fig.~\ref{subfig:TRACE_PRE_I}. This reduces the required number of iterations by $\approx 40\%$ from $65$ to $37$ for a small rate loss of $\approx\SI{0.1}{bpcu}$. 

\begin{table}
    \centering
    \pgfplotstableset{
    columns={M,SNRdB,ratemincplx,nitmin,mindampvec,minannealvec,mpib},
    col sep = comma,
    every head row/.style={before row=\toprule,after row=\midrule},
    every last row/.style={after row=\bottomrule},
    columns/M/.append style={column type = {r|},column name={$M$}},
    columns/SNRdB/.append style={column type = {c},fixed, precision=1,column name={SNR},
      postproc cell content/.append style={
            /pgfplots/table/@cell content/.add={}{$\,$dB},
        }, 
    },
    columns/ratemincplx/.append style={column type = {r},fixed zerofill, precision=1,column name={rate},
          postproc cell content/.append style={
            /pgfplots/table/@cell content/.add={}{$\,$bpcu},
        }, },
    columns/nitmin/.append style={column type = {c}, fixed, precision=0, column name={$n_\text{it}$}},
    columns/mindampvec/.append style={column type = {|c}, column name={$n_\text{W}$}},
    columns/minannealvec/.append style={column type = {l|},  fixed zerofill, precision=2, column name={$\alpha_\text{anneal}$}},
    columns/mpib/.append style={column type = {r}, fixed, precision=0, column name={mpib}},
}

    \caption{Optimized mpib for optical amplification, SDD ($\ell=1$), $M$-ASK-0.2, over $\SI{4}{\kilo\meter}$ of fiber at $\SI{300}{\giga Bd}$ (C-band).}
    \label{tab:mpib_optimized}
    {\IfTwoCol{\normalsize}
    \begin{filecontents*}{Fig/r1_optim_complexity.txt}
M,SNRdB,ratemincplx,nitmin,mindampvec,minannealvec,mpib
4,13.5,1.567577,33.54615,2,0.6,219.4163
8,18.5,2.465849,27.36154,1,0.3832653,121.1674
16,23.5,3.374421,32.06154,2,0.4193878,108.4344
32,28.5,4.200265,34.09231,2,0.3712245,95.76833
64,33.5,5.035605,37.21538,2,0.2869388,92.8175
\end{filecontents*}
\pgfplotstabletypeset[
    columns={M,ratemincplx,SNRdB,nitmin,mindampvec,minannealvec,mpib},
    ]{Fig/r1_optim_complexity.txt}
}
\end{table}

For example, operating at \SI{3.3}{bpcu} and \SI{300}{\giga Bd} achieves $\approx\SI{1}{\tera bit \per\second}$ net data rate. Here, GVAMP reduces the mpib by more than a factor of $300$ compared to NNs. The cost is even reduced by $\approx 3$ orders of magnitude at \SI{5}{bpcu}. For these cases, GVAMP requires 108 and 93 mpib, or $\approx 12$ and $6$ times the complexity of linear equalization. GVAMP has $\approx 4$ times the complexity of L-FBA at $M=8$. The complexity of GVAMP and L-FBA is roughly the same at $M=16$.

\subsection{Simulations with Polar-Coded Modulation}
To verify the AIRs, we simulated polar-coded modulation (PCM) \cite{arikan_channel_2009,Seidl-IT13} for an optically-amplified link with the parameters in Table~\ref{tab:simparams}. We compare the performance of bipolar 32-ASK-0.2 and unipolar 32-ASK-1 for SDD. The transmitter uses multilevel polar codes and an outer 16-bit cyclic redundancy check code to enable successive cancellation list (SCL) decoding \cite{tal_list_2015}. The receiver applies multi-stage SCL decoding with a list size of 32 and list passing across stages \cite{Prinz-Yuan-ISTC18,Karakchieva-SCC19}.

The target rate is $R = \SI{4.2}{bpcu}$, which 32-ASK-0.2 and 32-ASK-1 achieve at the SNRs $\approx\SI{27.9}{dB}$ and $\approx\SI{33.3}{dB}$, respectively; see Fig.~\ref{fig:vamp_opt_rates}. We simulate frames with $n=4096$ symbols and evaluate the frame-error rate (FER) after observing at least 50 frame errors. 32-ASK has five bit levels and thus five component polar codes of length 4096 each. For $R=\SI{4.2}{bpcu}$, there are $n R \approx 3440$ data bits per transmitted frame. We optimized the polar codes using Monte Carlo methods \cite{arikan_channel_2009,bocherer_efficient_2017}. The optimization identified reliable bit positions by decoding multiple frames during a training run.

Fig.~\ref{fig:fer_optical_noise_4km} plots the FERs and SNR decoding thresholds. We observed that GVAMP did not converge for all frames; these failures could be identified via outliers in the metrics \eqref{eq:variational_bayes}. To reduce their number, we chose $n_\mathrm{W}=14$ with $\alpha_\text{anneal} = 0.45$ for 32-ASK-0.2 and $\alpha_\text{anneal} = 0.35$ for 32-ASK-1. A second GVAMP was run with a random initialization if \eqref{eq:variational_bayes} indicated an outlier, namely at least 10 times worse than the average metric of previously detected blocks. Such attempts were initiated approximately every $500^\text{th}$ block. The damping and annealing increased the number of GVAMP iterations to $n_\text{it} = 71$ for 32-ASK-0.2 and $n_\text{it} = 92$ for 32-ASK-1.

\begin{figure}
\centering
\begin{tikzpicture}%

  \begin{axis}[%
      STDPLOT,
      width=1.1*\figwidth,
      height=1.3*\figheight,
      Rate_VS_PRX_NuN1_HALVE,
      ylabel={FER},
      yminorticks=true,
      xmajorgrids,
      ymajorgrids,
      yminorgrids,
      minor x tick num=3,
      minor y tick num=3,
      grid=both,
      ymode=log,
      ymin=1E-5,
      xmin=23,
      xmax=37,
      legend style={at={(0,0)}, anchor=south west, font=\footnotesize}
    ]

    \addplot[ASK,mark=diamond*,mark options={fill=white},mark size=1.2pt,name path global=ASKFER] table [x=SNR, y=FER, col sep=comma,OPTSNRreal]{plots/vamp_v11_r1/CODED/r1_ASK_SDD_R=4.2bpcu_FER_20251222.txt};
    \addlegendentry{32-ASK-0.2};

    \addplot[PAM,mark=*,mark size=1.0pt,mark options={fill=white},name path global=PAMFER] table [x=SNR, y=FER, col sep=comma,OPTSNRreal]{plots/vamp_v11_r1/CODED/r1_PAM_SDD_R=4.2bpcu_FER_20251222.txt};
    \addlegendentry{32-ASK-1};

    \addplot [ASK,draw=none,abovecurve={threshold}{mycolor6}{0.55ex}{0.52\columnwidth}] coordinates {(27.8950,1) (27.8950,1E-8)}; %
    \addplot [ASK,dashed,name path global=ASKthres,] coordinates { 
      (27.8950,1)
      (27.8950,1E-8)
    };

    \addplot [PAM,draw=none,abovecurve={threshold}{mycolor1}{-1.8ex}{0.52\columnwidth}] coordinates {(33.2830,1) (33.2830,1E-8)}; %
    \addplot [PAM,dashed,name path global=PAMthres] coordinates {
      (33.2830,1)
      (33.2830,1E-8)
    };

    \MeasureXDistance{8E-4}{ASKFER}{PAMFER}{anchor=east,yshift=-0.2cm,xshift=-1cm,font=\normalsize};
    \MeasureXDistance{1E-3}{ASKthres}{PAMthres}{anchor=east,yshift=+0.2cm,xshift=-1cm,font=\normalsize};

  \end{axis}

\end{tikzpicture}%
\caption{FERs for SDD with PCM. The end-to-end rate is $\SI{4.2}{bpcu}$.}
\label{fig:fer_optical_noise_4km}
\end{figure}

\begin{figure}
\centering
\begin{tikzpicture}%

  \begin{axis}[%
      STDPLOT,
      width=1.1*\figwidth,
      height=1.3*\figheight,
      Rate_VS_PRX_NuN1_HALVE,
      ylabel={BER},
      xmajorgrids,
      ymajorgrids,
      yminorgrids,
      minor x tick num=3,
      yminorticks=true,
      grid=both,
      xmin=23,
      xmax=37,
      ymax=0.1,
      ymin=1E-16,
      ymode=log,
      extra y ticks={1e-2,1e-3,1e-4,1e-5,1e-6,1e-7,1e-8,1e-9,1e-10,1e-11,1e-12,1e-13,1e-14,1e-15,1e-16},     
      extra y tick labels={},
      legend style={at={(0,0)}, anchor=south west, font=\footnotesize}
    ]
    \addplot[ASK,densely dotted,mark=diamond*,mark options={solid,fill=white},mark size=1.2pt,name path global=BERoutASK] table [x=SNR, y=BERin, col sep=comma,OPTSNRreal, ]{Fig/r1_coarse_ASK_a=0.29_d=2_trial=6_20260116.txt};
    \addlegendentry{PCM};

    \addplot[ASK,line width=0.25mm,mark=diamond*,mark options={solid},mark size=1.2pt,name path global=BERoutASK] table [x=SNR, y=BERout, col sep=comma,OPTSNRreal]{Fig/r1_fine_ASK_a=0.29_d=2_trial=6_20260116.txt};
    
    \addlegendentry{Concatenated};

    \addplot[PAM,densely dotted,mark=*,mark size=1.0pt,mark options={solid,fill=white},name path global=BERoutPAM] table [x=SNR, y=BERin, col sep=comma,OPTSNRreal]{Fig/r1_coarse_PAM_a=0.7_d=4_trial=4_20260115.txt};
    \addlegendentry{PCM};

    \addplot[PAM,mark=*, mark size=1.0pt, mark options={solid},line width=0.25mm,name path global=BERoutPAM] table [x=SNR, y=BERout, col sep=comma,OPTSNRreal]{Fig/r1_fine_PAM_a=0.7_d=4_trial=4_20260115.txt};
    \addlegendentry{Concatenated};

    \addplot [ASK,draw=none,abovecurve={threshold}{mycolor6}{0.55ex}{0.12\columnwidth}] coordinates {(27,1E-3) (27,1E-18)}; %
    \addplot [ASK,dashed,name path global=ASKthres,] coordinates { 
      (27,0.003)
      (27,1E-18)
    };

    \addplot [PAM,draw=none,abovecurve={threshold}{mycolor1}{-1.8ex}{0.42\columnwidth}] coordinates {(32.4326,1E-3) (32.4326,1E-18)}; %
    \addplot [PAM,dashed,name path global=PAMthres] coordinates {
      (32.4326,0.003)
      (32.4326,1E-18)
     };

    \MeasureXDistance{2E-11}{BERoutASK}{BERoutPAM}{anchor=east,yshift=-0.2cm,xshift=-0.8cm,font=\normalsize};
    \MeasureXDistance{3.5E-11}{ASKthres}{PAMthres}{anchor=east,yshift=+0.2cm,xshift=-0.8cm,font=\normalsize};

  \end{axis}

\end{tikzpicture}%
\caption{BERs for SDD with PCM and an outer BCH-BCH product code. The end-to-end rate is $(1-0.069) \cdot \SI{4.2}{bpcu} \approx \SI{3.94}{bpcu}$.}
\label{fig:fer_optical_noise_4km_concatenated}
\end{figure}

Fig.~\ref{fig:fer_optical_noise_4km} shows that the SNR gap between the code curves is similar to that of the AIR thresholds. Bipolar 32-ASK-0.2 gains $\approx \SI{5.5}{dB}$ over unipolar 32-ASK-1 at FER $\SI{1e-3}{}$. The FER error floor is $\approx\SI{2e-4}{}$ for 32-ASK-0.2 and less than $\SI{1e-4}{}$ for 32-ASK-1. The FER error floor can be reduced by lowering the outlier threshold for repeated GVAMP attempts or increasing the number of attempts. Tuning these parameters is left for future work.

We remark that fiber applications require end-to-end error rates below $10^{-15}$. This is achieved by serially concatenating a soft-decision inner code with an interleaver and a hard-decision outer code, e.g., a Bose-Chaudhuri-Hocquenghem (BCH), product, or staircase code; see \cite{ITU-04,Mizuochi-09,Magarini-10,Smith-12}. 

Fig.~\ref{fig:fer_optical_noise_4km_concatenated} shows BERs for concatenated coded modulation with inner polar codes and the outer BCH-BCH product code in \cite[I.9]{ITU-04} with 6.9\% overhead. The end-to-end rate is reduced to \SI{3.94}{bpcu}. 
To reduce complexity, 32-ASK-0.2 uses the optimized damping and annealing parameters of Table~\ref{tab:mpib_optimized}, while 32-ASK-1 uses $n_\mathrm{W}=4$ with $\alpha_\text{anneal} = 0.7$. We allow up to $6$ GVAMP attempts per received block to lower the error floor. These are initiated based on the detection of outliers in the metrics \eqref{eq:variational_bayes}, as described above. The number of required GVAMP iterations for 32-ASK-0.2 is $34$, while 32-ASK-1 requires $49$ iterations. The slightly increased GVAMP complexity for 32-ASK-1 helps avoid larger rate losses. 

\subsection{EXIT Charts}
\label{subsec:EXIT-Charts}
Fig.~\ref{fig:64-ASK-0.2_exit_trajectory} plots an EXIT chart for the functions $T_1$ and $T_2$ in \eqref{eq:variance_tfs} and SDD. The chart uses the same parameters as in Fig.~\ref{subfig:TRACE_PRE_I}, except for a slightly larger block length of $n=4096$. The dashed and solid red curves show the $T_2$ function with and without annealing, respectively. During annealing, the LMMSE denoiser overestimates the noise. We iterate between the modules in Fig. \ref{fig:variance_exit}, using the means from \eqref{eq:awgn_perturbed_extrinsics_W} and \eqref{eq:awgn_perturbed_extrinsics_U}, and estimate the noise power via Monte Carlo simulation. We approximate the EXIT functions by taking the convex hull of the decoding trajectory. When annealing is disabled, the EXIT functions are computed as described in Sec.~\ref{subsec:vEXIT-functions}.

The figure also shows the GVAMP trajectories of two transmit blocks, i.e., the MSEs $\nu_{W_i} = \frac{1}{m}\lVert \mathbf{p}_i^{(n_\text{it})} - \mathbf{w}\rVert^2$ for $i=1,2$ at each iteration.
Both trajectories closely follow the EXIT function corridor. We relate the $\nu_{W_2}$ component of the EXIT chart fixed point (the red dot) to mutual information using \eqref{eq:ext_simpl_nuU1} and $\mathbf{R}_1$ from \eqref{eq:awgn_perturbed_extrinsics_U}. The predicted rate of the EXIT analysis differs from the actual GVAMP rate by less than \SI{0.1}{bpcu}. For higher SIC stages, e.g., $\ell=2,3,4$, the EXIT function corridor widens, resulting in faster convergence.

Note that EXIT charts can be used to tune GVAMP's convergence. For example, one can shape the $T_1$ function by optimizing the transmit symbol constellation.
\begin{figure}
    \centering
    \input{Fig/VTC_final}
    \caption{EXIT chart for SDD ($\ell=1$), $64$-ASK-$0.2$, $L_\text{fib}=\SI{4}{\kilo\meter}$, $B=\SI{300}{\giga Bd}$, $\text{SNR}=\SI{33.5}{dB}$ and two  blocks with each $n = 4096$.}
    \label{fig:64-ASK-0.2_exit_trajectory}
\end{figure}

\subsection{Constellation Shaping}
\label{subsec:shaping}
The surrogate channel \eqref{eq:awgn_perturbed_extrinsics_U} motivates using constellation shaping \cite{bocherer2015bandwidthefficient}. Let $\mathcal{M}_0$ be the bipolar ASK alphabet \eqref{eq:alphabet_A} with $o=0$. The iid channel inputs are $X_\kappa = c \cdot (X_0 + o)$, where $X_0$ is uniformly distributed over $\mathcal M_0$ and $c$ is the constellation scaling. Define the shaped iid inputs
\begin{align}
    X_{\kappa,\text{sh}} &= c \cdot (\sqrt{b} X_0' + o)
\end{align}
where
\begin{align}
    P_{X_0'}(x) = c' \cdot \exp{(-\nu x^2)},
    \quad x \in \mathcal{M}_0
\end{align}
and $c'$ normalizes to a PMF. Let $\var{(X)}$ be the variance of $X$. We choose $b =\var{(X_0)} / \var(X_0')$ so the shaped $X_{\kappa,\text{sh}}$ and unshaped $X_\kappa$ have the same mean and average power.

\subsubsection{Optically amplified links}
Fig.~\ref{fig:shaped_airs} compares the AIRs, which were optimized with a grid search over the pairs $(o,\nu)$. We found that $o=0.25$ is near-optimal across the depicted SNR range. Shaping with $\nu\approx 3.3$ reduces the SNR gap to capacity to roughly \SI{1.55}{dB} at \SI{4.2}{bpcu}, corresponding to a rate loss of  $\approx \SI{0.25}{bpcu}$. The gap to the equiprobable-ASK capacity is less than \SI{0.1}{bpcu}. The remaining gaps are likely because the offset $o$ wastes power and $\mA$ has structure. Shaping gains $\approx \SI{1}{bpcu}$ over equiprobable 64-ASK-1.

We also compare to the capacity with intensity modulation (IM) and symbol-rate sampling \cite{mecozzi_2001_im_capacity,lapidoth2002phase}. Consider sinc pulses and zero fiber dispersion. Processing the $n$ baud-rate samples $[Y_1, Y_3, \dots, Y_{m-1}]$ in \eqref{eq:yvec} gives the \textit{asymptotic} capacity approximation~\cite[Eq.~(23)]{lapidoth2002phase}
\begin{align}
    C^\infty_{\text{IM/DD,opt}} = \frac{1}{2}
    \log_2\left(1 + \frac{\Prx}{2 B \nu_1} \right)
\end{align}
which is accurate at high SNR~\cite[Sec.~IV-B]{katz2004noncoherent}. The curve $R_\text{BA}$ shows a capacity lower bound~\cite{keang2005exact,katz2004noncoherent,keykhosravi2020whentouse} computed by the Blahut-Arimoto algorithm~\cite{blahut1972computation,arimoto1972algorithm,varnica2002capacity}. Shaping gains $\approx \SI{4.2}{dB}$ over $R_\text{BA}$ near $\SI{4.2}{bpcu}$.

\begin{figure}
    \centering
    \pgfkeys{
  /pgf/number format/.cd,
  precision=2,
}

\pgfdeclarelayer{background}
\pgfdeclarelayer{foreground}
\pgfsetlayers{background,main,foreground}

\begin{tikzpicture}%

  \begin{axis}[%
      STDPLOT,
      yminorticks=true,
      xmajorgrids,
      ymajorgrids,
      yminorgrids,
      minor x tick num=3,
      minor y tick num=3,
      ytick={0,1,2,3,4,5,6},
      grid=both,
      legend style={legend cell align=left,  draw=white!15!black, %
        legend pos=north west
      },
      xlabel style={font=\color{white!15!black}},
      ylabel style={font=\color{white!15!black}},
      axis background/.style={fill=white},
      scale only axis,
      width=1.01*\figwidth,
      height=1.1*\figheight,
      xmin=21,
      xmax=33.1,
      Rate_VS_PRX_NuN1_HALVE,
      ymin=3.5,
      ymax=5,
      minor y tick num=4,
      legend cell align=left,
      legend style={at={(0,1)},anchor=north west}
      ,
    ]

    \addplot[draw=none,line width = 1.2, black,domain=-10:40,name path=Cr,forget plot] (x,{1/2*log2(1 + 10^(x/10))});

    \addplot[line width = 0.7, black,domain=20:32,decoration={
        text align={left, left indent=1.7cm},
        text along path,
        raise=2pt,
        text={|\small|Coherent Capacity},
      },
      mark options={},
    postaction={decorate}] (x,{1/2*log2(1 + 10^(x/10))});
    \addlegendentry{$C_{\mathbb{R},\text{coh}}$};

    \addplot[line width = 0.6, black,densely dotted,name path global=Rr,] table [x=SNR, y=IqYX, col sep=comma,x expr= \thisrow{SNR}]{Fig/256-ASK-AWGN.txt};
    \addlegendentry{$R_\text{ASK}$};

    \addplot[mark=star, mark size=3pt, mark repeat=2,mark phase=2, name path global=BA,colG] table [x=SNR, y=Imax, col sep=comma,x expr= \thisrow{SNR},x expr={\thisrow{SNR}+3}]{Fig/Blahut_Arimoto_IMDD_opt.txt};
    \addlegendentry{$R_{\text{BA}}$};

    \addplot[domain=-10:40, samples=51, name path global=symbolspaced, colG, mark=square*,mark options={solid}] (x,{1/2*log2(1+ 1/4*10^(x/10))}); 
    \addlegendentry{$C^\infty_{\text{IM/DD},\text{opt}}$};

    \addplot[QAM,colC!80!black,mark=diamond*,mark size=1.8pt,mark options={solid,fill=colC}, solid,name path global=MB,restrict x to domain=20:32] table [x=SNRdB, y=Maxrate, col sep=comma,x expr= \thisrow{SNRdB}]{Fig/r1_64-MB-nu_shaped.txt};
    \addlegendentry{Shaped};

    \pgfplotstableread[col sep=comma]{Fig/r1_64-MB-nu_shaped.txt}\datatable

    \pgfplotstablecreatecol[
      create col/assign/.code={
        \pgfmathparse{\thisrow{nu}}
        \pgfmathprintnumberto[fixed, precision=1]{\pgfmathresult}\roundedB
        \edef\temp{\roundedB}
        \pgfkeyslet{/pgfplots/table/create col/next content}\temp
      }
    ]{mergedlabel}{\datatable}

    \addplot[
      only marks,
      forget plot,
      mark=none,
      nodes near coords,
      point meta=explicit symbolic,
      node near coord style=
      {font=\footnotesize, yshift=-0.2cm, xshift=-0.1cm, color=colC!50!black, anchor=center, inner sep=0,
      opacitylabel,}   %
    ] table [x=SNRdB, y=Maxrate, x expr= \thisrow{SNRdB}, col sep=comma, meta=mergedlabel] {\datatable};

    \addplot[ASK,mark=*,mark size=1.5, name path global=64-ASK-0.2,] table [x=SNR, y=IqYX, col sep=comma,OPTSNRreal]{plots/vamp_v11_r1/PRE/r1,64-ASK-0.2,S=4,C=0,Rb=1,n1=1,n2=0,n1m=0,n2m=0,P=cO,ps=RRC,a=0.01,Nsp=250,Npi=0,n=2048,L=4,Rs=300,Nr=1,Nb=128,I=250.txt};
    \addlegendentry{64-ASK-0.2};

    \addplot[PAM,mark=triangle*,name path global=64-ASK-1] table [x=SNR, y=IqYX, col sep=comma,OPTSNRreal]{plots/vamp_v11_r1/PRE/r1,64-ASK-1.0,S=4,C=0,Rb=1,n1=1,n2=0,n1m=0,n2m=0,P=cO,ps=RRC,a=0.01,Nsp=250,Npi=0,n=2048,L=4,Rs=300,Nr=1,Nb=128,I=250.txt};
    \addlegendentry{64-ASK-1};

    \MeasureYDistancePrecise{4.2+0.255}{Cr}{MB}{anchor=west,yshift=0.8cm,xshift=-0.8cm,font=\normalsize};

    \MeasureXDistancePrecise{4.2}{Cr}{MB}{anchor=east,yshift=0cm,xshift=-1.1cm,font=\normalsize};
    \MeasureXDistancePrecise{4.2}{MB}{64-ASK-0.2}{anchor=west,yshift=-0.25cm,xshift=-1cm,font=\normalsize};
    \MeasureXDistancePrecise{4.2}{64-ASK-0.2}{BA}{anchor=west,yshift=0cm,xshift=-0.13cm,font=\normalsize};

  \end{axis}

\end{tikzpicture}%
    \caption{AIRs for optically amplified links, $M=64$ from Fig.~\ref{fig:vamp_opt_rates}, and with shaping. The optimized shaping has $o = 0.25$ and values $\nu$ shown below the diamond markers. The curve $R_\text{ASK}$ shows the equiprobable-ASK capacity with a coherent receiver \cite[Fig.~1]{forney1998modulation}.
    }
    \label{fig:shaped_airs}
\end{figure}

\subsubsection{Unamplified links}
Fig.~\ref{fig:shaped_airs_el_noise} compares the AIRs, which were again optimized with a grid search over the $(o,\nu)$. The offset $o=0.25$ was again near-optimal. Shaping with $\nu\approx 3.1$ improves the SNR by roughly $\SI{0.4}{dB}$ over uniform 64-ASK-0.2, and shaping gains $\SI{1}{bpcu}$ over equiprobable 64-ASK-1.

We again compare to the IM/DD capacity with symbol-rate sampling, sinc pulses, and zero fiber dispersion. Processing the samples $[Y_1, Y_3, \dots, Y_{m-1}]$ in \eqref{eq:yvec} gives the capacity lower bound~\cite[Eq.~(26)]{lapidoth2009capacity}
\begin{align}
    R_{\text{IM/DD,el}} = \frac{1}{2}
    \log_2\left( 1 + \frac{\e}{2\pi} \frac{\Prx^2}{\nu_{N_2}} \right)
\end{align}
which is accurate at high SNR~\cite[Fig.~5]{lapidoth2009capacity}. The curve $R_\text{BA}$ again shows a capacity lower bound computed by the Blahut-Arimoto algorithm; see~\cite[Fig.~2]{keykhosravi2020whentouse}. Shaping gains $\approx \SI{2.6}{dB}$ over $R_\text{BA}$ near $\SI{4.2}{bpcu}$. We remark that shaping can be optimized for each SIC stage.

\begin{figure}
    \centering
    \pgfkeys{
  /pgf/number format/.cd,
  precision=2,
}

\pgfdeclarelayer{background}
\pgfdeclarelayer{foreground}
\pgfsetlayers{background,main,foreground}

\begin{tikzpicture}%

  \begin{axis}[%
      STDPLOT,
      yminorticks=true,
      width=1.01*\figwidth,
      height=1.1*\figheight,
      xmajorgrids,
      ymajorgrids,
      yminorgrids,
      minor x tick num=4,
      minor y tick num=4,
      ytick={0,1,2,3,4,5,6},
      grid=both,
      legend style={legend cell align=left,  draw=white!15!black, %
        legend pos=north west
      },
      xlabel style={font=\color{white!15!black}},
      ylabel style={font=\color{white!15!black}},
      axis background/.style={fill=white},
      scale only axis,
      xmin=9.0,
      xmax=15.5,
      Rate_VS_PRX_NuN2,
      ymin=3.5,
      ymax=5.001,
      minor y tick num=4,
      legend cell align=left,
      legend style={at={(0,1)},anchor=north west},
    ]

    \addplot[mark=star, mark size=3pt, name path global=BA,colG] table [x=SNR, y=Imax, col sep=comma,x expr= \thisrow{SNR}]{Fig/Blahut_Arimoto_IMDD_el.txt};
    \addlegendentry{$R_\text{BA}$};

    \addplot[colG,mark=square*,domain=-10:40, samples=51, name path global=symbolspaced, mark options={solid}] (x,{1/2*log2(1 + e/(2*pi)*(10^(x/10))^2 )});
    \addlegendentry{$R_{\text{IM/DD},\text{el}}$};

    \addplot[QAM,colC!80!black,mark=diamond*,mark size=1.8pt,mark options={solid,fill=colC}, solid,name path global=MB,] table [x=SNRdB, y=Maxrate, col sep=comma,x expr= \thisrow{SNRdB}]{Fig/r1_64-MB-nu_shaped_elect_dacfilt_ON.txt};
    \addlegendentry{Shaped};

    \addplot[ASK,mark=*,name path global=64-ASK-0.2] table [x=SNR, y=IqYX, col sep=comma,x expr= \thisrow{SNR}]{plots/vamp_v11_r1/POST/r1,64-ASK-0.2,S=4,C=0,Rb=0,n1=0,n2=1,n1m=0,n2m=1,P=cO,ps=RRC,a=0.01,Nsp=250,Npi=0,n=2048,L=4,Rs=300,Nr=1,Nb=128,I=250.txt};
    \addlegendentry{64-ASK-0.2};

    \addplot[PAM,mark=triangle*,name path global=64-ASK-1] table [x=SNR, y=IqYX, col sep=comma,x expr= \thisrow{SNR}]{plots/vamp_v11_r1/POST/r1,64-ASK-1.0,S=4,C=0,Rb=0,n1=0,n2=1,n1m=0,n2m=1,P=cO,ps=RRC,a=0.01,Nsp=250,Npi=0,n=2048,L=4,Rs=300,Nr=1,Nb=128,I=250.txt};
    \addlegendentry{64-ASK-1};

    \pgfplotstableread[col sep=comma]{Fig/r1_64-MB-nu_shaped_elect_dacfilt_ON.txt}\datatable

    \pgfplotstablecreatecol[
      create col/assign/.code={
        \pgfmathparse{\thisrow{nu}}
        \pgfmathprintnumberto[fixed, precision=1]{\pgfmathresult}\roundedB
        \edef\temp{\roundedB}
        \pgfkeyslet{/pgfplots/table/create col/next content}\temp
      }
    ]{mergedlabel}{\datatable}

    \addplot[
      only marks,
      forget plot,
      mark=none,
      nodes near coords,
      point meta=explicit symbolic,
      domain=9:14,
      node near coord style=
      {font=\footnotesize, yshift=-0.05cm, xshift=-0.35cm, color=colC!50!black, anchor=center, inner sep=0,
      opacitylabel,}   %
    ] table [x=SNRdB, y=Maxrate, x expr= \thisrow{SNRdB}, col sep=comma, meta=mergedlabel] {\datatable};

    \MeasureXDistancePrecise{4.2}{MB}{64-ASK-0.2}{anchor=east,xshift=-0.8cm,font=\normalsize};
    \MeasureXDistancePrecise{4.2}{64-ASK-0.2}{BA}{anchor=west,xshift=-0.12cm,font=\normalsize};

  \end{axis}

\end{tikzpicture}%
    \caption{AIRs for unamplified links, $M=64$ from Fig.~\ref{fig:vamp_el_rates}, and with shaping. The optimized shaping has $o = 0.25$ and values $\nu$ shown beside the diamond markers.}
    \label{fig:shaped_airs_el_noise}
\end{figure}

\section{Conclusions}
\label{sec:conclusions}
We studied EP-based receivers for real-valued modulation in bandlimited channels with DD. The receivers operate within \SI{0.26}{bpcu} of the coherent capacity for optically amplified SSMF links of \SI{2}{\kilo\meter}-\SI{10}{\kilo\meter} at \SI{300}{GBd} in the C-band. The rates are achieved by ASK with a small offset, resulting in significant power gains of $\approx\SI{6}{dB}$ and $\approx\SI{3}{dB}$ over unipolar ASK with and without optical amplification, respectively. The receiver complexity is $\mathcal{O}(n_\text{it} \log_2 n + M)$ multiplications per symbol, with only a few tens of iterations typically needed. For example, SDD requires \SI{93}{mpib} to achieve \SI{5}{bpcu} at \SI{300}{GBd} (giving \SI{1.5}{Tbit/s}) over \SI{4}{\kilo\meter} of SSMF in the C-band.

There are many directions for future work. For example, one may optimize symbol constellations using EXIT charts. One may modify the channel matrix, e.g., via precoding and other means, under realistic hardware constraints \cite{ma_toward_2024,liu2025random}. One may also study low-cost ADCs with smaller sampling rates; see Sec.~\ref{sec:practical_tx_excess_bw}. Another idea is to improve the EP updates in Appendix~\ref{appendix-EP} and EXIT surrogates \eqref{eq:awgn_perturbed_extrinsics_WU} by including correlations across the EP messages. Finally, it is interesting to evaluate how closely the GVAMP detector approaches the performance predicted by the replica method.

\section*{Acknowledgment}
\noindent The authors wish to thank the reviewers for their comments, which improved the manuscript.

\begin{appendices}
\renewcommand{\thesectiondis}[2]{\Alph{section}:}
\section{EP Messages}
\label{appendix-EP}
This Appendix describes how the functions $\widetilde{t}_1,\widetilde{t}_2,\widetilde{t}_3$ in \eqref{eq:post_approx_fact} are updated. First, choose some $\mathbf{r}_1$, $\mathbf{p}_1$ and $\nu_{U_1},\nu_{W_1}$ and initialize  
\begin{align}
 &\widetilde{t}_{21} = \mathcal{N}(\mathbf{r}_1, \nu_{U_1} \mathbf{I}), &&
 \widetilde{t}_{23} =
 \mathcal{CN}(\mathbf{p}_1, \nu_{W_1} \mathbf{I}).
 \label{eq:ttilde_21_23}
\end{align}
\subsection{Update \texorpdfstring{$\widetilde{t}_1$}{t1} and \texorpdfstring{$\widetilde{t}_3$}{t3}}
In each iteration, EP computes
\begin{subequations}
\begin{align}
    \widetilde{t}_1  &= \Proj\left[\, t_1\, \widetilde{t}_{21} \right] / \widetilde{t}_{21}
    \label{eq:projection_ep_I_II-1}
    \\
    \widetilde{t}_3  &=    \CProj\left[\, t_3\,\widetilde{t}_{23} \right] / \widetilde{t}_{23}.
    \label{eq:projection_ep_I_II-2}
\end{align}
\end{subequations}
The projection in \eqref{eq:projection_ep_I_II-1} gives $\mathcal{N}(\vdIh, \alpha_1 \mathbf{I})$ where $\vdIh$ has entries $\hat{u}_{1,\kappa}$ given in \eqref{eq:x_den_mean} of Table~\ref{tab:explicit_denoisers}, and
\begin{align}
    \alpha_1 & = \frac{1}{n'} \sum\nolimits_{\kappa=1}^{n'} \alpha_{1,\kappa}
    \label{eq:alpha1_avg}
\end{align}
where the $\alpha_{1,\kappa}$ are given in \eqref{eq:x_den_var} of Table~\ref{tab:explicit_denoisers}. Similarly, the projection in \eqref{eq:projection_ep_I_II-2} gives 
$\mathcal{CN}(\zIh,\beta_1)$ where $\zIh$ has entries $\hat{w}_{1,k}$ given by \eqref{eq:z_den_mean_final} in Appendix~\ref{appendix-denoising-w1}, and
\begin{align}
    \beta_1 & = \frac{1}{m}\sum\nolimits_{k=1}^m \beta_{1,k}
\end{align}
where $\beta_{1,k}$ is given by \eqref{eq:z_den_var_final} in Appendix~\ref{appendix-denoising-w1}. We thus have
\begin{subequations}
\begin{align}
    \widetilde{t}_1 %
    & = \frac{\mathcal{N}(\vdIh,\alpha_1)}{ \mathcal{N}(\mathbf{r}_1, \nu_{U_1} \mathbf{I})} \propto \mathcal{N}(\mathbf{r}_2, \nu_{U_2} \mathbf{I})
    \label{eq:extrinsic_u1_updated} \\[0.1cm]
    \widetilde{t}_3 %
    & = \frac{\mathcal{CN}(\zIh, {\beta}_1)}{ \mathcal{CN}(\mathbf{p}_1, \nu_{W_1} \mathbf{I})}
    \propto \mathcal{CN}(\mathbf{p}_2, \nu_{W_2}\mathbf{I})
    \label{eq:extrinsic_z1_updated}
\end{align}
\end{subequations}
where
\begin{subequations}
\begin{align}
    \mathbf{r}_2 &= \displaystyle\frac{\nu_{U_1} \vdIh - {\alpha}_1  \mathbf{r}_1}{\nu_{U_1} - {\alpha}_1}, &&
    \nu_{U_2} = \displaystyle\frac{\nu_{U_1} {\alpha}_1}{\nu_{U_1}- {\alpha}_1}, 
    \label{eq:ext_input}
    \\[0.4cm]
    \mathbf{p}_2 &= \displaystyle\frac{\nu_{W_1}\hat{\z}_{1}  - {\beta}_1 \mathbf{p}_1}{\nu_{W_1}- {\beta}_1}, &&
    \nu_{W_2} = \displaystyle\frac{{\beta}_1 \nu_{W_1}}{\nu_{W_1}-{\beta}_1}.    
    \label{eq:ext_output}
\end{align}
\end{subequations}

We remark that the projection in \eqref{eq:projection_ep_I_II-1} can be based on the surrogate model 
\begin{align}
    \mathbf{R} = \mathbf{U} + \mathbf{N}
    \label{eq:surrogate_r_u}
\end{align}
with prior $\mathbf{U} \sim P_\mathbf{U}$, and independent $\mathbf{N} \sim \mathcal{N}(\mnull{}, \nu_{U_1}\mident{})$, and approximating the PMF $P(\mathbf{U}|\mathbf{R} = \mathbf{r}_1)$ by 
$\mathcal{N}(\vdIh, {\alpha}_1 \mathbf{I})$, where $\vdIh$ is the conditional mean and ${\alpha}_1$ is the average conditional variance. 
Similarly, the complex projection in \eqref{eq:projection_ep_I_II-2} is based on the surrogate model
\begin{align}
    \mathbf{Y} = |\mathbf{W}|^2 + \mathbf{N}_2 %
    \label{eq:surrogate_y_w}
\end{align}
with $\mathbf{W} \sim \mathcal{CN}(\mathbf{p}_1, \nu_{W_1}\mident{})$ and noise $\mathbf{N}_2$ in \eqref{eq:discrete_n1_n2}, and results in $\mathcal{CN}(\zIh, {\beta}_1)$. The projections \eqref{eq:projection_ep_I_II-1} and \eqref{eq:projection_ep_I_II-2} perform MMSE estimation for the input model \eqref{eq:surrogate_r_u} and output model \eqref{eq:surrogate_y_w}, respectively. 

\subsection{Update \texorpdfstring{$\widetilde{t}_2$}{t2}}
Similar to the belief propagation example in Sec.~\ref{sec:EP}, calculate the vector update:
\begin{align}
    \widetilde{t}_2  
    &= (\Proj \times \CProj)\!\left[\, t_2 (\widetilde{t}_1 \widetilde{t}_3) \right] \,/\, \big(\widetilde{t}_1 \widetilde{t}_3\big)
    \label{eq:projection_ep_III}
\end{align}
where the projection $(\Proj \times \CProj)$ returns the Gaussian product
\begin{align}
    q_2(\mathbf{u},\mathbf{w}) = 
    \mathcal{N}(\mathbf{u}; \vdIIh,  \alpha_2 \mathbf{I})
    \cdot
    \mathcal{CN}(\mathbf{w}; \zIIh,  \beta_2 \mathbf{I})
    \label{eq:q_2_EP}
\end{align} that minimizes
$D( t_2 (\widetilde{t}_1 \widetilde{t}_3) / c \,||\, q_2)$
where $c$ normalizes to a density. Using \eqref{eq:factor_c_approx}, we obtain the EP functions
\begin{subequations}
\begin{align}
    \widetilde{t}_{21}(\mathbf{u})  & = \frac{\mathcal{N}(\vdIIh,  \alpha_2 \mathbf{I})}{\mathcal{N}(\mathbf{r}_2, \nu_{U_2} \mathbf{I})}
    \propto \mathcal{N}(\mathbf{r}_1, \nu_{U_1} \mathbf{I})
    \\[0.1cm]
    \widetilde{t}_{23}(\mathbf{w}) & = \frac{\mathcal{CN}(\zIIh,  \beta_2 \mathbf{I})}{\mathcal{CN}(\mathbf{p}_2, \nu_{W_2}\mathbf{I})}
    \propto \mathcal{CN}(\mathbf{p}_1, \nu_{W_1} \mathbf{I})
\end{align}
\end{subequations}
where
\begin{subequations}
\begin{align}
    \mathbf{r}_1 &= \displaystyle\frac{\nu_{U_2}\vdIIh - \alpha_2 \mathbf{r}_2}{\nu_{U_2}-\alpha_2}, &&
    \nu_{U_1} =  \displaystyle\frac{\nu_{U_2} {\alpha}_2}{\nu_{U_2}- {\alpha}_2},
    \label{eq:ext_lmmse_u}
     \\[0.4cm]
    \mathbf{p}_1 &= \displaystyle\frac{\nu_{W_2}\hat{\z}_{2}  - {\beta}_2 \mathbf{p}_2}{\nu_{W_2}- {\beta}_2}, &&
    \nu_{W_1} = \displaystyle\frac{{\beta}_2 \nu_{W_2}}{\nu_{W_2}-{\beta}_2}.
    \label{eq:ext_lmmse_w}
\end{align}
\end{subequations}

It remains to solve \eqref{eq:projection_ep_III}. Since \eqref{eq:q_2_EP} is separable in $\mathbf{u}$ and $\mathbf{w}$, the projection \eqref{eq:projection_ep_III} applies Gaussian projections to the marginals of $t_2 (\widetilde{t}_1 \widetilde{t}_3)$:
\begin{subequations}
\begin{align}
    \mathcal{N}(\mathbf{u}; \vdIIh,  \alpha_2 \mathbf{I}) &= \Proj\left[\,  \medint\int   t_2(\mathbf{u},\mathbf{w}) \,\widetilde{t}_1(\mathbf{u}) \, \widetilde{t}_3(\mathbf{w}) \,\mathrm{d} \mathbf{w} \,\right] 
    \label{eq:projection_ep_III_1}
    \\
    \mathcal{CN}(\mathbf{w}; \zIIh,  \beta_2 \mathbf{I}) &= \CProj\left[\,  \medint\int  t_2(\mathbf{u},\mathbf{w}) \, \widetilde{t}_1(\mathbf{u}) \, \widetilde{t}_3(\mathbf{w})  \,\mathrm{d} \mathbf{u} \,\right]
    \label{eq:projection_ep_III_2}
    \!.
\end{align}
\end{subequations}
These LMMSE estimates are described in Appendix~\ref{appendix-LMMSE-denoiser}, and Table~\ref{tab:explicit_denoisers} summarizes the results. Finally, to derive the EP updates for complex-valued inputs $\mathbf{U}$, replace $\mathcal{P}$ with $\mathcal{CP}$ in \eqref{eq:projection_ep_I_II-1} and \eqref{eq:projection_ep_III}; this requires minor modifications in Table~\ref{tab:explicit_denoisers}. We remark that one can use more sophisticated approximations in \eqref{eq:min_div_proj}, e.g., diagonal or full covariance matrices.

\renewcommand{\thesectiondis}[2]{\Alph{section}:}
\section{LMMSE Denoiser}
\label{appendix-LMMSE-denoiser}
We study Gaussian models of the form
\begin{align}
    \mathbf{Y} = \mathbf{A} \mathbf{X} + \mathbf{N} + \mathbf{s}
\end{align}
where $\mathbf{A}$ is a complex-valued matrix, $\mathbf{s}$ is a vector, $\mathbf{X}$ and $\mathbf{N}$ are statistically independent, and $\mathbf{N} \sim \mathcal{CN}(\mnull{},\mathbf{\Sigma}_\mathbf{N})$. 

\subsection{CSCG-\texorpdfstring{$\boldsymbol{\mu}$}{μ} Inputs}
\label{subsec:LMMSE-Estimation_cscg}
If $\mathbf{X} \sim \mathcal{CN}(\boldsymbol{\mu}_\mathbf{X}, \mathbf{\Sigma}_\mathbf{X})$ then $\mathbf{X}$, $\mathbf{N}$, $\mathbf{Y}$ are jointly Gaussian and
$\mathbf{Y} \sim \mathcal{CN}(\boldsymbol{\mu}_\mathbf{Y}, \mathbf{\Sigma}_\mathbf{Y})$
where
\begin{align}
    \boldsymbol{\mu}_\mathbf{Y} &= \mathbf{A} \boldsymbol{\mu}_\mathbf{X} + \mathbf{s}, &&
    \mathbf{\Sigma}_\mathbf{Y} = \mathbf{A} \mathbf{\Sigma}_\mathbf{X} \mathbf{A}\herm + \mathbf{\Sigma}_\mathbf{N}.
\end{align}
Next, define the zero-mean $\mathbf{X}_0=\mathbf{X}-\boldsymbol{\mu}_\mathbf{X}$ and $\mathbf{Y}_0=\mathbf{Y}-\boldsymbol{\mu}_\mathbf{Y}$. The conditional expectation 
\begin{align}
    \E[\mathbf{X}_0 | \mathbf{Y}_0]
    & = \E[ \mathbf{X}_0 \mathbf{Y}_0\herm ]\, 
    \mathbf{\Sigma}_\mathbf{Y}^{-1}\, \mathbf{Y}_0 %
    = \mathbf{\Sigma}_\mathbf{X} \mathbf{A}\herm \mathbf{\Sigma}_\mathbf{Y}^{-1}\, \mathbf{Y}_0
\end{align}
is the LMMSE estimate of $\mathbf{X}_0$ given $\mathbf{Y}_0$; see~\cite[Sec.~10.6]{kay1993fundamentals}. This is also the MMSE estimate for Gaussian vectors. Let $\mathbf{X}|\mathbf{y}$ be the random vector $\mathbf{X}$ conditioned on the event $\mathbf{Y}=\mathbf{y}$ and suppose $\mathbf{\Sigma}_\mathbf{X}$ and $\mathbf{\Sigma}_\mathbf{N}$ are invertible. We compute
\begin{align}
    \mathbf{X}|\mathbf{y} \sim
    \mathcal{CN}(\E[\mathbf{X}|\mathbf{y}],  \mathbf{\Sigma}_{\mathbf{X}|\mathbf{y}})
\end{align}
where
\begin{subequations}
\begin{align}
     \E[\mathbf{X}|\mathbf{y}] & = 
     \E[\mathbf{X}_0 | \mathbf{Y}_0 = \mathbf{y} - \boldsymbol{\mu}_\mathbf{Y}] + \boldsymbol{\mu}_\mathbf{X} \nonumber \\
     & = \mathbf{\Sigma}_\mathbf{X} \mathbf{A}\herm \mathbf{\Sigma}_\mathbf{Y}^{-1}\,
     (\mathbf{y} - \boldsymbol{\mu}_\mathbf{Y}) + \boldsymbol{\mu}_\mathbf{X}
     \label{eq:nota_cond_mean_alt} \\
     & = \mathbf{\Sigma}_{\mathbf{X}|\mathbf{y}} \big( \mathbf{A}\herm \mathbf{\Sigma}_\mathbf{N}^{-1} (\mathbf{y}-\mathbf{s}) + \mathbf{\Sigma}_\mathbf{X}^{-1} \boldsymbol{\mu}_\mathbf{X}  \big) 
     \label{eq:nota_cond_mean}
\end{align}
\end{subequations}
and
\begin{subequations}
\begin{align}  \mathbf{\Sigma}_{\mathbf{X}|\mathbf{y}}
     & = \mathbf{\Sigma}_\mathbf{X} - \mathbf{\Sigma}_\mathbf{X} \mathbf{A}\herm \mathbf{\Sigma}_\mathbf{Y}^{-1} \mathbf{A} \mathbf{\Sigma}_\mathbf{X} 
     \label{eq:nota_cond_var_alt}
     \\
     & = \big( \mathbf{A}\herm \mathbf{\Sigma}_\mathbf{N}^{-1} \mathbf{A} + \mathbf{\Sigma}_\mathbf{X}^{-1} \big)^{-1} .
    \label{eq:nota_cond_var}
\end{align}
\end{subequations}
The steps \eqref{eq:nota_cond_mean} and \eqref{eq:nota_cond_var} use the matrix inversion lemma
\begin{align}
    & (\mathbf{A} + \mathbf{B} \mathbf{C} \mathbf{D})^{-1} \IfTwoCol{\nonumber \\}
    \IfTwoCol{&} = \mathbf{A}^{-1} - \mathbf{A}^{-1} \mathbf{B}
    \big( \mathbf{C}^{-1} + \mathbf{D} \mathbf{A}^{-1} \mathbf{B} \big)^{-1}
    \mathbf{D} \mathbf{A}^{-1} .
    \label{eq:matrix-inversion-lemma}
\end{align}
Note that $\mathbf{\Sigma}_{\mathbf{X}|\mathbf{y}}$ does not depend on $\mathbf{y}$.

\subsection{Real Gaussian Inputs}
\label{subsec:LMMSE-Estimation_real}
Suppose $\mathbf{X} \sim \mathcal{N}(\boldsymbol{\mu}_\mathbf{X}, \mathbf{\Sigma}_\mathbf{X})$ and construct the composite-real Gaussian model: 
\begin{align}
    \bar{\mathbf{Y}}
    = \bar{\mathbf{A}} \mathbf{X} + \bar{\mathbf{N}} + \bar{\mathbf{s}}
    \label{eq:composite_system_lmmse}
    .
\end{align}
where overbars denote stacking, e.g., $\bar{\mathbf{Y}} = [\Re\{\mathbf{Y}\}\tran, \Im\{\mathbf{Y}\}\tran]\tran$ and $\bar{\mathbf{A}} = [\Re\{\mathbf{A}\}\tran, \Im\{\mathbf{A}\}\tran]\tran$.
The vectors $\mathbf{X}$, $\bar{\mathbf{N}}$ and $\bar{\mathbf{Y}}$ are jointly Gaussian and $\bar{\mathbf{N}}\sim \mathcal{N}(\mnull{}, \mathbf{\Sigma}_{\bar{\mathbf{N}}})$ with $\mathbf{\Sigma}_{\bar{\mathbf{N}}} = \frac{1}{2}(\mident{2} \otimes \mathbf{\Sigma}_\mathbf{N})$.
We may directly use \eqref{eq:nota_cond_mean} and \eqref{eq:nota_cond_var} to obtain 
\begin{align}
      \E[\mathbf{X}|\bar{\mathbf{y}}] 
     & = \mathbf{\Sigma}_{\mathbf{X}|\bar{\mathbf{y}}} \big(\bar{\mathbf{A}}\tran \mathbf{\Sigma}_{\bar{\mathbf{N}}}^{-1} (\bar{\mathbf{y}}-\bar{\mathbf{s}}) + \mathbf{\Sigma}_\mathbf{X}^{-1} \boldsymbol{\mu}_\mathbf{X}  \big)  \nonumber \\
     &= \mathbf{\Sigma}_{\mathbf{X}|\bar{\mathbf{y}}} \big( \Re\big\{\mathbf{A}\herm \left(\mathbf{\Sigma}_{\mathbf{N}} \big/ 2\right)^{-1}   (\mathbf{y}-\mathbf{s})\big\} + \mathbf{\Sigma}_\mathbf{X}^{-1} \boldsymbol{\mu}_\mathbf{X}  \big) 
     \label{eq:nota_cond_mean_realX}
\end{align}
and 
\begin{align}  \mathbf{\Sigma}_{\mathbf{X}|\bar{\mathbf{y}}}
     & = \big( \bar{\mathbf{A}}\tran \mathbf{\Sigma}_{\bar{\mathbf{N}}}^{-1}  \bar{\mathbf{A}} + \mathbf{\Sigma}_\mathbf{X}^{-1} \big)^{-1} \nonumber \\
     & = \big( \Re\big\{\mathbf{A}\herm \left(\mathbf{\Sigma}_{\mathbf{N}} \big/ 2\right)^{-1}  \mathbf{A}\big\} + \mathbf{\Sigma}_\mathbf{X}^{-1} \big)^{-1}
    \label{eq:nota_cond_var_realX}
    .
\end{align}
The noise power in \eqref{eq:nota_cond_mean_realX} and \eqref{eq:nota_cond_var_realX} is effectively halved.

\subsection{LMMSE Messages}
\label{subsec:LMMSE-Messages}
Eqns.~\eqref{eq:projection_ep_III_1} and~\eqref{eq:projection_ep_III_2} require computing marginals of
\begin{align}
    \mathcal{N}(\mathbf{u}; \mathbf{r}_2, \nu_{U_2}\mathbf{I})  \,
    \mathcal{CN}(\mathbf{w}; \Ad \mathbf{u} + \mathbf{s}, \mathbf{\Sigma}_{\mathbf{N}_1})\, 
    \mathcal{CN}(\mathbf{w}; \mathbf{p}_2, \nu_{W_2}\mathbf{I}) 
    \label{eq:product_factors_ep_III}
\end{align}
with respect to $\mathbf{u}$ and $\mathbf{w}$. The marginal with respect to $\mathbf{u}$ is 
\begin{align}
    \mathcal{N}(\mathbf{u}; \mathbf{r}_2, \nu_{U_2}\mathbf{I}) \,
    \mathcal{CN}\left( \Ad \mathbf{u} + \mathbf{s}; \mathbf{p}_2, \widetilde{\mathbf{\Sigma}}_{\mathbf{N}_1} \right) 
    \label{eq:lmmse_marg_w}
\end{align}
where $\widetilde{\mathbf{\Sigma}}_{\mathbf{N}_1} = \mathbf{\Sigma}_{\mathbf{N}_1} + \nu_{W_2}\mathbf{I}$. Note that~\eqref{eq:lmmse_marg_w} is proportional to $\mathbf{U} | \mathbf{W}=\mathbf{p}_2$ where 
\begin{align}
\mathbf{W} = \Ad \mathbf{U} + \mathbf{s} + \mathbf{N}
\label{eq:surrogate_y_ad_u_appdx}
\end{align}
with $\mathbf{U} \sim \mathcal{N}(\mathbf{r}_2, \nu_{U_2}\mathbf{I})$ and $\mathbf{N} \sim \mathcal{CN}(\mnull{}, \widetilde{\mathbf{\Sigma}}_{\mathbf{N}_1})$. We refer to Appendix~\ref{subsec:LMMSE-Estimation_real}, which shows that the posterior is real and Gaussian with mean (see~\eqref{eq:nota_cond_mean_realX})
\begin{align}
     \vdIIh
     &= \mathbf{Q}_{\mathbf{U}} \big( \Re\big\{\Ad\herm \big(\widetilde{\mathbf{\Sigma}}_{\mathbf{N}_1} \big/ 2\big)^{-1}   (\mathbf{p}_2-\mathbf{s})\big\} + \tfrac{1}{\nu_{U_2}} \mathbf{r}_2  \big) 
   \label{eq:cond_mean_u}
\end{align}
and covariance matrix (see~\eqref{eq:nota_cond_var_realX})
\begin{align} 
     \mathbf{Q}_{\mathbf{U}}
     & = \big( \Re\big\{\Ad\herm \big(\widetilde{\mathbf{\Sigma}}_{\mathbf{N}_1} \big/ 2 \big)^{-1}  \Ad\big\} + \tfrac{1}{\nu_{U_2}}\mident{} \big)^{-1}
     \label{eq:cond_cov_u}
     .
\end{align}
Observe that computing $\mathbf{Q}_{\mathbf{U}}$ requires two matrix inversions. These inversions simplify by applying the approximation \eqref{eq:noise_cov1_approx} to obtain the expression \eqref{eq:sigma_u_giv_y_simp}.
Finally, carrying out the projection~\eqref{eq:projection_ep_III_1} gives
\begin{align}
    \alpha_2 = (1/n') \, \trace\left(\mathbf{Q}_{\mathbf{U}} \right)
    .
    \label{eq:alpha2_cov_u}
\end{align}

We compute the marginal of~\eqref{eq:product_factors_ep_III} with respect to $\mathbf{w}$
by using composite-real representations; see Appendix.~\ref{subsec:LMMSE-Estimation_real}. The marginal is proportional to the density
of $\bar{\mathbf{W}} | \bar{\mathbf{W}}' = \Adbar \bar{\mathbf{r}}_2 + \bar{\mathbf{s}}$ with the composite surrogate channel
\begin{align}
    \bar{\mathbf{W}}' = \bar{\mathbf{W}} + \bar{\mathbf{N}}
    \label{eq:Wbartilde}
\end{align}
where $\bar{\mathbf{W}} \sim \mathcal{N}(\bar{\mathbf{p}}_2, \tfrac{\nu_{W_2}}{2}\mathbf{I})$ and  $\bar{\mathbf{N}} \sim \mathcal{N}(\mnull{}, \nu_{U_2} \Adbar {\Adbar}\tran 
+ \mathbf{\Sigma}_{\bar{\mathbf{N}}_1})$.
Define $\widetilde{\mathbf{\Sigma}}_{\bar{\mathbf{N}}_1} =  \mathbf{\Sigma}_{\bar{\mathbf{N}}_1} + \tfrac{\nu_{W_2}}{2}\mathbf{I}$. The composite-mean of the posterior is calculated from~\eqref{eq:nota_cond_mean_alt}:
\begin{align}
    \hat{\bar{\z}}_{2} 
    &\overset{(a)}= \tfrac{\nu_{W_2}}{2} \widetilde{\mathbf{\Sigma}}_{\bar{\mathbf{N}}_1}^{-1}
    \big( 
    \mident{} -  \Adbar \,
    \mathbf{Q}_{\mathbf{U}} \,
    \Adbar\tran  \widetilde{\mathbf{\Sigma}}_{\bar{\mathbf{N}}_1}^{-1} 
    \big) 
    (\Adbar \bar{\mathbf{r}}_2 + \bar{\mathbf{s}} - \bar{\mathbf{p}}_2) + \bar{\mathbf{p}}_2 \nonumber\\ 
    &\overset{(b)}= 
    \tfrac{\nu_{W_2}}{2} \widetilde{\mathbf{\Sigma}}_{\bar{\mathbf{N}}_1}^{-1} 
    \left(
    \Adbar
    \left\{ 
        \mathbf{Q}_{\mathbf{U}} \Adbar\tran \widetilde{\mathbf{\Sigma}}_{\bar{\mathbf{N}}_1}^{-1} (\bar{\mathbf{p}}_2 - \bar{\mathbf{s}}) \nonumber \right.\right.\\
         &\left.\left. \qquad\quad+
        \big(
            \mident{} - \mathbf{Q}_{\mathbf{U}} \Adbar\tran \widetilde{\mathbf{\Sigma}}_{\bar{\mathbf{N}}_1}^{-1}  \Adbar 
        \big)  \, \bar{\mathbf{r}}_2
    \right\}
    +  \bar{\mathbf{s}}  - \bar{\mathbf{p}}_2 
    \right) + \bar{\mathbf{p}}_2 \nonumber \\
    &\overset{(c)}=  \tfrac{\nu_{W_2}}{2} \widetilde{\mathbf{\Sigma}}_{\bar{\mathbf{N}}_1}^{-1}  
    (\Adbar \hat{\bar{\mathbf{u}}}_2  +  \bar{\mathbf{s}}  - \bar{\mathbf{p}}_2 ) + \bar{\mathbf{p}}_2
    \label{eq:compos_real_mean_w}
\end{align}
where $(a)$ follows by \eqref{eq:matrix-inversion-lemma} and inserting~\eqref{eq:cond_cov_u}, step $(b)$ distributes the product and refactors, and step $(c)$ uses~\eqref{eq:cond_mean_u}. 
One may express~\eqref{eq:compos_real_mean_w} by the complex vector 
\begin{align}
\hat{\z}_{2} =
\nu_{W_2} \widetilde{\mathbf{\Sigma}}_{\mathbf{N}_1}^{-1}  
(\Ad \hat{\mathbf{u}}_2  +  \mathbf{s}  - \mathbf{p}_2 ) + \mathbf{p}_2
.
\label{eq:cplx_mean_w}
\end{align}

Finally, the composite covariance matrix of the posterior follows from~\eqref{eq:nota_cond_var_alt}:
\begin{align}
\mathbf{Q}_{\bar{\mathbf{W}}} &= 
   \tfrac{\nu_{W_2}}{2} \mident{} - \left(\tfrac{\nu_{W_2}}{2}\right)^2 
   \left(\nu_{U_2} \Adbar {\Adbar}\tran 
+ \left(\mathbf{\Sigma}_{\bar{\mathbf{N}}_1} + \tfrac{\nu_{W_2}}{2}  \mident{}\right)\right)^{-1} 
\nonumber
    \\
&\overset{(a)}=   \tfrac{\nu_{W_2}}{2} \mident{} - \left(\tfrac{\nu_{W_2}}{2}\right)^2 
\big( \widetilde{\mathbf{\Sigma}}_{\bar{\mathbf{N}}_1}^{-1} - \widetilde{\mathbf{\Sigma}}_{\bar{\mathbf{N}}_1}^{-1} \Adbar \,
\mathbf{Q}_{\mathbf{U}} \,
\Adbar\tran  \widetilde{\mathbf{\Sigma}}_{\bar{\mathbf{N}}_1}^{-1} \big) 
     \label{eq:cond_cov_w}
\end{align}
where step $(a)$ uses~\eqref{eq:matrix-inversion-lemma} and~\eqref{eq:cond_cov_u}. Finally,~\eqref{eq:projection_ep_III_2} projects onto a CSCG-$\boldsymbol{\mu}$ with variance:
\begin{align}
    \beta_2
    &= (1/m) \, \trace\left\{ \mathbf{Q}_{\bar{\mathbf{W}}} \right\}  \nonumber \\
    &= (1/m) \, \trace\Big\{ \nu_{W_2} \mident{m} -  \nu_{W_2}^2
\big( \widetilde{\mathbf{\Sigma}}_{\mathbf{N}_1}^{-1} \IfTwoCol{\nonumber\\
& \qquad\quad\qquad \qquad} -  
\widetilde{\mathbf{\Sigma}}_{\mathbf{N}_1}^{-1} \Re\left\{\Ad \,
\mathbf{Q}_{\mathbf{U}} \,
\Ad\herm \right\}  \widetilde{\mathbf{\Sigma}}_{\mathbf{N}_1}^{-1} \big) 
\Big\}
\label{eq:beta2_cov_w}
\end{align}
where the last step uses $\widetilde{\mathbf{\Sigma}}_{\bar{\mathbf{N}}_1} = \tfrac{1}{2}(\mident{2} \otimes \widetilde{\mathbf{\Sigma}}_{\mathbf{N}_1})$, the definition of $\Adbar$ and that $\mathbf{Q}_{\mathbf{U}}$ is real-valued.

\renewcommand{\thesectiondis}[2]{\Alph{section}:}
\section{Chi-Square Distributions}
\label{appendix-chi2}
\subsection{Non-Central Chi-Square Distribution}
Consider a sum of $d$ independent RVs:
\begin{align}
    & Z = \sum\nolimits_{i=1}^d X_i^2, && X_i \sim \mathcal{N}(\mu_i, 1).
    \label{eq:appdx_z_sum_gauss}
\end{align}
$Z$ is called a non-central chi-square random variable with $d$ degrees of freedom and non-centrality parameter $\lambda := \sum_{i=1}^d \mu_i^2$. Its density is
\begin{align}
    p_Z(z) = \mathbbm{1}_{z\geq 0}\, \frac{1}{2} e^{-(\lambda + z)/2} (z/\lambda)^{\frac{d}{4}-\frac{1}{2}}  I_{\frac{d}{2}-1}(\sqrt{\lambda z})
    \label{eq:appdx_chi_square}
\end{align}
where $I_{k}(z)$ is the modified Bessel function of the first kind of order $k$ and $\mathbbm{1}_{z\geq 0}$ is the indicator function, equal to $1$ if $z\geq 0$ and zero otherwise. The \textit{scaled} chi-square random variable $U := \gamma Z$, $\gamma>0$, has the density 
\begin{align}
p_U(u) =\gamma^{-1} \, p_Z(u/\gamma)
\label{eq:appdx_scaled_chi_square}
.
\end{align}

\subsection{Generalized Chi-Square Distribution}
Let $Z$ be a chi-square random variable with $d$ degrees of freedom and non-centrality parameter $\lambda$. Consider the sum
\begin{align}
    Y = \gamma Z + \sigma N
    \label{eq:appdx_gchi2_model}
\end{align}
where $\gamma>0$, $\sigma>0$, and $N \sim \mathcal{N}(0,1)$ is independent of $Z$.
$Y$ is called a \textit{generalized} chi-square random variable in~\cite[Eq.~(10)]{davies1973numerical}. The density of $Y$ is the convolution (see~\cite{davies1973numerical})
\begin{align}
    p_Y(y) &= \medint\int_{u \geq 0} p_{\gamma Z}(u) \cdot \mathcal{N}(y; u, \sigma^2) \,\mathrm{d} u 
    \nonumber %
    \\
    &=:  \chig{y}{\gamma}{d}{\lambda}{\sigma}
    .
    \label{eq:appdx_gchi2}
\end{align}
The parameters $d$ and $\lambda$ are the degrees of freedom and non-centrality parameter of $Z$, respectively. Methods to compute the density are given in~\cite{davies1973numerical,das2024newmethods}.

The moment generating function (MGF)  of~\eqref{eq:appdx_gchi2_model} is 
\begin{align}
    M(t) =  \left(1 - 2 \gamma t \right)^{-d/2} \exp\left( \frac{\lambda \gamma t }{1 - 2 \gamma t} +\frac{\sigma^2 t^2}{2}\right)
    \label{eq:xig_mgf}
\end{align}
with domain $t < \frac{1}{2\gamma}$. We use~\eqref{eq:xig_mgf} to approximate~\eqref{eq:appdx_gchi2} in Appendix~\ref{appendix-saddle_point}.

\renewcommand{\thesectiondis}[2]{\Alph{section}:}
\section{Output Denoiser}
\label{appendix-denoising-w1}
The projection~\eqref{eq:projection_ep_I_II-2} considers the non-linear model $\mathbf{Y} = |\mathbf{W}|^{2} + \mathbf{N}$ 
with $\mathbf{W} \sim \mathcal{CN}(\mathbf{p}_1, \nu_{W_1} \mident{})$ and $\mathbf{N} \sim \mathcal{N}(0, \nu_{N_2}\mident{})$. We may compute the second-order statistics component-wise because the model is memoryless. Consider the scalar model
\begin{align}
    Y = |W|^2 + N  
    \label{eq:appdx_scalar_model_y_w_n}
\end{align}
where $W \sim \mathcal{CN}(p, \nu_W)$ and $N \sim \mathcal{N}(0,\nu_{N_2})$. We have
\begin{subequations}
\begin{align}
    \E[W | Y = y] &= \frac{1}{p(y)} \medint\int_{\mathbb{C}} w \, p_{Y|W}(y|w) \, p_W(w) \,\mathrm{d} w  \label{eq:appc-mean} \\
    \E[|W|^2 | Y = y] &= \frac{1}{p(y)} \medint\int_{\mathbb{C}} |w|^2 \, p_{Y|W}(y|w) \, p_W(w) \,\mathrm{d} w . \label{eq:appc-moment2}
\end{align}
\end{subequations}

Now write~\eqref{eq:appdx_scalar_model_y_w_n} as 
\begin{align}
    Y = \frac{\nu_W}{2} \, |W'|^2 + \sqrt{\nu_{N_2}}\, N'
    \label{eq:appdx_y_wp_np}
\end{align}
where $W' \sim \mathcal{CN}(p \sqrt{2/\nu_W} , 2)$ and $N' \sim \mathcal{N}(0, 1)$. Using~\eqref{eq:appdx_gchi2_model}, we see that $Y$ is a generalized chi-square random variable with density  
\begin{align}
    p_Y(y)= \chig{y}{\frac{\nu_W}{2}}{2}{\frac{2|p|^2}{\nu_W}}{\sqrt{\nu_{N_2}} }
    \label{eq:appdx_normalization_const}
\end{align}
where $d =2$, $\gamma = \frac{\nu_W}{2}$, $\lambda = \frac{2|p|^2}{\nu_W}$ and $\sigma = \sqrt{\nu_{N_2}}$. 

\subsection{Conditional Mean }
The integral in \eqref{eq:appc-mean} is
\begin{equation}
\begin{aligned}
    &\medint\int_{0}^{\infty} \medint\int_{0}^{2\pi} r^2 e^{\mathrm{j}\theta} \, \mathcal{N}(y; r^2, \nu_{N_2}) \, \mathcal{CN}(r e^{\mathrm{j}\theta}; p, \nu_{W})  \, \mathrm{d} r \mathrm{d}\theta \\
    &\overset{(a)}= \medint\int_{0}^{\infty}  \frac{2 e^{\mathrm{j} \angle p }}{\nu_{W}}  r^2  \, \mathcal{N}(y; r^2, \nu_{N_2}) \, e^{-\frac{r^2 + |p|^2}{\nu_{W}}} I_1\left(\frac{2r|p|}{\nu_W}\right)  \, \mathrm{d}r \\
    &\overset{(b)}= 
    e^{\mathrm{j} \angle p }    \medint\int_{0}^{\infty}  \mathcal{N}(y; u, \nu_{N_2}) \, \frac{\sqrt{u}}{\nu_W}   \, e^{-\frac{u + |p|^2}{\nu_{W}}} I_1\left(\frac{2 \sqrt{u} |p|}{\nu_W}\right)  \, \mathrm{d} u
    \label{eq:appdx_cme_deriv}
\end{aligned}
\end{equation}
where step $(a)$ uses~\cite[Eq.~(35)]{schniter2012phase} and step $(b)$ $u = r^2$. 

Consider a chi-square random variable $Z$ with $d=4$ degrees of freedom and non-centrality parameter $\lambda = 2|p|^2/\nu_W$. The density of $U = \gamma Z$ where $\gamma = \nu_W/2$, is (see~\eqref{eq:appdx_scaled_chi_square}): 
\begin{align}
    p_U(u)  = \mathds{1}_{u\geq 0}\; \frac{1}{|p|} \frac{\sqrt{u}}{ \nu_W}  e^{-\frac{u + |p|^2}{\nu_W}} I_1\left(\frac{2 \sqrt{u} |p|}{\nu_W}\right)
    .
    \label{eq:nxi_d4}
\end{align}
The integral in~\eqref{eq:appdx_cme_deriv} is the scaled convolution of~\eqref{eq:nxi_d4} with $\mathcal{N}(u; 0, \nu_{N_2})$. We may thus write~\eqref{eq:appdx_cme_deriv} as 
\begin{align}
    p \cdot \chig{y}{\frac{\nu_W}{2}}{4}{\frac{2|p|^2}{\nu_W}}{\sqrt{\nu_{N_2}} }
    .
\end{align}
Taking~\eqref{eq:appdx_normalization_const} into account, one obtains the conditional mean
\begin{align}
    \E[W | Y = y]
    = p \,
    \frac{
    f_{\chi,4}(y)
    }{ 
    f_{\chi,2}(y)
    }
    \label{eq:z_den_mean_final}
\end{align}
where we abbreviate $f_{\chi,d}(y) = f_{\chi,d}\left(y; \, \frac{\nu_W}{2}, \, \frac{2|p|^2}{\nu_W}, \,\sqrt{\nu_{N_2}}\right)$. 

\subsection{Conditional Variance }
The integral in \eqref{eq:appc-moment2} is 
\begin{align}
    &\medint\int_{0}^{\infty} r^3  \, \mathcal{N}(y; r^2, \nu_{N_2}) \, \medint\int_{0}^{2\pi}  \mathcal{CN}(r e^{\mathrm{j}\theta}; p, \nu_{W})  \, \mathrm{d} r \mathrm{d}\theta \nonumber \\
    &\overset{(a)}= \medint\int_{0}^{\infty}  \frac{2}{\nu_{W}}  r^3  \, \mathcal{N}(y; r^2, \nu_{N_2}) \, e^{-\frac{r^2 + |p|^2}{\nu_{W}}} I_0\left(\frac{2r|p|}{\nu_W}\right)  \, \mathrm{d}r \nonumber \\
    &\overset{(b)}= \medint\int_{0}^{\infty}  \frac{u}{\nu_{W}} \, \mathcal{N}(y; u, \nu_{N_2}) \, e^{-\frac{u + |p|^2}{\nu_{W}}} \bigg(\frac{\nu_W}{\sqrt{u} |p|} I_1\left(\frac{2 \sqrt{u}|p|}{\nu_W}\right) \IfTwoCol{\nonumber  \\
    &\qquad} + I_2\left(\frac{2 \sqrt{u}|p|}{\nu_W}\right)\bigg)  \, \mathrm{d}u
     \label{eq:appdx_cvar_deriv}
\end{align}
where step $(a)$ uses~\cite[Eq.~(26)]{schniter2012phase} and step $(b)$ substitutes $u = r^2$ and $I_0(x) =  \frac{2}{x} I_1(x) + I_2(x)$. Using similar steps as above, $\var[W | Y = y]$ simplifies to
\begin{align}
    \frac{
    \nu_W \, f_{\chi,4}(y) + |p|^2   \, f_{\chi,6}(y)
    }{
    f_{\chi,2}(y)
    } - \big|\E[W | Y = y]\big|^2
    \label{eq:z_den_var_final}
\end{align}
where we abbreviate $f_{\chi,d}(y) = f_{\chi,d}\left(y; \, \frac{\nu_W}{2}, \, \frac{2|p|^2}{\nu_W}, \,\sqrt{\nu_{N_2}}\right)$. 

We remark that for $\nu_{N_2} \rightarrow 0$ one may apply the sifting property of the delta function to  simplify~\eqref{eq:appdx_cme_deriv} and~\eqref{eq:appdx_cvar_deriv}.

\renewcommand{\thesectiondis}[2]{\Alph{section}:}
\section{Saddlepoint Approximation}
\label{appendix-saddle_point}
We use a saddle point approximation for $f_{\chi,d}(y; \, \gamma, \lambda, \sigma)$ in \eqref{eq:appdx_gchi2} (see~\cite{daniel1954saddle,davies1973numerical}):
\begin{align}
    \hat{f}_\chi(y) = \frac{1}{\sqrt{2 \pi K''\big(\hat{t}\big)}} \exp\left( K\big(\hat{t}\big) - \hat{t} \, y \right)
\end{align}
where $K(t) := \log M(t)$ is the cumulant generating function, and $M(t)$ is the MGF~\eqref{eq:xig_mgf}. The saddle point $\hat{t}$ is the solution of $K'(t) = y$. The first and second derivatives of $K(t)$ are: 
\begin{subequations}
\begin{align}
    K'(t) &= \frac{2 \gamma^2 \lambda t}{(1 - 2 \gamma t)^2} +  \frac{\gamma (\lambda+d)}{1 - 2 \gamma t}  + t \nu_{N_2} \\
    K''(t) &= \frac{ 8 \gamma^3 \lambda t}{(1 - 2 \gamma t)^3} +  \frac{2 \gamma^2 (2 \lambda+d)}{(1 - 2 \gamma t)^2}  + \nu_{N_2} .
\end{align}
\end{subequations}
Solving $K'(t) - y = 0$  requires finding the roots of a cubic function. Cardano's formula gives a single real root 
$\hat{t}$ in the domain of~\eqref{eq:xig_mgf}, i.e., $\hat{t} < 1/(2\gamma)$.

Fig.~\ref{fig:spa} plots an example. The saddle-point approximation matches the true density in the tails but deviates slightly in the center.

\begin{figure}[!t]
    \centering
    \input{sub/appdx_spa_figure}
    \caption{Exact PDF and saddle point approximation of the generalized chi-square RV~\eqref{eq:appdx_gchi2_model} with $d=2$ and $\gamma = \lambda= \sigma=1$.}
    \label{fig:spa}
\end{figure}

\end{appendices}

\bibliographystyle{IEEEtran}
\bibliography{IEEEabrv,lit}
\end{document}